%% file: EXOT-2016-16-PAPER.tex
\pdfoutput=1
\newcommand*{\ATLASLATEXPATH}{atlaslatex-05-06-00/latex/}
\documentclass[UKenglish,cernpreprint,texlive=2016]{\ATLASLATEXPATH atlasdoc}

\usepackage[backend=biber]{\ATLASLATEXPATH atlaspackage}
\usepackage{multirow}
\usepackage{\ATLASLATEXPATH atlasbiblatex}

\usepackage{\ATLASLATEXPATH atlascontribute}

\usepackage{\ATLASLATEXPATH atlasphysics}
\usepackage{siunitx}
\makeatletter
\DeclareOldFontCommand{\rm}{\normalfont\rmfamily}{\mathrm}
\DeclareOldFontCommand{\sf}{\normalfont\sffamily}{\mathsf}
\DeclareOldFontCommand{\tt}{\normalfont\ttfamily}{\mathtt}
\DeclareOldFontCommand{\bf}{\normalfont\bfseries}{\mathbf}
\DeclareOldFontCommand{\it}{\normalfont\itshape}{\mathit}
\DeclareOldFontCommand{\sl}{\normalfont\slshape}{\@nomath\sl}
\DeclareOldFontCommand{\sc}{\normalfont\scshape}{\@nomath\sc}
\makeatother

\graphicspath{{atlaslatex-05-06-00/logos/}{figures/}}

\def\gSM{\ensuremath{g_{\text{SM}}}}
\def\gDM{\ensuremath{g_{\text{DM}}}}

\def\C4t{\ensuremath{C_{4t}}}

\newcolumntype{C}[1]{>{\centering\let\newline\\\arraybackslash\hspace{0pt}}m{#1}}

\input{EXOT-2016-16-PAPER-metadata}

\hypersetup{pdftitle={ATLAS document},pdfauthor={The ATLAS Collaboration}}

\begin{document}

\maketitle


\section{Introduction}
\label{sec:intro}

One of the primary goals of the ATLAS experiment at the CERN Large Hadron Collider (LHC) is to search for physics beyond the Standard Model (BSM).  The existence of dark matter, the matter--antimatter asymmetry of the universe, and the high degree of fine tuning required to stabilise the Higgs boson mass at 125~\GeV{} are among the motivations for the existence of BSM physics. In this analysis, events with two leptons of the same electric charge or three leptons and at least one jet identified as originating from a $b$-hadron are considered.  This is a promising final state to search for new phenomena, since the backgrounds from known processes are small.  In addition to the lepton requirements, kinematic criteria are imposed to select events containing objects with large transverse momenta to further suppress the background.  After applying these criteria, the largest background sources are $t\bar{t}W$, $t\bar{t}Z$, $t\bar{t}H$, and diboson production. There is also substantial background from events that appear to have the targeted final state only because one or more objects is misidentified. Three potential BSM sources of events in this final state are considered: production of vector-like quarks (VLQ), anomalous four-top-quark production ($t\bar{t}t\bar{t}$), and same-sign top-quark pair production ($tt$).  Four-top-quark production in the context of the Standard Model (SM) is also studied, since this process has not yet been observed.
Throughout this paper, `lepton' is taken to mean electron or muon and is denoted by $\ell$ in formulae and tables, and a particular set of electrons and muons in the final state is referred to as a `lepton flavour combination'. 

This final state represents one of the  most sensitive channels for VLQ searches with a top quark involved in the decay, especially for masses below $\SI{1}{TeV}$, and is also one of the most sensitive channels for four-top-quark production. An earlier ATLAS analysis using this signature at $\sqrt{s} = 8$~\TeV{}~\cite{EXOT-2013-16} placed limits on the models considered in this paper, including $m_B > 0.62$ \TeV{} and $m_T > 0.59$ \TeV{} in the context of the singlet model of Ref.~\cite{AguilarSaavedra:2009es}, where $B$ and $T$ indicate the VLQ with the same charges as the SM $b$- and $t$-quarks, respectively.  That analysis also placed an  upper limit of $70$ fb on the cross-section of four-top-quark production with SM kinematics. Limits on same-sign top-quark pair production were also set; in the context of a flavour-changing neutral current (FCNC) model with a mediator similar to a Higgs boson of mass 125~\GeV{}  the cross-section for $uu \rightarrow tt$ was found to be $< 35$ fb.  In addition, there have been prior searches for BSM effects in similar final states at $\sqrt{s} = 13$~\TeV{}: Ref.~\cite{Aaboud:2017dmy} reports an ATLAS search in the context of supersymmetric (SUSY) models, and Ref.~\cite{Sirunyan:2017uyt} reports a search by the CMS Collaboration where SUSY models and other BSM models are considered.  
The CMS Collaboration performed searches for pair production of the vector-like $T_{5/3}$ quark, which has charge $5/3$, using events with either a single lepton or a same-charge lepton pair~\cite{Sirunyan:2017jin}, resulting in limits of $m_{T_{5/3}} > 1.02$~\TeV{} (0.99~\TeV{}) for right-handed (left-handed) couplings.  The CMS Collaboration reported on a search for SM four-top-quark production in Ref.~\cite{Sirunyan:2017roi}, resulting in a measured cross-section of $16.9^{+13.8}_{-11.4}$ fb and a limit on the Yukawa coupling between the top quark and the Higgs boson of less than 2.1 times its expected SM value.

\section{Signals considered}
\label{sec:signals}

\subsection {Vector-like $T$, $B$, and $T_{5/3}$ quarks}
\label{subsec:vlq}

\begin{figure}[!ht]
  \centering
    \subfloat[]{\label{fig:VLT}  \includegraphics[height=3.2cm]{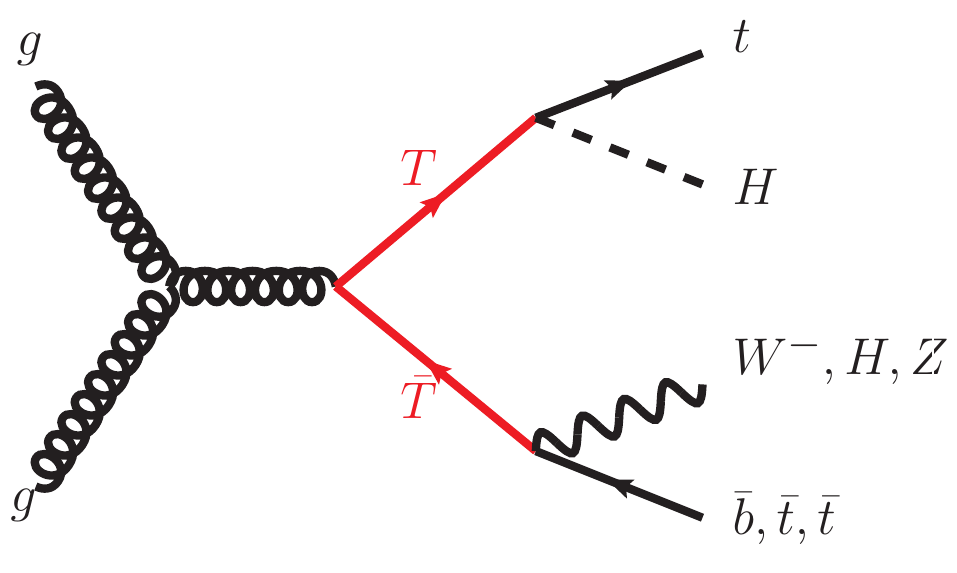}} \hfill
    \subfloat[]{\label{fig:T53sp}\includegraphics[height=3.2cm]{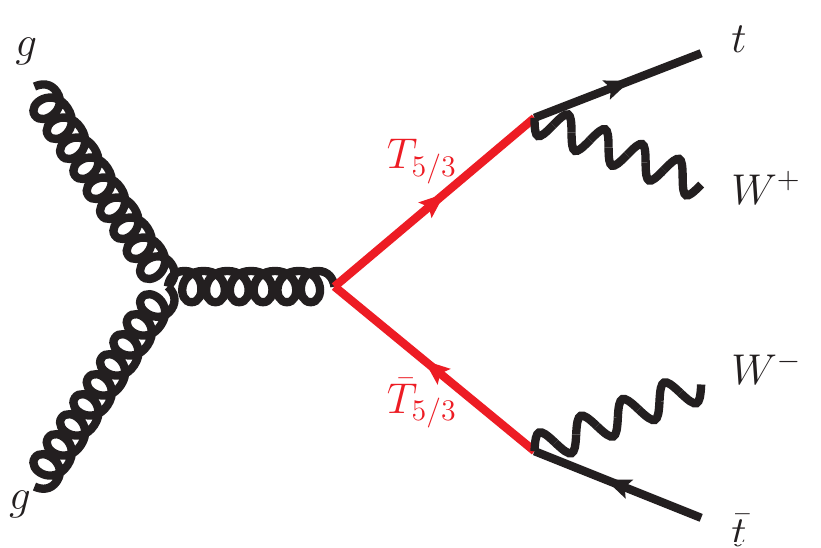}} \hfill
     \subfloat[]{\label{fig:VLB}  \includegraphics[height=3.2cm]{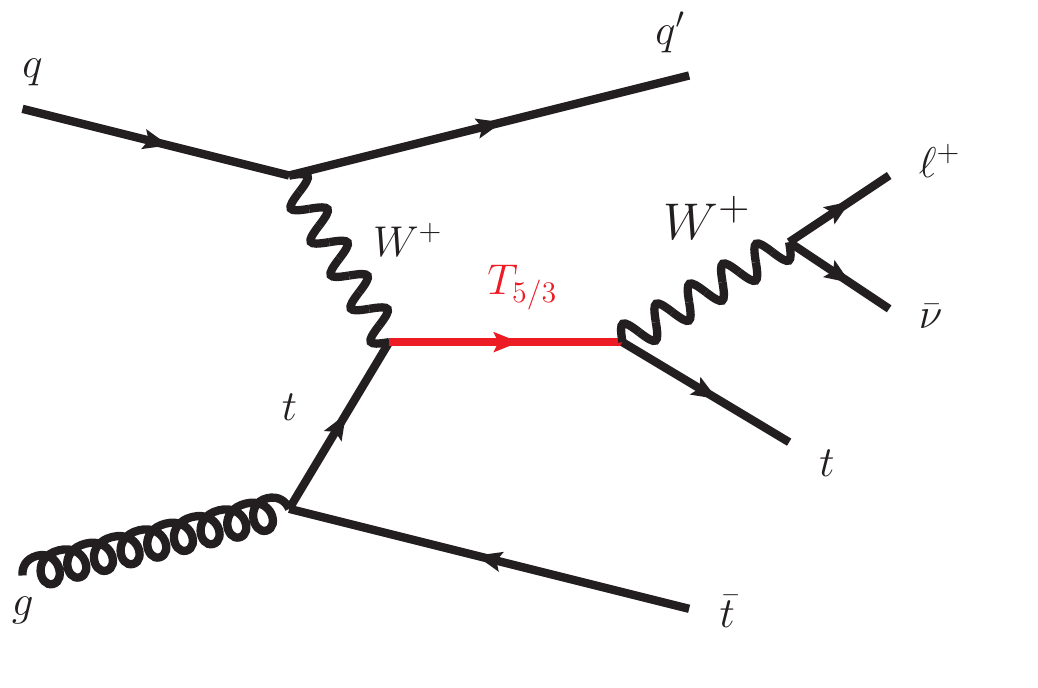}} \hfill
    \caption{Three examples of VLQ production with (a) pair-produced $T$, (b) pair-produced $T_{5/3}$, and (c) singly produced $T_{5/3}$.}
  \label{fig:VLQ_processes}
\end{figure}

Vector-like quarks are fractionally charged, coloured fermions whose right- and left-handed components transform identically under weak isospin.  Their existence is predicted in many BSM models that address the Higgs boson mass fine-tuning problem~\cite{delAguila:1982fs, Dobrescu:1997nm,Hill:2002ap,Agashe:2004rs, Anastasiou:2009rv,PhysRevLett.60.1813,PhysRevD.41.1286,ArkaniHamed:2002qy,Schmaltz:2005ky,Matsumoto:2008fq}. VLQ may come in several varieties, including the aforementioned $B$-, $T$-, and $T_{5/3}$-quarks and the $B_{-4/3}$-quark that has charge $-4/3$.\footnote{The $B_{-4/3}$ quark can only decay into $Wb$, and therefore pair-production of these quarks does not result in same-charge prompt lepton pairs.  Thus the $B_{-4/3}$ quark is not considered in this analysis.} They may appear as singlets, doublets, or triplets under SU(2). In many models, the VLQ couple predominantly to third-generation SM quarks in order to address the naturalness problem, mostly driven by the couplings between the top quark and the Higgs boson~\cite{AguilarSaavedra:2009es}. Therefore, in this paper it is assumed that couplings to first- and second-generation SM quarks are negligible.  Several production and decay scenarios could lead to an enhanced rate of multilepton events~\cite{ Contino:2008hi, AguilarSaavedra:2009es, Kong:2011aa,}. The $B$- and $T$-quarks could decay via both the charged and neutral current channels: $B \rightarrow Wt, Hb, Zb$, and $T \rightarrow Wb, Ht, Zt$, with model-dependent branching ratios.  The most likely scenarios resulting in same-charge lepton pair or trilepton production are
\begin{itemize}
  \item $B\bar{B} \rightarrow W^-tW^+\bar{t} \rightarrow W^-W^+bW^+W^-\bar{b} $
  \item $B\bar{B}\rightarrow W^-tZ\bar{b} \rightarrow W^-W^+bZ\bar{b} $
  \item $T\bar{T}\rightarrow ZtZ\bar{t} \rightarrow ZW^+bZW^-\bar{b} $
  \item $T\bar{T}\rightarrow ZtH\bar{t} \rightarrow ZW^+bHW^-\bar{b} $
\end{itemize}
where two or more of the vector bosons decay leptonically.
Results are given for the SU(2) singlet models of Ref.~\cite{AguilarSaavedra:2009es}, as well as in a model-independent framework where all branching ratios are considered.   The only decay mode of the $T_{5/3}$ quark is into $W^+t \rightarrow W^+W^+b$.  If both $W$ bosons decay leptonically, then a same-charge lepton pair is produced from a single $T_{5/3}$ decay.  Therefore, results for the $T_{5/3}$ are presented for both pair and single production.  The single production results depend on the assumed strength of the $T_{5/3}tW$ coupling. Figure~\ref{fig:VLQ_processes} shows typical Feynman diagrams leading to the signature considered in this paper.

\subsection{Four-top-quark production}
\label{subsec:4top}
\begin{figure}[!ht]
  \centering
    \subfloat[]{\label{fig:4topCI}\includegraphics[height=3.4cm]{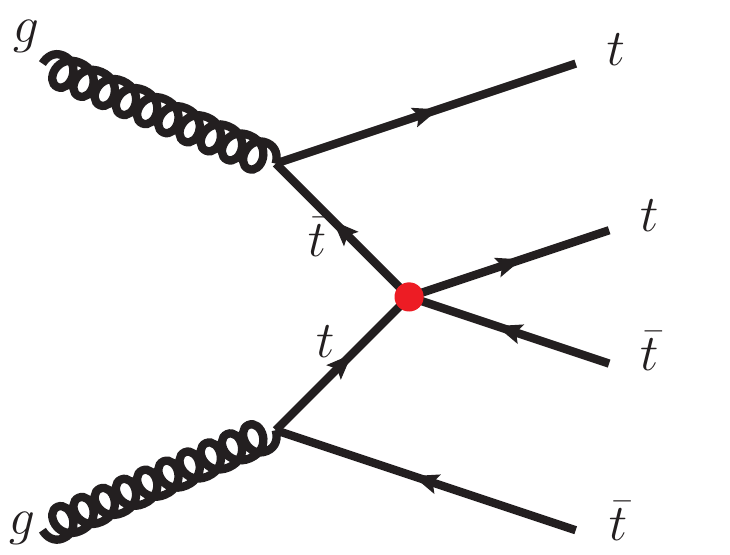}} \hfill
    \subfloat[]{\label{fig:4top2UED}\includegraphics[height=3.7cm]{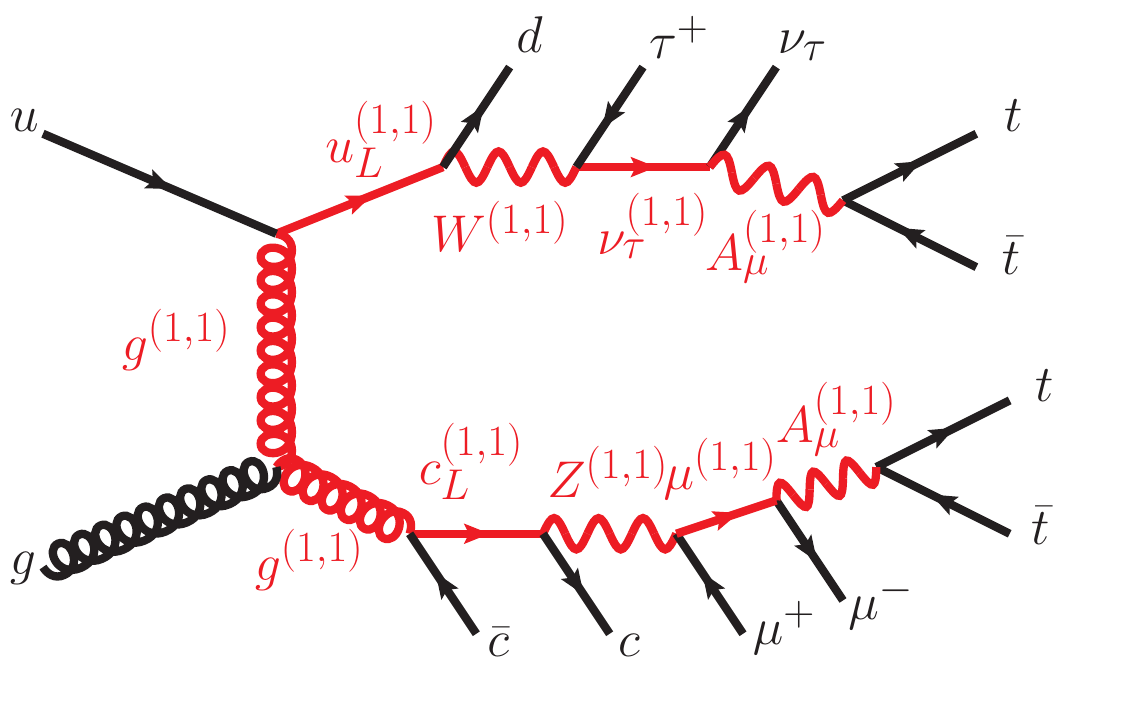}} \hfill
    \subfloat[]{\label{fig:4top2HDM}\includegraphics[height=3.4cm]{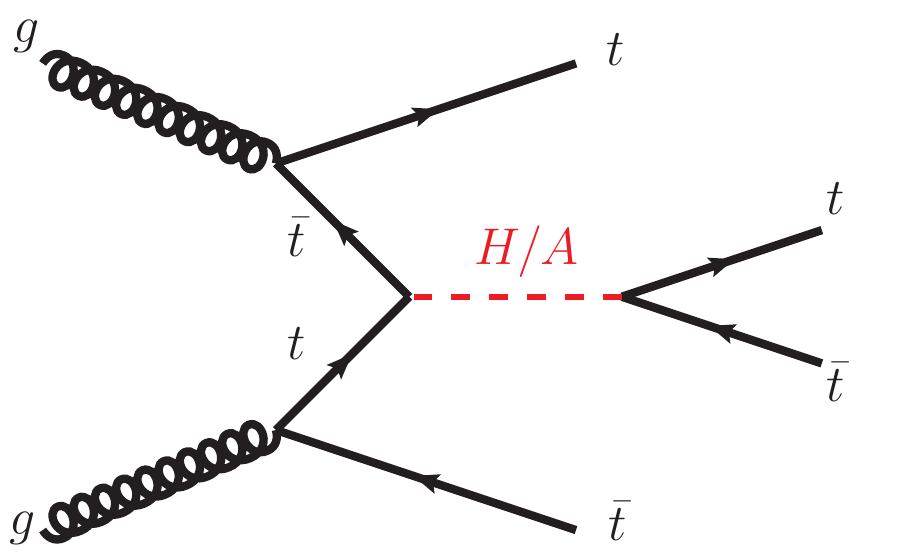}}
    \caption{Three examples of four-top-quark production in the context of (a) a four-fermion contact interaction (CI), (b) two compactified
      universal extra-dimensions (2UED), and (c) two-Higgs-doublet model (2HDM).}
  \label{fig:4top_processes}
\end{figure}

Four-top-quark production is expected to occur in the SM with a next-to-leading-order cross-section of \SI{9.2}{\fb} at $\sqrt{s} = 13$ TeV~\cite{Alwall:2014hca} and leads to a same-charge lepton pair or a trilepton final state with a branching ratio of $12.1\%$, including leptonically decaying $\tau$-leptons. In addition, the four-top-quark production rate could be enhanced in several BSM scenarios. Three benchmarks are considered in this paper. The first is based on an effective field theory (EFT) approach where the BSM contribution is represented via a contact interaction (CI) independently of the details of the underlying theory: \\
\begin{linenomath*}
\begin{equation*}
\label{eq:lagrangian4tops}
{\cal L}_{4t} = \frac{\C4t}{\Lambda^2} \left(\bar{t}_R \gamma^\mu t_R\right)\left(\bar{t}_R \gamma_\mu t_R\right)
\end{equation*}
\end{linenomath*}
where $t_R$ is the right handed top spinor, $\gamma_\mu$ are the Dirac matrices, $\C4t$ is a dimensionless constant and $\Lambda$ is the new-physics energy scale. Only the contact interaction operator with right-handed top quarks is considered as left-handed top operators are already strongly constrained by electroweak precision data \cite{PhysRevD.51.3888}. The four-top-quark production mechanism in this model is shown in Figure~\ref{fig:4topCI}.

The second BSM four-top-quark production model is one with two universal extra dimensions (2UED) that are compactified in the real projective plane geometry (RPP), as described in Ref.~\cite{Lyon09}.  The compactification of the two extra dimensions, characterised by the radii $R_4$ and $R_5$, leads to the discretisation of the momenta along these directions with the allowed values labelled by the integers $i$ and $j$. Each momentum state appears as a particle  called a Kaluza--Klein (KK) excitation with a mass $m$, defined by $(i,j)$ values and later referenced as a `tier'. At leading order, the mass of a KK excitation of a particle with a mass $m_0$ is
\begin{linenomath*}
\begin{equation}
  m^2 = \frac{i^2}{R_4^2} + \frac{j^2}{R_5^2} + m^2_0.
 \label{eq:signals:2UEDmass}
\end{equation}
\end{linenomath*}
The additional mass differences within a given tier $(i,j)$ are due to next-to-leading-order corrections and are small compared with the masses \cite{Lyon09}. By using the notations $m_{\text{KK}} = 1/R_4$ and $\xi = R_4/R_5$, Eq.~\eqref{eq:signals:2UEDmass} reads as
\begin{linenomath*}
\begin{equation*}
  m^2 = m^2_{\text{KK}} \left( i^2 + j^2 \xi^2 \right) + m^2_0.
  \label{eq:signals:2UEDmass2}
\end{equation*}
\end{linenomath*}

The four-top-quark signal of the model considered in this paper arises from pair-produced particles of tier $(1,1)$, which then chain-decay into the lightest particle of this tier, the KK excitation of the photon, $A^{(1,1)}$, by emitting SM particles~\cite{Cacciapaglia:2011kz}, as shown in Figure~\ref{fig:4top2UED}. This heavy photon $A^{(1,1)}$ decays into $\ttbar$ with a branching ratio assumed to be $100\%$. Therefore, additional quarks and leptons are expected to be produced in association with the four-top-quark system, which makes this signature quite different from the other considered benchmarks, as shown in Figure~\ref{fig:4top_processes}. In addition, cosmological observations constrain $m_{\text{KK}}$ between 600~\GeV{} and 1000~\GeV~\cite{Cacciapaglia:2011kz, Lyon12}, leading to typical resonance masses between $0.6~\TeV{}$ and $2~\TeV{}$ depending on the ratio $\xi$ of the two compactification radii. This analysis probes different scenarios varying both $m_{\text{KK}}$ and $\xi$, where the four-top-quark signal arises from particles of tier $(1,1)$~\cite{Cacciapaglia:2011kz}. 

The third BSM four-top-quark production model is one with two Higgs doublets $\Phi_1$ and $\Phi_2$ (2HDM),
which spontaneously break the electroweak symmetry $\mathrm{SU(2)}_\mathrm {L} \times \mathrm{U(1)}_\mathrm{Y}$~\cite{Branco:2011iw}.
In this model, $\Phi_{1}$ couples only to down-type quarks and leptons, and $\Phi_{2}$ couples only to up-type quarks and neutrinos~\cite{Dev:2014yca}. The parameter space is  constrained to avoid large FCNC at tree level, resulting in four different sets of Yukawa couplings between the Higgs doublets and SM fermions. Among these, the Type-II 2HDM is considered. Measurements of the properties of the SM Higgs boson constrain all 2HDM types to be in the so-called alignment limit~\cite{Dev:2014yca}, where the mass eigenstates are aligned with the gauge eigenstates in the new scalar sector. In this model, the $t\bar{t}t\bar{t}$ final state arises from the production of heavy neutral Higgs bosons $H$ (scalar) and $A$ (pseudo-scalar) in association with a $t\bar{t}$ pair, with the $H$ or $A$ boson decaying into $t\bar{t}$ as shown in Figure~\ref{fig:4top2HDM}:
\begin{linenomath*}
\begin{equation*}
\label{eq:2HDM4tprocess}
g g \rightarrow t \bar{t} H/A \rightarrow t \bar{t} t \bar{t}.
\end{equation*}
\end{linenomath*}
In the alignment limit, the scalar and the pseudo-scalar Higgs boson have the same mass $m_{H/A}$ and both contribute to the four-top-quark production with similar kinematics.
The cross-section predicted by this model depends on $m_{H/A}$ and the ratio $\tan\beta$ of vacuum expectation values of the two Higgs doublets.
This benchmark is particularly interesting since the four-top-quark kinematics are rather soft compared with the CI signature,
especially at low masses where the direct search for $H/A \to t\bar{t}$ loses sensitivity due to interference effects with
the SM $t\bar{t}$ production~\cite{EXOT-2016-04}.

\subsection{Same-sign top-quark pair production}
\label{subsec:sstop}

\begin{figure}[!ht]
  \centering
    \subfloat[]{\label{fig:exclusive_tt}\includegraphics[height=3.2cm]{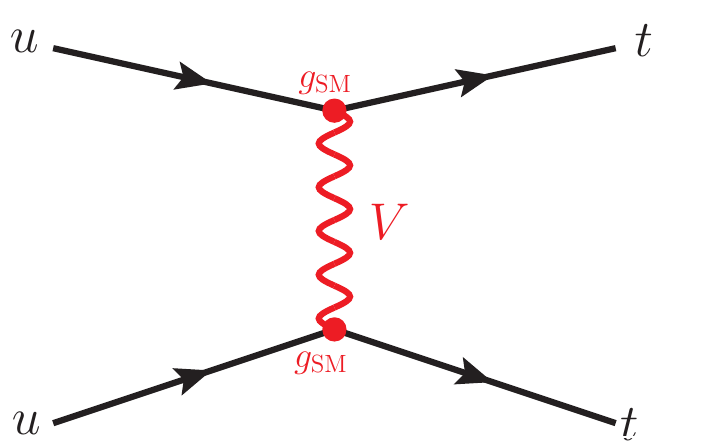}} \hfill
    \subfloat[]{\label{fig:onshellV}\includegraphics[height=3.2cm]{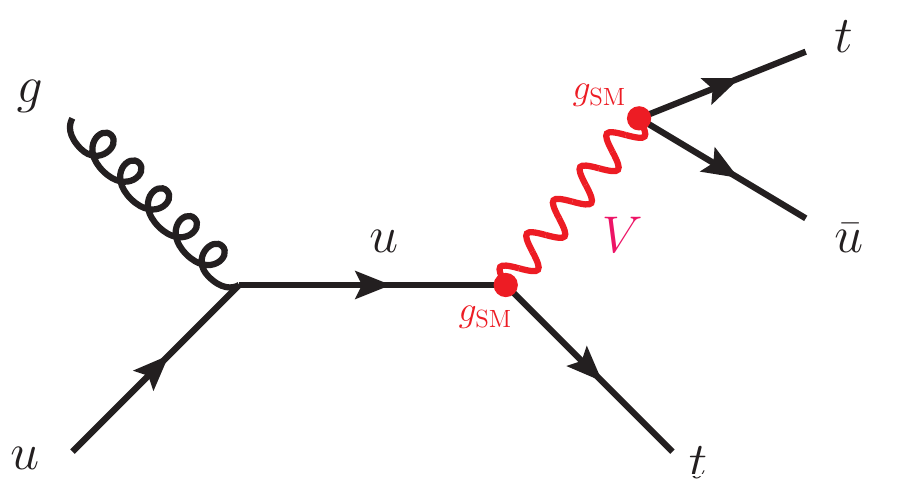}} \hfill
    \subfloat[]{\label{fig:offshellV}\includegraphics[height=3.2cm]{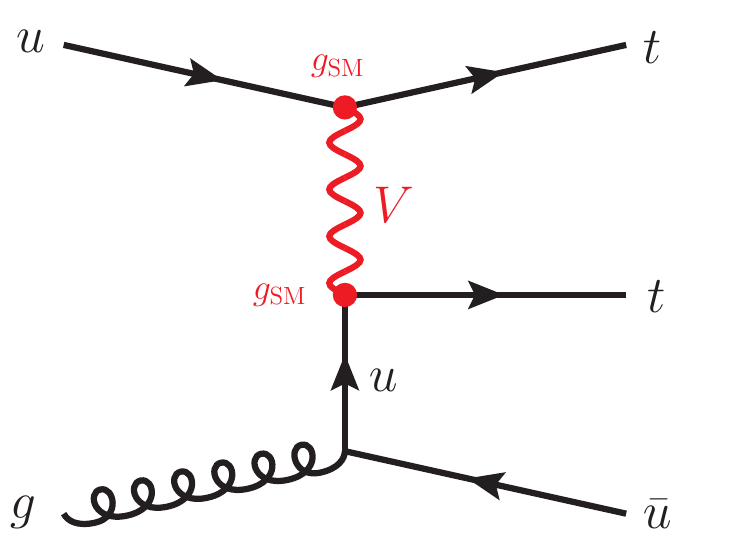}}
    \caption{Three examples of same-sign top quark pair signatures in the context of the dark-matter mediator model: (a) prompt $tt$ production,
      (b) via an on-shell mediator, (c) via an off-shell mediator. The mediator is denoted by $V$ and its coupling to SM particles is denoted by \gSM{}.}
  \label{fig:tt_subprocesses}
\end{figure}

Same-sign top-quark pair production ($tt$) is suppressed to a negligible level in the SM but allowed in BSM models. This signature is distinct from VLQ
or $t\bar{t}t\bar{t}$ production, and is treated separately in the analysis. In particular, only positively charged lepton pairs are
considered for this signal (since $tt$  production has a cross-section higher by a typical factor $100$ than $\bar{t}\bar{t}$ production at the LHC due to the charge asymmetry in the initial state).
The kinematic criteria also differ from those applied in the VLQ and four-top-quark searches. The considered benchmark is a generic dark-matter model relying on an effective theory
invariant under $SU(2)_{L} \times U(1)_{Y}$~\cite{Monotop_theory_2015}. In this model, a top quark is produced
in association with an FCNC mediator $V$ which could then decay into dark-matter $\chi$ or SM particles $t\bar{u}/\bar{t}u$:
\begin{linenomath*}
\begin{equation*}
  \label{eq:Signals:SStopLagrangian}
\mathcal{L}_{\mathrm{DM}} \; = \;  \mathcal{L}_{\mathrm{kin}} [\chi, V_\mu] \; + \; \gSM \: V_\mu \: \bar{t}_R \gamma^\mu u_R 
                          \;+ \; \gDM \: V_\mu \: \bar{\chi} {\gamma^\mu} \chi
\end{equation*}
\end{linenomath*}
where $\gSM$ and $\gDM$ represent the coupling strengths of the mediator to SM and dark-matter particles, respectively, and $\mathcal{L}_{\mathrm{kin}} [\chi, V_\mu]$
represents the kinetic term of the mediator and the dark-matter fields.
The $tt$ final state could arise if $V$ couples to the top quark, in both the $t$- and $s$-channels, leading to the three processes
shown in Figure~\ref{fig:tt_subprocesses}, with a relative contribution which depends on the total width of the mediator.

The results are interpreted for each process in Figure~\ref{fig:tt_subprocesses} independently, allowing constraints to be placed on generic FCNC via the process $uu\to tt$ as well as specific resonances decaying into $t\bar{u}$. The results are also interpreted for different values of the mass of the mediator $m_V$, $\gSM$, and $\gDM$, taking into account width effects.
This provides additional sensitivity to dark-matter mediators
when its branching ratio into SM particles is sizeable,\footnote{This is a realistic scenario since the mediator must have visible partial width
  in order to be produced in proton--proton collisions.} where a direct search based on final states with missing transverse energy and a top quark~\cite{1801.08427}
might lose sensitivity.

\section{ATLAS detector}
\label{sec:detector}
The ATLAS detector~\cite{PERF-2007-01} at the LHC covers nearly the entire solid angle around the collision point.\footnote{ATLAS uses a right-handed coordinate system with its origin at the nominal interaction point (IP) in the centre of the detector and the $z$-axis along the beam pipe. The $x$-axis points from the IP to the centre of the LHC ring, and the $y$-axis points upward. Cylindrical coordinates $(r,\phi)$ are used in the transverse plane, $\phi$ being the azimuthal angle around the $z$-axis. The pseudorapidity is defined in terms of the polar angle $\theta$ as $\eta=-\ln\tan(\theta/2)$, and the rapidity $y$  is defined as $y = {1 \over 2} \ln { E + p_z \over E - p_z }$.}
It consists of an inner tracking detector surrounded by a thin superconducting solenoid, electromagnetic and hadronic calorimeters,
and a muon spectrometer incorporating three large superconducting toroidal magnets.
The inner-detector system is immersed in a \SI{2}{\tesla} axial magnetic field 
and provides charged-particle tracking in the range $|\eta| < 2.5$.

A high-granularity silicon pixel detector covers the vertex region and typically provides four three-dimensional measurements per track, 
the innermost being in the  
insertable B-layer~\cite{Capeans:1291633}.
It is followed by a silicon microstrip tracker, which provides four two-dimensional measurement points per track.
These silicon detectors are complemented by a transition radiation tracker,
which enables radially extended track reconstruction up to $|\eta| = 2.0$. 
The transition radiation tracker also provides electron identification information 
based on the fraction of hits (typically 30 in total) above a higher energy-deposit threshold corresponding to transition radiation.

The calorimeter system covers the pseudorapidity range $|\eta| < 4.9$.
Within the region $|\eta|< 3.2$, electromagnetic calorimetry is provided by barrel and 
endcap high-granularity lead/liquid-argon (LAr) sampling calorimeters,
with an additional thin LAr presampler covering $|\eta| < 1.8$
to correct for energy loss in material upstream of the calorimeters.
Hadronic calorimetry is provided by a steel/scintillator-tile calorimeter,
segmented into three barrel structures within $|\eta| < 1.7$, and two copper/LAr hadronic endcap calorimeters.
The solid angle coverage is completed with forward copper/LAr and tungsten/LAr calorimeter modules
optimised for electromagnetic and hadronic measurements, respectively.

The muon spectrometer comprises separate trigger and
high-precision tracking chambers measuring the deflection of muons in a magnetic field generated by the superconducting air-core toroidal magnets.
The field integral of the toroidal magnets ranges between \num{2.0} and \SI{6.0}{\tesla\metre}
across most of the acceptance. 
A set of precision chambers covers the region $|\eta| < 2.7$ with three layers of monitored drift tubes,
complemented by cathode strip chambers in the forward region, where the background is highest.
The muon trigger system covers the range $|\eta| < 2.4$ with resistive plate chambers in the barrel, and thin gap chambers in the endcap regions.

The ATLAS detector has a two-level trigger system to select events for offline analysis~\cite{Aaboud:2016leb}. The first-level trigger is implemented in hardware and uses a subset of detector information
to reduce the event rate to a design value of \SI{100}{\kHz}.
This is followed by a software-based high-level trigger which  reduces the event rate to about \SI{1}{\kHz}.

\section{Data sample and trigger requirements}
\label{sec:datatrig}
The data were recorded in LHC proton--proton ($pp$)
collisions at $\sqrt{s} = 13$ \TeV{} in 2015 and 2016, 
corresponding to an integrated luminosity of $36.1 \pm 0.8$~\ifb{}.  The luminosity and its uncertainty are
 derived, following a methodology similar to that detailed in Ref.~\cite{Aaboud:2016hhf}, from a calibration
of the luminosity scale using $x$ -- $y$ beam-separation scans.
In this dataset the average number of simultaneous $pp$ interactions per bunch crossing in addition to the triggered hard-scatter interaction, pile-up, was approximately 24.
Data-quality requirements were applied to ensure that events were selected only from periods where all subdetectors were operating at nominal conditions, and where the LHC beams were in stable-collision mode.  The events used in the analysis were required to have at least one primary vertex formed from at least two charged-particle tracks with transverse momentum $\pt > 0.4$~\GeV{}, and to have been triggered either by two leptons or a single high-\pt lepton. Only triggers with loose lepton quality and isolation requirements were used, since tight requirements at the trigger level would complicate the estimation of the background originating from fake or non-prompt leptons.  The dilepton triggers provide sensitivity at low lepton \pt values, and the single-lepton triggers provide additional efficiency for high-\pt leptons.  The \pt\ thresholds for the dilepton triggers varied from 8 to 24~\GeV{} depending on the lepton flavours and the year in which the event was recorded.  The single-muon trigger had a \pt threshold of 50~\GeV{}; the corresponding single-electron trigger had a \pt threshold of 24~\GeV{} for data recorded in 2015 and 60~\GeV{} for data recorded in 2016.  The trigger efficiency depends on the lepton flavour combination, but in all cases is $>95\%$ for events of interest in this analysis.

\section{Object selection criteria}
\label{sec:obj}
This analysis makes use of reconstructed electrons, muons, jets, $b$-tagged jets, and missing transverse momentum. 
Their selection is described in this section and summarised in Table~\ref{tab:Object:Summary}.

Electrons are reconstructed from clusters of energy deposits in  electromagnetic calorimeter cells with a matching inner detector track~\cite{ATL-PHYS-PUB-2015-041}. The candidate electrons are required to have $\pt > 28~\GeV$ and be in the $|\eta| < 2.47$ region, excluding the transition region between the barrel and endcap calorimeters ($1.37 < |\eta| < 1.52$). For events with two electrons or one electron and one muon, electrons with $|\eta| > 1.37$ are not considered since such events are subject to backgrounds from electron charge misidentification, which has a substantially higher probability of occurring for electrons at high $|\eta|$, as detailed in Section~\ref{sec:bkg}.
Muons are reconstructed from tracks in the muon spectrometer and inner detector~\cite{PERF-2015-10}. They must have $\pt > 28~\GeV$ and $|\eta| < 2.5$. 

Electrons and muons are required to be consistent with originating from the primary event vertex, using the quantities 
$|d_0/\sigma_{d_0}|$, where $d_0$ is the impact parameter relative to the primary vertex in the $x$--$y$ plane and $\sigma_{d_0}$ is its uncertainty, and $|z_0\sin\theta|$, where $z_0$ is the difference between the $z$ coordinate of the point of closest approach of the lepton track to the beamline and the $z$ coordinate of the primary vertex. Electrons are required to satisfy $|d_0/\sigma_{d_0}| < 5$ and $|z_0\sin\theta| $ < 0.5~mm, while muons are required to satisfy $|d_0/\sigma_{d_0}| < 3$ and $|z_0\sin\theta |$ < 0.5~mm.  

All leptons are required to satisfy either relaxed or nominal identification criteria.  The nominal sample, which is a subset of the relaxed sample, is used in the final analysis, and the relaxed sample is used to estimate one component of the reducible background as described in Section~\ref{sec:bkg}.  
For electrons, the relaxed (nominal) selection requires that the electron satisfies the likelihood medium (tight) requirements defined in Ref.~\cite{ATL-PHYS-PUB-2015-041}, while for muons both the relaxed and nominal selections require that the muon satisfies the medium criteria defined in Ref.~\cite{PERF-2015-10}.  
Nominal leptons are required to be isolated from other activity in the detector: the scalar sum of the \pt of tracks within a variable-size cone around the lepton (excluding its own track), must be less than 6\% of the lepton \pt. The track isolation cone size for electrons (muons) $\Delta R=\sqrt{(\Delta\eta)^2+(\Delta\phi)^2}$ is given by the smaller of $\Delta R = \SI{10}{~GeV}/\pt$ and $\Delta R = 0.2\,(0.3)$. In addition, in the case of electrons the sum of the transverse energy of the calorimeter energy clusters in a cone of $\Delta R = 0.2$ around the electron (excluding the energy from the electron itself) must be less than 6\% of the electron \pt.

Jets are reconstructed from clusters of energy in the calorimeter using the anti-\kt algorithm~\cite{Cacciari:2008gp} with a radius parameter of 0.4. Jets are considered if $\pt > 25~\GeV$ and $|\eta| < 2.5$. Quality criteria are applied to jets to ensure that they are not reconstructed from detector noise, beam losses, or cosmic rays~\cite{ATLAS-CONF-2015-029}.  If any jet fails to satisfy these criteria, the event is vetoed.
To reject jets from pile-up, an observable called the jet vertex tagger (JVT) is formed by
combining variables that discriminate pile-up jets from hard-scattering jets~\cite{ATLAS-CONF-2014-018}. Jets with \pt~$< 60$~\GeV{} and $|\eta| < 2.4$ that have associated tracks are subject to a requirement on JVT that is $92\%$ efficient for hard-scattering jets while rejecting $98\%$ of pile-up jets.  If such jets have no associated tracks they are removed.
Jets containing a $b$-hadron are identified using a multivariate technique~\cite{PERF-2012-04}.   An operating point is defined by a threshold in the range of discriminant output values, and is chosen to provide specific $b$-, $c$-, and light-jet efficiencies in simulated $t\bar{t}$  events. The operating point used in this analysis has a 77\% $b$-jet efficiency  with rejection factors of 6 and 134 for $c$- and light-jets, respectively. 

The missing transverse momentum is calculated as the negative vectorial sum of the transverse momenta of reconstructed calibrated objects in the event.
Its magnitude is denoted \met, and is computed using electrons, photons, hadronically decaying $\tau$-leptons, jets and muons as well as a soft term calculated with tracks matched to the primary vertex which are not associated with any of these objects~\cite{Aaboud:2018tkc}. 

A set of requirements are applied to resolve overlaps between reconstructed objects.
This procedure is applied to the leptons satisfying the relaxed selection criteria.
In the first step, electrons which share a track with a muon are removed, to avoid cases where  muon radiation would mimic an electron.
Next, the jet closest to an electron within $\Delta R_{y} \equiv \sqrt{(\Delta y)^2 + (\Delta \phi)^2}$ = 0.2 is removed to avoid double counting.
Then, to reduce the contributions from non-prompt electrons originating from heavy-flavour decays, electrons within $\Delta R_{y}$ = 0.4 of any remaining jets are removed.
Finally, the overlap between muons and jets is considered: jets with less than three tracks and within $\Delta R_{y}$ = 0.4 of a muon are removed. Muons are then removed if they are within $\Delta R_{y} = 0.04 + 10~\GeV{}/p_{{\text T}, \mu}$ of remaining jets.

The events are preselected if they contain at least one jet, and at least two leptons that satisfy the nominal selection criteria.   
If exactly two of the three highest-\pt\ leptons satisfy the nominal criteria, they must have the same electric charge, and if these two leptons are electrons, a quarkonia/$Z$-veto is applied to their invariant mass: 
$m_{ee} > 15~\GeV{}$ and $|m_{ee} - 91~\GeV{}| > 10~\GeV{}$.  Events that satisfy these criteria are called `same-charge lepton' events.
If the three highest-\pt\ leptons satisfy the nominal criteria, no requirement is imposed on their charge or on the invariant mass of any pair.  These events are called `trilepton' events.  Same-charge lepton and trilepton events are treated separately in the analysis.

\begin{table}[!h]
  \small
  \begin{center}
    \caption{Summary of object identification and definition. `ID quality' refers to the identification criteria used for each object type.  For electrons, `mediumLH' and `tightLH' refer to the likelihood medium and tight requirements defined in Ref.~\cite{ATL-PHYS-PUB-2015-041}; for muons the criteria for `medium'  ID quality are defined in Ref.~\cite{PERF-2015-10}.  For jets, `cleaning' means applying a  procedure to reduce  contamination from spurious jets~\cite{ATLAS-CONF-2015-029}, and `JVT' means applying criteria to select jets that are consistent with being produced at the primary vertex rather than from pile-up~\cite{ATLAS-CONF-2014-018}.  For $b$-jets, `MVA77' refers to placing a requirement on the multivariate discriminant defined in Ref.~\cite{PERF-2012-04} that is 77\% efficient for $b$-jets in simulated $t\bar{t}$ events.
    }\label{tab:Object:Summary}
    \begin{tabular}{|l|cC{3cm}|cc|c|c|}
      \hline\hline
                              & \multicolumn{2}{c|}{Electrons} & \multicolumn{2}{c|}{Muons} & Jets & $b$-jets \\
                              & relaxed & nominal           & relaxed  & nominal        &     &     \\
      \hline
      $p_{\text{T}}$ [\GeV{}] &      \multicolumn{2}{c|}{$>28$}  &   \multicolumn{2}{c|}{$> 28$} & $>25$ & $>25$ \\
      $|\eta|$         &   \multicolumn{2}{c|}{$<1.37$ or  $1.52$ -- $2.47$}  &  \multicolumn{2}{c|}{$<2.5$} & $<2.5$ & $<2.5$  \\
                            &   \multicolumn{2}{c|}{($< 1.37$ for $ee$ and $e\mu$)} &                &                    &              &             \\
      &&&&&&\\
      ID quality                  & mediumLH & tightLH  &   \multicolumn{2}{c|}{medium}& cleaning   & MVA77 \\
                                       &                    &              &                       &                       &   + JVT           &    \\
      &&&&&&\\
      Isolation        &   none    &  track- and calorimeter-based &  none & track-based &     &  \\
      Track vertex :      &       &             &        &                             &    &    \\
      ~~$-$ $|d_0/\sigma_{d_0}|$ & \multicolumn{2}{c|}{$<5$}  &    \multicolumn{2}{c|}{$<3$}                       &   &    \\
      ~~$-$ $|z_{0}\sin\theta|$ [mm]  & \multicolumn{2}{c|}{$<0.5$}    &  \multicolumn{2}{c|}{$<0.5$}  &   &   \\
      \hline\hline
    \end{tabular}
  \end{center}
\end{table}

\section{Simulation}
\label{sec:sim}

Monte Carlo (MC) simulation was used to model the signals and the irreducible backgrounds.  
EvtGen~v1.2~\cite{EvtGen} was used to model charm and bottom hadron decays for all samples, except those generated with \textsc{Sherpa}~\cite{Gleisberg:2008ta}, and the A14 set of tuned parameters~\cite{ATL-PHYS-PUB-2014-021} was used for all samples unless stated otherwise.

The production of $T\bar{T}$, $B\bar{B}$ and $T_{5/3}\bar{T}_{5/3}$ pairs was modelled by {\sc Protos} v2.2~\cite{AguilarSaavedra:2009es}, with \textsc{Pythia}~v8.186~\cite{Sjostrand:2014zea} for showering and hadronisation,\footnote{Throughout this analysis it is assumed that pair production of vector-like quarks occurs only via QCD.  There are models, such as the model in Ref.~\cite{Chala:2014mma}, that allow additional production mechanisms.} using the NNPDF2.3LO set~\cite{Ball:2012cx} of parton distribution functions (PDF).  VLQ masses from 0.50 to 1.40~\TeV{} were simulated.  Standard Model production of four top quarks was simulated using  {\sc MG5\_aMC@NLO}~v2.2.2~\cite{Alwall:2014hca} with \textsc{Pythia}~v8.186, using the NNPDF2.3LO PDF set.  In the 2UED/RPP scenario, four-top-quark production was modelled with \textsc{Pythia}~v8.186, using the NNPDF2.3LO PDF set.  For the contact interaction model, four-top-quark production was modelled with {\sc MG5\_aMC@NLO}~v2.2.3 and \textsc{Pythia}~v8.205 using the NNPDF2.3LO PDF set.  Same-sign top-quark pair production was modelled by {\sc MG5\_aMC@NLO}~v2.3.3 and \textsc{Pythia}~v8.210 using the NNPDF2.3LO PDF set.

The main sources of irreducible backgrounds are $t\bar{t}V$ production (where $V$ represents either a $W$ or $Z$ boson), $t\bar{t}H$ production, and diboson production. Smaller contributions from triboson, $VH$, $tt\bar{t}$, $t\bar{t}WW$, $tZW$, and $tZ$ production are shown in the tables and figures as `Other bkg'. The SM four-top-quark production is included as a background for all BSM searches, but is considered as the signal in the search for SM four-top-quark production. The matrix elements for $t\bar{t}V$, $t\bar{t}H$, $tt\bar{t}$, $t\bar{t}t\bar{t}$, $t\bar{t}WW$, and $tZW$ production processes were modelled with {\sc MG5\_aMC@NLO}~v2.2.2 and \textsc{Pythia}~v8.186 for hadronisation and showering, using the NNPDF3.0NLO PDF set. Next-to-leading-order (NLO) matrix-element calculation was used for $t\bar{t}V$, $t\bar{t}H$ and $tZW$ while leading-order (LO) calculation was used for $tt\bar{t}$, $t\bar{t}t\bar{t}$ and $t\bar{t}WW$.  The $tZ$ process was modelled by {\sc MG5\_aMC@NLO}~v2.2.2 based on LO matrix-element calculation with \textsc{Pythia} v6.428~\cite{Sjostrand:2006za} for showering and hadronisation. The CTEQ6L1 PDF set~\cite{Pumplin:2002vw} and Perugia2012 set of tuned parameters~\cite{Skands:2010ak} were used. Diboson and triboson production was modelled with the \textsc{Sherpa}~v2.2.1 generator, which uses the Comix~\cite{Gleisberg:2008fv} and OpenLoops~\cite{Schumann:2007mg} matrix-element generators merged with the \textsc{Sherpa} parton shower~\cite{Hoeche:2012yf} using the ME+PS@NLO prescription~\cite{Hoeche:2009rj}. The $VH$ production process was modelled using \textsc{Pythia}~v8.186, with the  NNPDF2.3LO PDF set.  The cross-sections for all processes are calculated at NLO in QCD, except for $tZ$ where the leading-order calculation is used.

Simulated \textsc{Pythia}~v8.186 minimum-bias events were overlaid on each simulated event to model the effects of pile-up; the generated events were then reweighted so that the distribution of the number of interactions per bunch crossing matched the distribution observed in the data. The response of the ATLAS detector for most samples was modelled using {\sc Geant4}~\cite{Agostinelli:2002hh} within the ATLAS simulation infrastructure~\cite{SOFT-2010-01}.  The $tt\bar{t}$, single $T_{5/3}$, and same-sign top-quark pair production samples were processed with a fast simulation that relies on a parameterisation of the calorimeter response~\cite{ATL-PHYS-PUB-2010-013}.  Events were reconstructed using the same algorithms as used for the collider data.  Corrections were applied to the simulated events to account for differences observed in trigger efficiencies, object identification efficiencies and resolutions when comparing the simulation with data.

\section{Estimation of reducible backgrounds}
\label{sec:bkg}

In addition to the irreducible backgrounds described above, there are reducible backgrounds where a jet or lepton from heavy-flavour hadron decay mimics a
 prompt lepton\footnote{Prompt leptons are leptons which do not originate from hadron decays or conversion processes.} (called `fake/non-prompt lepton background' in the following), or the charge of a lepton is misidentified.  These backgrounds are estimated using data-driven techniques.

The fake/non-prompt lepton background yield is estimated with the matrix method~\cite{Abbott:1999tt,ATLAS-CONF-2014-058}, which uses the relaxed and nominal lepton categories defined in Table~\ref{tab:Object:Summary}.  The fraction of prompt leptons satisfying the relaxed
criteria that also satisfy the nominal criteria is referred to as $r$. Similarly,
the fraction of fake/non-prompt leptons satisfying the relaxed requirements that also satisfy the nominal requirements is referred to as $f$.  Using the measured values of $r$ and
$f$, the number of events with at least one non-prompt/fake lepton in the nominal sample can be inferred from the numbers of relaxed and nominal leptons in the relaxed sample, and this number is taken as the fake/non-prompt yield. 
A Poisson likelihood approach is used to estimate the final fake/non-prompt yield and its statistical uncertainty.
This approach guarantees that the estimated yield is not negative, and provides a more reliable estimate of the statistical uncertainty in regions with a small number of selected events.

Single-lepton control regions enriched in prompt and fake/non-prompt leptons are used to measure $r$ and $f$.  The criteria used to select the single-lepton
events are different for electrons and muons due to the different
sources of fake/non-prompt leptons for each flavour. For electrons, $r$ is
measured using events with $\met > 150$~\GeV{}, where the dominant
contribution is from $W\rightarrow e\nu$, and $f$ is measured using events
with the transverse mass of the \met{}--lepton system\footnote{The transverse
  mass of the \met{}--lepton system is defined as $m_{\rm T}(W) \equiv
  \sqrt{2p_{{\rm T}\ell} \met(1-\cos\Delta \phi)}$ where $p_{{\rm T}\ell}$
  is the lepton transverse momentum and $\Delta \phi$ is the azimuthal angle between the
  lepton and the missing transverse momentum.}
$m_{\rm T}(W)< 20$~\GeV{} and $\met + m_{\rm T}(W) < 60$~\GeV{}, where the
dominant contribution is from multijet production (including heavy-flavour production) where one or more jets are
misidentified as electrons.  For muons, $r$ is measured using events with
$m_{\rm T}(W) > 100$~\GeV{}, a sample dominated by $W\rightarrow \mu\nu$, and
$f$ is measured using events where the transverse impact parameter of the muon relative
 to the primary vertex is more than five standard deviations away from
zero, consistent with muons originating from heavy-flavour hadron decays.  The
small contribution of prompt leptons in the control samples used to measure
$f$ is estimated from simulation and this contribution is subtracted from
the sample. The values of $r$ and $f$ are parameterised in terms of variables of the leptons ($| \eta |$, \pt, and the angular distance to the nearest jet) and the number of $b$-tagged jets.  For muons, $r$ ranges from 55\% to 97\% while $f$ ranges from 7\% to 30\%.  For electrons, $r$ ranges from 70\% to 95\% while $f$ ranges from 8\% to 30\%.  In general, the values of $r$ and $f$ are smaller for leptons near a jet, and larger for high-\pt leptons.

The second reducible background, corresponding to events where the charge of a lepton is misidentified, is considered only for electrons since the probability of muon  charge misidentification is negligibly small.  There are two primary mechanisms by which the electron charge can be misidentified: the first is the `trident' process in which an electron emits an energetic bremsstrahlung photon, which subsequently produces an $e^+e^-$ pair. This can result in a track of the incorrect charge being associated with the electron.  The second is the mismeasurement of the curvature of the electron track. 
The probability for an electron to have its charge incorrectly reconstructed is measured using a sample of dielectron events with invariant mass consistent with the $Z$ boson. The trident process can result in misidentified charge for an electron that is also likely to be considered fake/non-prompt due to the presence of nearby charged tracks.  To avoid double-counting the background contribution from such electrons,  the matrix method is used to subtract the fake/non-prompt electron yield from the $Z$ sample.  The charge misidentification probability is calculated in bins of electron $|\eta|$ and \pt, using a likelihood fit that adjusts these binned probabilities to find the best agreement with the observed numbers of same-charge and opposite-charge electron pairs.  The charge misidentification probability varies from $2 \times 10^{-5}$ (for electrons at low \pt and small $|\eta|$) to $10^{-2}$ for electrons at high \pt and $|\eta|$ near the edge of the barrel calorimeter; for electrons with larger values of $|\eta|$ the probability can reach 10\%. 

Since charge misidentification is negligible for muons and not relevant for trilepton events (for which no lepton charge requirements are imposed), the background from charge misidentification (called charge mis-ID hereafter) only appears in $ee$ or $e\mu$ events. To estimate its yield, $ee$ and $e\mu$ events are selected using all the criteria applied in the analysis, with the exception that the leptons are required to have opposite charge.  Then the charge misidentification probabilities are applied to this sample to determine the background yield. 

\section{ Signal and validation regions}
\label{sec:regions}

Several signal regions (SR) are defined to represent the broad range of BSM signals considered.  The selection criteria are designed to maximise the sensitivity to the signals.  The signal regions are separated into two categories: one category is designed for maximal sensitivity to VLQ and four-top-quark production, while the second category is optimised for the same-sign top-quark pair production searches. For the VLQ and four-top-quark searches, the preselected sample is first split according to the numbers of leptons (two or three) and $b$-tagged jets (one, two, or greater than two). Within each of the resulting subsamples, requirements are placed on \HT and \met, where \HT is the scalar sum of the \pT of all selected jets and leptons, to maximise the average sensitivity for the signal models considered. In addition, to fully exploit specific features of VLQ and four-top-quark signatures, the signal regions with at least three $b$-tagged jets are further split. Relaxed \HT  and high jet multiplicity requirements are sensitive to the four-top-quark signature, while  high \HT and low jet multiplicity requirements enhance sensitivity to the VLQ signature. For all the signal regions described above, lepton flavours are considered inclusively to increase the number of data events in the loosely selected samples used to estimate the reducible backgrounds.  The values of $r$ and $f$ appropriate to each lepton's flavour are used to estimate the fake/non-prompt lepton background. The selection criteria are summarised in Table~\ref{tab:general_SRs_VRs}, and the selection efficiencies for some signal models are shown in Table~\ref{tab:selection_eff}.

The same-sign top-quark selection requires exactly two leptons with positive charge, reflecting the preponderance of $tt$ over
$\bar{t}\bar{t}$ production in $pp$ collisions by a typical factor of $100$.
Additional criteria are imposed to maximise the sensitivity of the search: at least one $b$-tagged jet, \HT greater than 750~\GeV{}, \met greater than 40~\GeV{}, 
and the azimuthal separation between the two leptons $|\Delta \phi_{\ell\ell}|$ greater than 2.5. Since the optimal kinematic selection is looser than for VLQ and four-top-quark signal regions, more statistics are available for estimating the reducible backgrounds, so the lepton flavours ($ee$, $e\mu$, and $\mu\mu$) are treated separately in the search for same-sign top-quark pair production. These selection criteria are summarised in Table~\ref{tab:tt_SRs_VRs} and the selection efficiencies for the three same-sign top-quark pair signal processes are shown in Table~\ref{tab:selection_eff_SStop}.

\begin{table}[ht]
  \small
  \begin{center}
    \caption{Definitions of the validation and corresponding signal regions for the four-top-quark and VLQ searches, where $N_j$ is the number of jets, $N_b$ is the number of $b$-tagged jets,  and $N_{\ell}$ is the number of leptons.  The name of each signal (validation) region begins with ``SR'' (``VR''), with the rest of the name indicating the number of leptons and number of $b$-tagged jets required.  The suffix ``\_L'' denotes the signal regions with relaxed \HT\  but stricter $N_j$ requirements.  For regions that require  two leptons, the leptons must have the same charge.  Events that appear in any of the signal regions are vetoed in the validation regions.}
    \label{tab:general_SRs_VRs}
       \begin{tabular}{ l c c  c c  p{8cm}}
           \toprule\toprule

    Region name & $N_j$ & $N_b$  & $N_{\ell}$ & Lepton charges & Kinematic criteria \\ \midrule
       VR1$b$2$\ell$ & $\ge 1$ & 1 & 2 & $++$ or $--$ & $400 < \HT < 2400$ \GeV{} or $\met <  40$ \GeV{} \\ 
      SR1$b$2$\ell$    & $\ge 1$ & 1 & 2 &  $++$ or $--$ & $\HT >1000$  \GeV{} and  $\met > 180 $ \GeV{} 
  \vspace{0.3cm}  \\
      VR2$b$2$\ell$ & $\ge 2$ & 2 & 2 & $++$ or $--$  & $\HT > 400$ \GeV{} \\ 
       SR2$b$2$\ell$    & $\ge 2$ & 2 & 2  & $++$ or $--$ &  $\HT >1200$ \GeV{} and  $\met > 40 $ \GeV{} 
 \vspace{0.3cm} \\
      VR3$b$2$\ell$ & $\ge 3$ & $\ge 3$ & 2 & $++$ or $--$  & $400 < \HT < 1400$ \GeV{} or $\met <  40$ \GeV{} \\ 
      SR3$b$2$\ell$\_L & $\ge 7$ & $\ge 3$ & 2 &  $++$ or $--$  & $500 < \HT < 1200$ \GeV{} and  $\met > 40 $ \GeV{} \\
      SR3$b$2$\ell$    & $\ge 3$ & $\ge 3$ & 2 &  $++$ or $--$   & $\HT >1200$  \GeV{} and  $\met > 100 $ \GeV{} \vspace{0.3cm}  \\
       VR1$b$3$\ell$ & $\ge 1$ & 1 & 3 & any &  $400 < \HT < 2000$ \GeV{} or $\met <  40$ \GeV{}   \\
      SR1$b$3$\ell$    & $\ge 1$ & 1 & 3 & any & $\HT >1000$  \GeV{} and  $\met > 140 $ \GeV{}  \vspace{0.3cm} \\
      VR2$b$3$\ell$ & $\ge 2$ & 2 & 3 & any &$400 < \HT < 2400$ \GeV{} or $\met <  40$ \GeV{} \\ 
       SR2$b$3$\ell$    & $\ge 2$ & 2 & 3 & any & $\HT >1200$  \GeV{} and  $\met > 100 $ \GeV{}  \vspace{0.3cm} \\
      VR3$b$3$\ell$ & $\ge 3$ & $\ge 3$ & 3 & any & $\HT > 400$ \GeV{}  \\
     SR3$b$3$\ell$\_L & $\ge 5$ & $\ge 3$ & 3 & any & $ 500 < \HT < 1000$ \GeV{} and  $\met > 40 $ \GeV{} \\
      SR3$b$3$\ell$    & $\ge 3$ & $\ge 3$ & 3 & any & $\HT >1000$ \GeV{} and  $\met > 40 $ \GeV{}  \\ 
      \bottomrule\bottomrule
\end{tabular}
  \end{center}
\end{table} 

\begin{table}[ht]
  \small
  \begin{center}
    \caption{Signal selection and preselection efficiencies for events in various signal models, as estimated from MC simulation. VLQs are assumed to decay with the branching ratios expected in the singlet model of Ref.~\cite{AguilarSaavedra:2009es}.}
    \label{tab:selection_eff}
    \begin{tabular}{ l c  c c c c   }
      \toprule\toprule
    Signal   & Preselection  & \multicolumn{4}{c}{Signal region efficiencies [\%] }\\ 
                 &                  efficiency [\%]                    & SR1$b$2$\ell$ / 3$\ell$  & SR2$b$2$\ell$ / 3$\ell$ & SR3$b$2$\ell$\_L / 3$\ell$\_L & SR3$b$2$\ell$ / 3$\ell$   \\ \midrule
    $B\bar{B}$, $m_B = 800$ \GeV{} & 1.7  & 0.12 / 0.16 & 0.19 / 0.14 & 0.007 / 0.002 & 0.05 / 0.04   \\ 
    $B\bar{B}$, $m_B = 1200$ \GeV{} & 1.9  & 0.27 / 0.28 & 0.31/ 0.24 & $ 4\times 10^{-4}$  /  $ 4\times 10^{-4}$ & 0.07 / 0.05  \\
    $T\bar{T}$, $m_T = 800$ \GeV{} & 1.2  &  0.06 / 0.02 & 0.09 / 0.02 & 0.008 / 0.006 & 0.04 / 0.06 \\
    $T\bar{T}$, $m_T = 1200$ \GeV{} & 1.3  & 0.10 / 0.25 & 0.13 /  0.22  & 0.002 /  $ 9\times 10^{-4}$ & 0.06 / 0.11  \\
    $t\bar{t}t\bar{t}$ (SM)& 2.7 & 0.02 / 0.02 & 0.11 / 0.04 & 0.37 / 0.17 & 0.20 / 0.18 \\
    $t\bar{t}t\bar{t}$ (CI) & 3.0 & 0.06 / 0.05 & 0.23 / 0.08 & 0.30 / 0.16 & 0.33 / 0.27  \\

    $t\bar{t}t\bar{t}$ (2HDM,  & 3.1 & 0.02 / 0.03 & 0.11 / 0.03 & 0.62 / 0.24 & 0.19 / 0.17  \\
    $m_{H} = 700$ \GeV{})      &   &        &      &            & \\
    $t\bar{t}t\bar{t}$ (2UED/RPP,  & 3.3 & 0.27 / 0.16 & 0.62 / 0.31 & $ 8\times 10^{-4}$ / 0.0 & 0.89 / 0.51  \\ 
    $m_{\mathrm{KK}} = 1400$ \GeV{})  &   &        &      &            &            \\  \bottomrule\bottomrule
    \end{tabular}
  \end{center}
\end{table}

\begin{table}[ht]
  \small
  \begin{center}
    \caption{Definitions of the validation and signal regions for the same-sign top-quark pair production search, where $N_b$ is the number of $b$-tagged jets,  $N_{\ell}$ is the number of leptons, and $|\Delta \phi_{\ell\ell}|$ is the azimuthal angle between the leptons.  The name of each signal (validation) region begins with ``SR'' (``VR'').  The validation region is inclusive in lepton flavour. }
      \label{tab:tt_SRs_VRs}
     \begin{tabular}{ c c  c c  c c c } \toprule\toprule
   Region name & $N_b$   & $N_{\ell}$ & \HT\ [\GeV{}] & \met [\GeV{}] & $|\Delta \phi_{\ell\ell}|$ [radians] & Lepton flavour/charge  \\ \midrule
  VRtt                   & $\ge 1$ & 2          &    $> 750$    &    $>40$      & $>2.5$                     & $e^-e^-+e^-\mu^-+\mu^-\mu^-$\\ \midrule\midrule
   SRtt$ee$ & \multirow{3}{*}{$\ge 1$} & \multirow{3}{*}{2}  & \multirow{3}{*}{$> 750$} & \multirow{3}{*}{$> 40$ } & \multirow{3}{*}{$> 2.5$ }&  $e^+e^+$ \\
   SRtt$e\mu$ &                          &                     &                          &                          &                          &  $e^+\mu^+$ \\
   SRtt$\mu\mu$ &                          &                     &                          &                          &                          &  $\mu^+\mu^+$  \\          
   \bottomrule\bottomrule
\end{tabular}
  \end{center}
\end{table}

\begin{table}[ht]
  \small
  \begin{center}
    \caption{Signal selection and preselection efficiencies for events in the three same-sign top-quark pair production processes, as estimated from MC simulation.}
    \label{tab:selection_eff_SStop}
    \begin{tabular}{ l S[table-format=3.2]  S[table-format=3.2]  S[table-format=3.2] S[table-format=3.2]}
      \toprule\toprule
    Signal   & \multicolumn{1}{l}{Preselection}  & \multicolumn{3}{c}{Signal region efficiencies [\%] }\\ 
                                                    &   \multicolumn{1}{l}{efficiency [\%]} & \multicolumn{1}{c}{SRtt$ee$} & \multicolumn{1}{c}{SRtt$e\mu$} & \multicolumn{1}{c}{SRtt$\mu\mu$}  \\ \midrule
    $tt$, $m_V = 2000$ \GeV{}                           & 2.0         & 0.1    & 0.3      & 0.3    \\ 
    $tt\bar{u}$ off-shell, $m_V = 2000$ \GeV{}          & 1.7         & 0.1    & 0.2      & 0.2    \\ 
    $tV(\to t\bar{u})$ on-shell, $m_V = 2000$ \GeV{}    & 1.8         & 0.04   & 0.2      & 0.1    \\    \bottomrule\bottomrule
    \end{tabular}
  \end{center}
\end{table}

In addition to the signal regions, a set of validation regions (VR) with criteria similar to those used for the SR, in which the expected signal yield is small, are defined.  The VR are used to verify that the background is correctly modelled in regions that are kinematically similar to the signal regions. The definitions of the validation regions are presented in Tables~\ref{tab:general_SRs_VRs} and~\ref{tab:tt_SRs_VRs}, and the corresponding expected and observed yields are shown in Tables~\ref{tab:VR_exp_yields} and~\ref{tab:VRSStop_yields}.  These tables also report the probability for the expected background to fluctuate to equal or exceed the observed yield in each validation region; the smallest such probability is 0.10, which occurs in VR2$b$2$\ell$. The distributions of \MET and \HT\ in each validation region are shown in Figures~\ref{fig:VR_MET_plots}--\ref{fig:VR_plots_SSTop}.  The $\chi^2$ probabilities for compatibility of the observed and expected distributions are reasonable when all systematic uncertainties, including their bin-to-bin correlations, are considered.  The smallest probability is 2\%, which occurs for the \MET distribution in VR1$b$2$\ell$. The systematic uncertainties are described in Section~\ref{sec:syst}.

\begin{figure}[htbp]
  \begin{center}
    \subfloat[]{\includegraphics[width=0.46\textwidth]{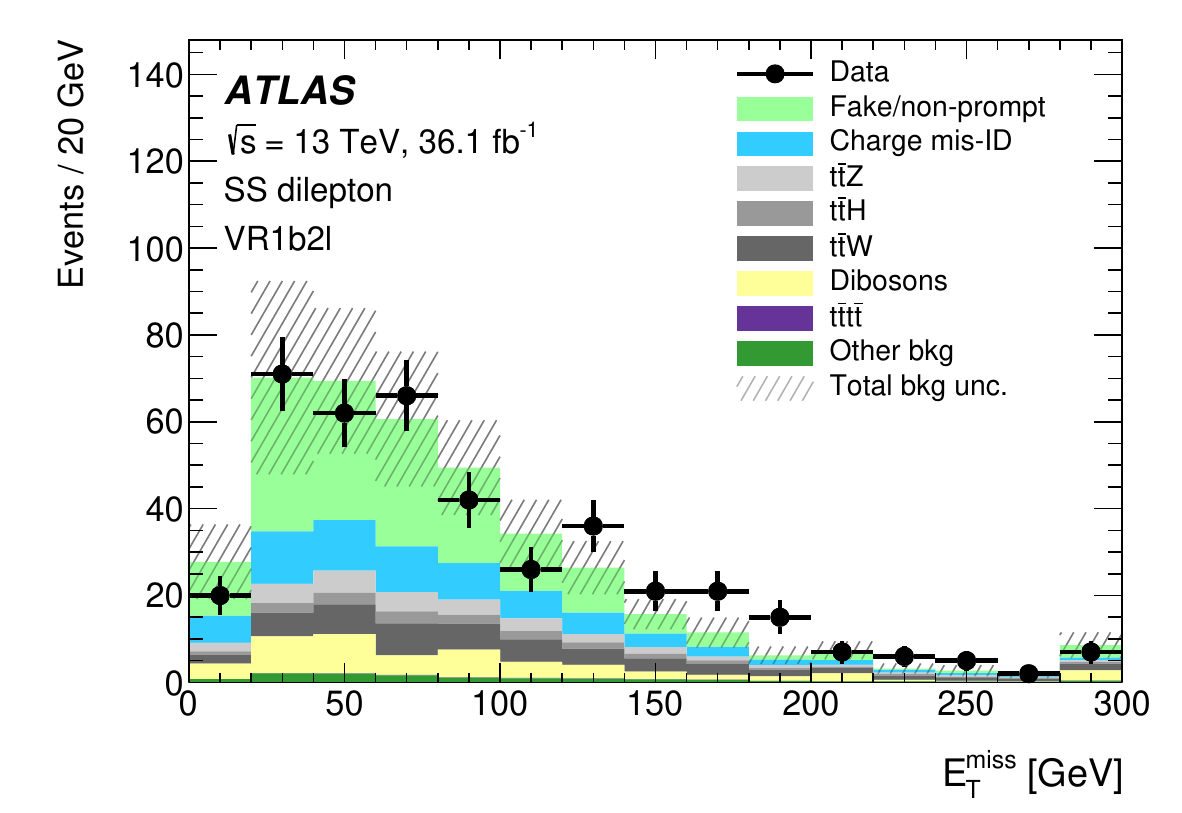}} \hfill
    \subfloat[]{\includegraphics[width=0.46\textwidth]{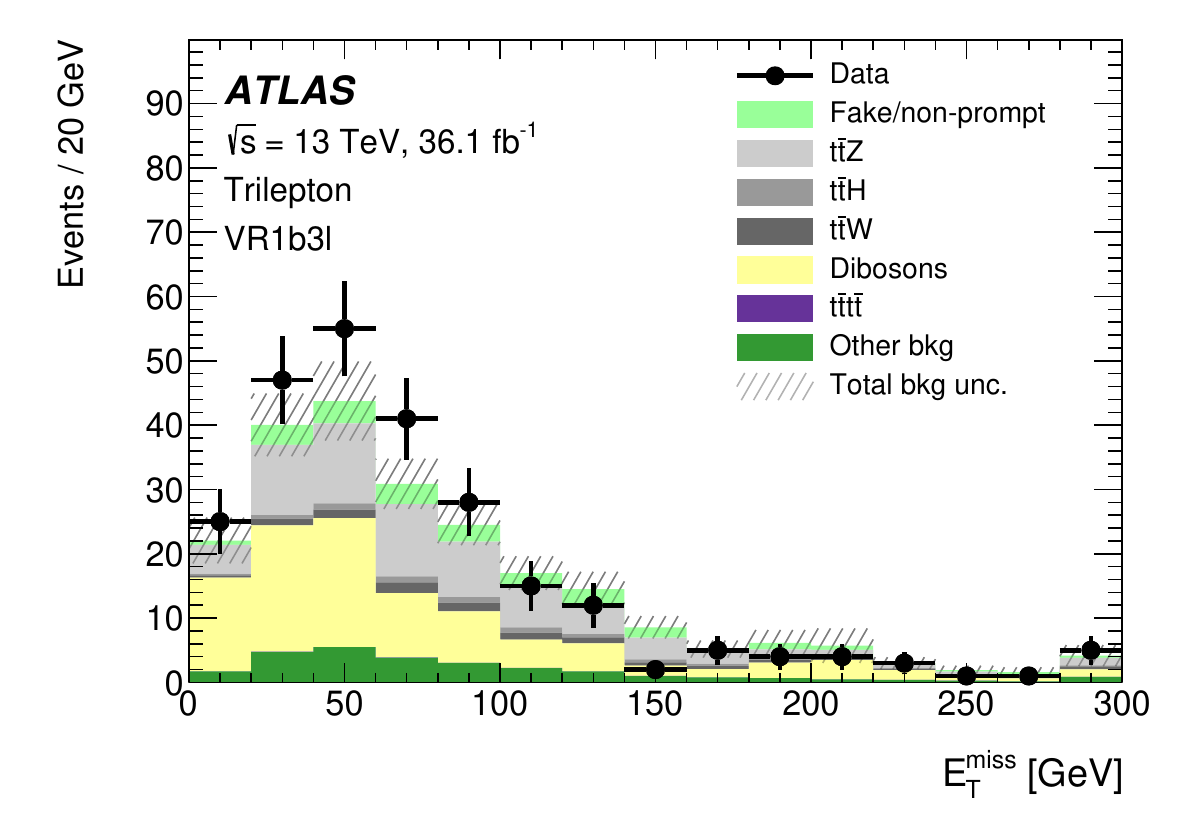}} \\ \vspace{-0.4cm}
    \subfloat[]{\includegraphics[width=0.46\textwidth]{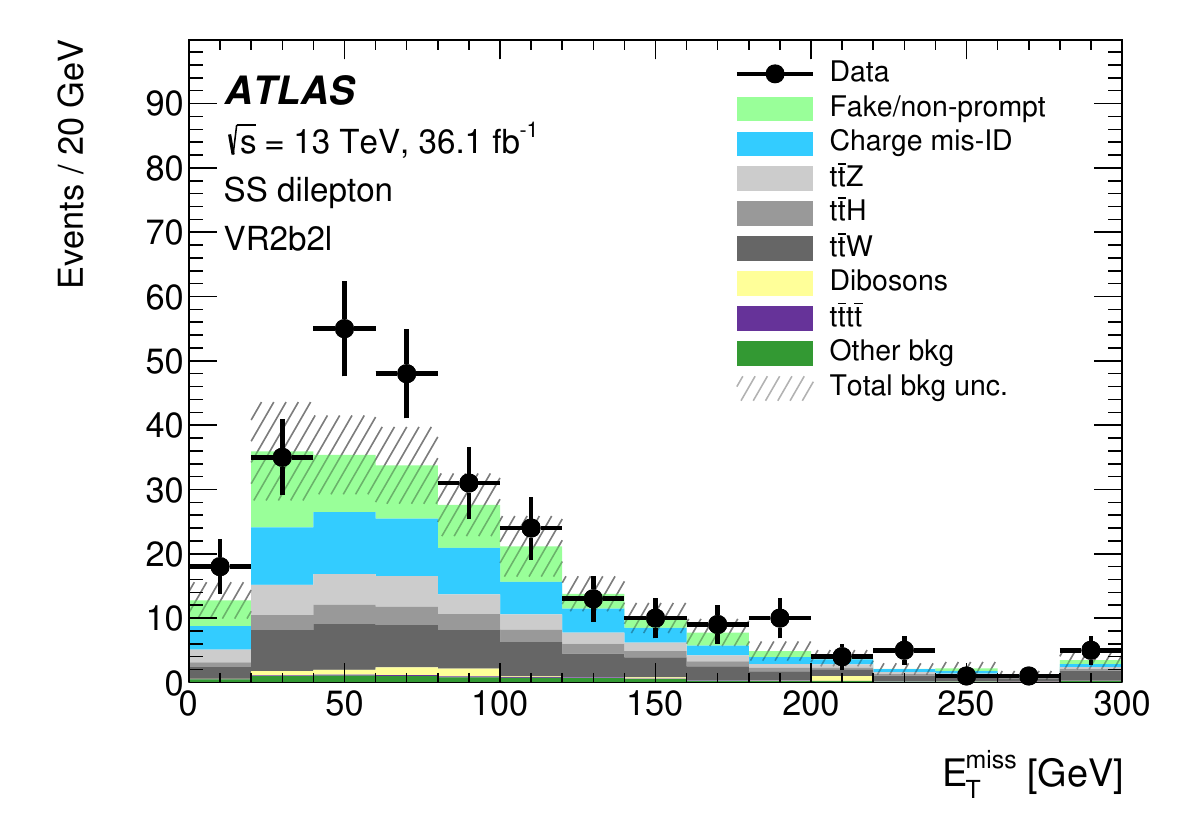}} \hfill
    \subfloat[]{\includegraphics[width=0.46\textwidth]{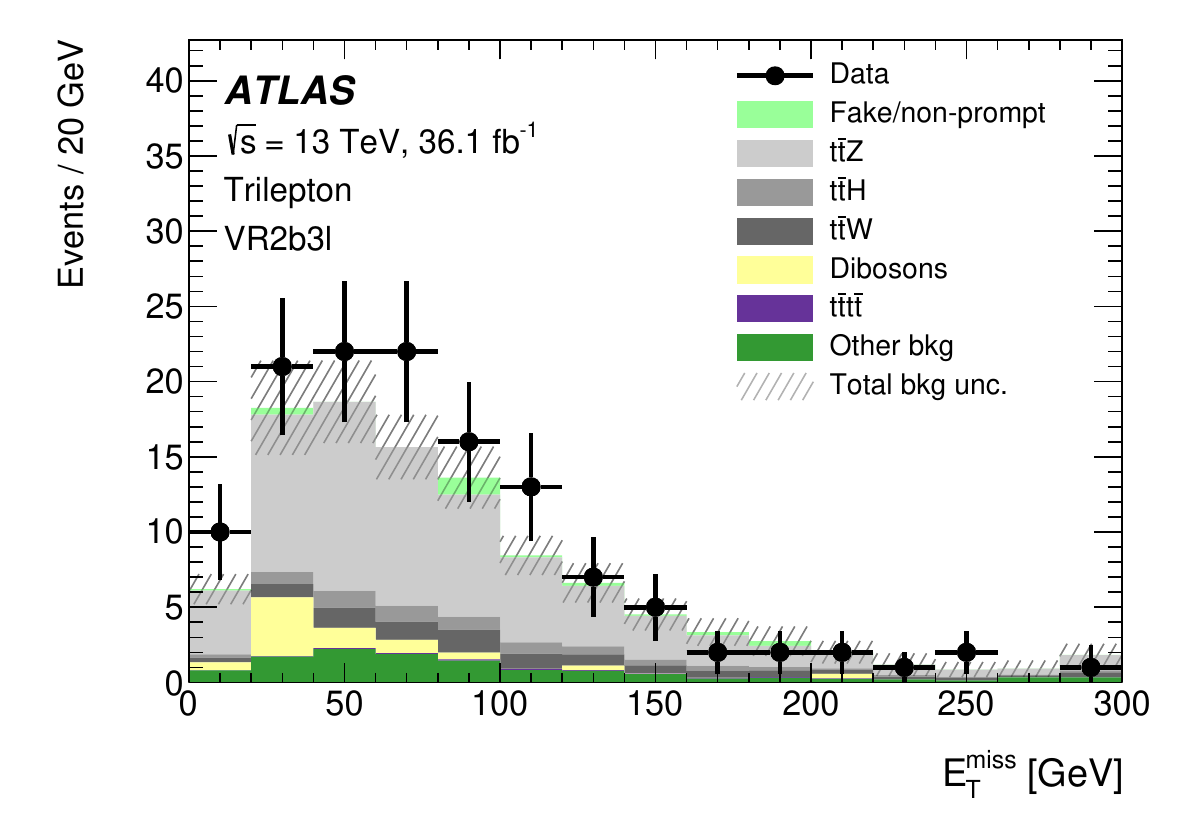}} \\ \vspace{-0.4cm} 
    \subfloat[]{\includegraphics[width=0.46\textwidth]{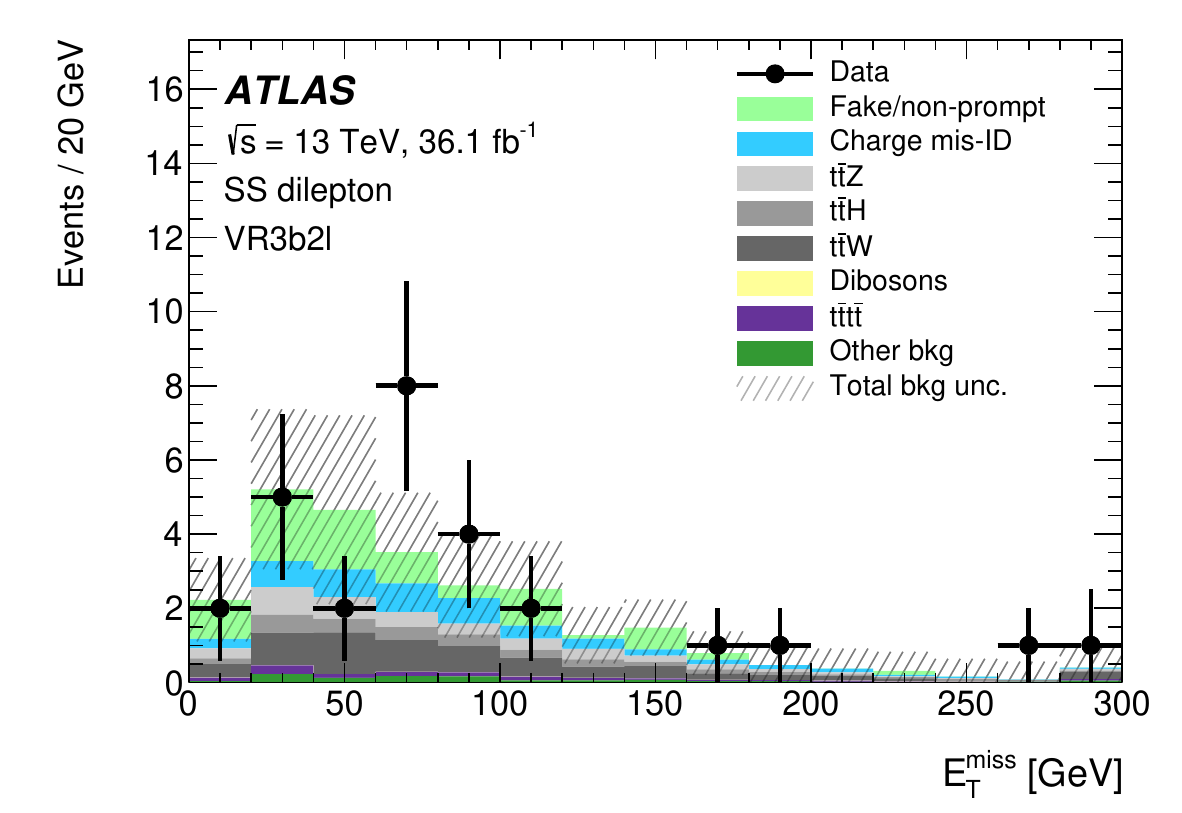}} \hfill
    \subfloat[]{\includegraphics[width=0.46\textwidth]{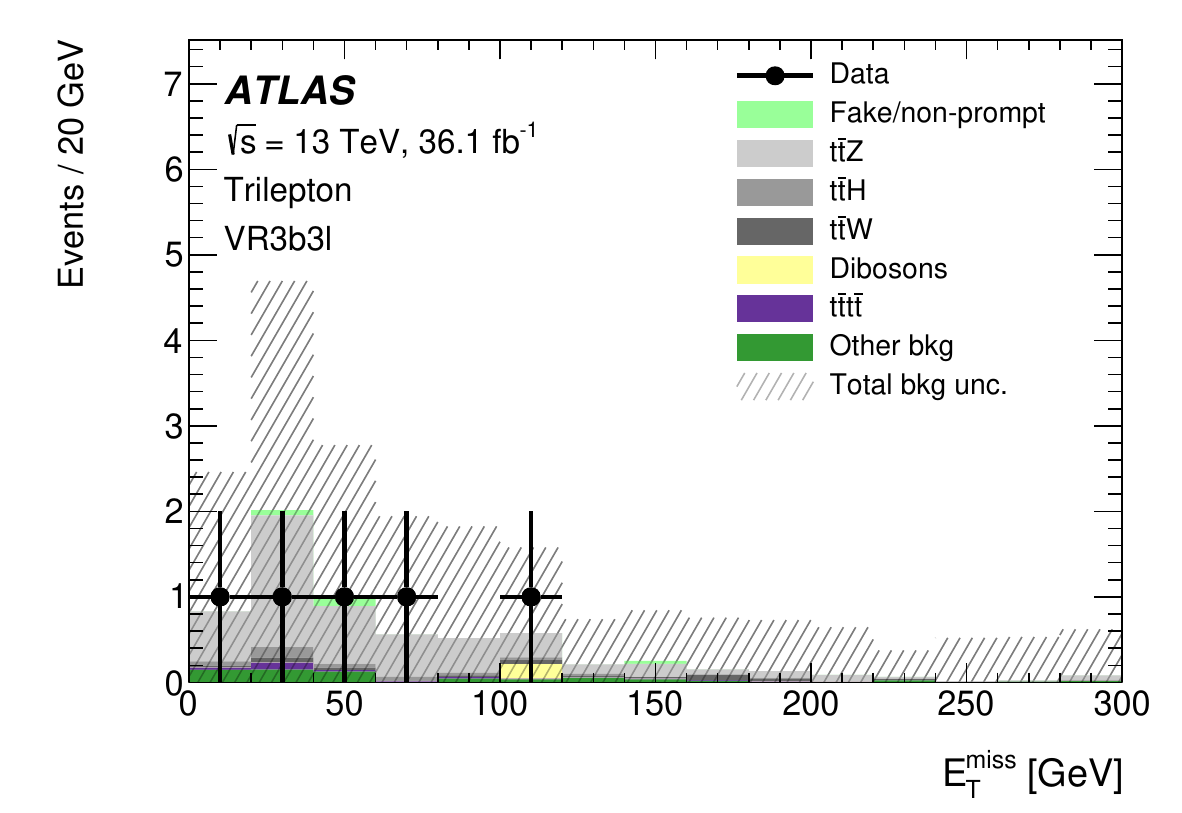}} 
    \caption{Distributions of \MET\ in each of the validation regions used for the four-top-quark and VLQ searches. The first (second) column shows distributions of dilepton (trilepton) events while each row corresponds to a given $b$-tagged jet multiplicity.
      The uncertainty, shown as the hashed region, includes both the statistical and systematic uncertainties from each background source.
         \label{fig:VR_MET_plots}}
  \end{center}
\end{figure}

\begin{figure}[htbp]
  \begin{center}
    \subfloat[]{\includegraphics[width=0.46\textwidth]{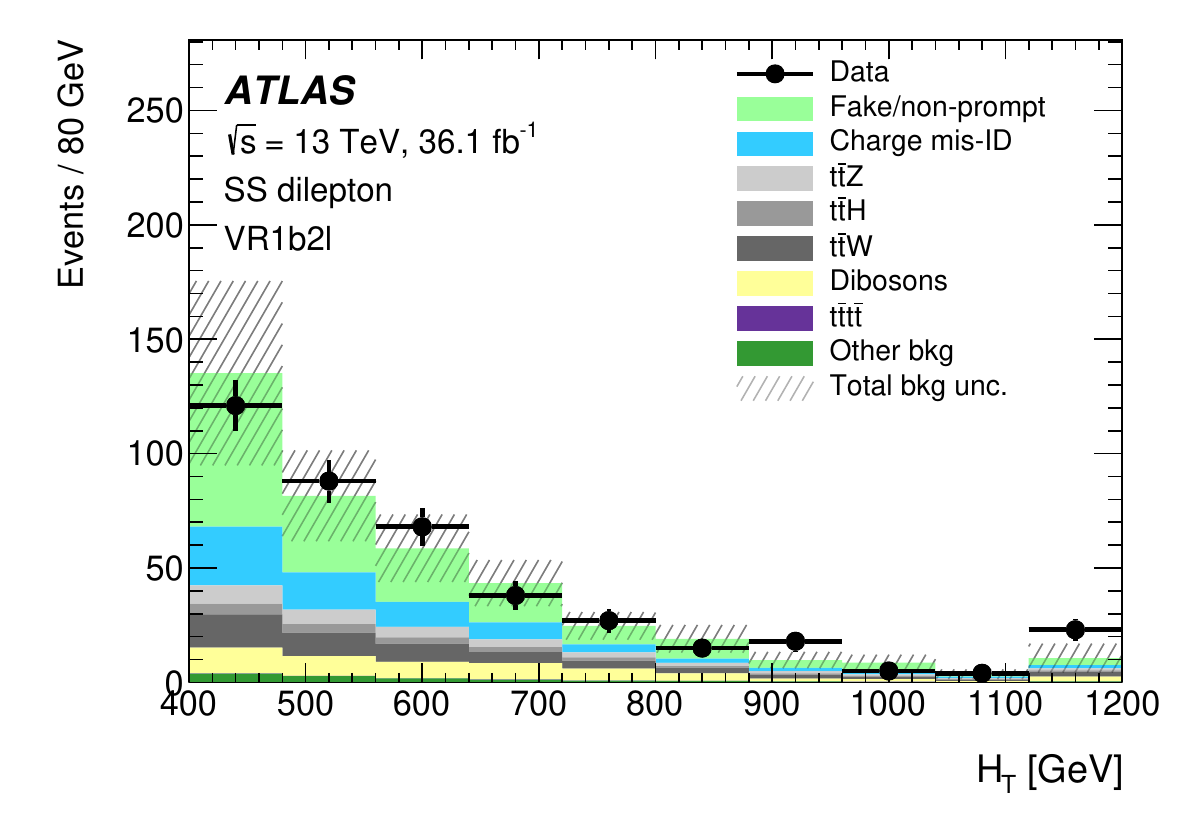}} \hfill
    \subfloat[]{\includegraphics[width=0.46\textwidth]{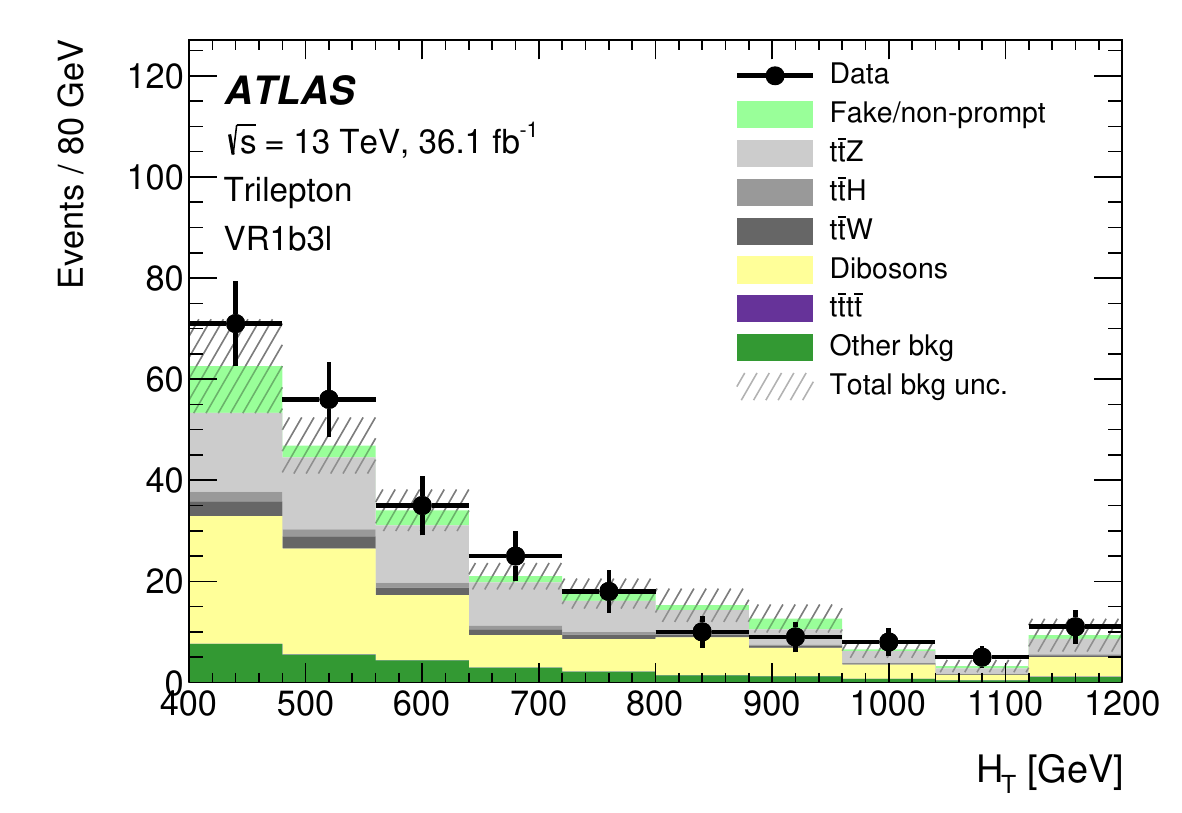}} \\ \vspace{-0.4cm} 
    \subfloat[]{\includegraphics[width=0.46\textwidth]{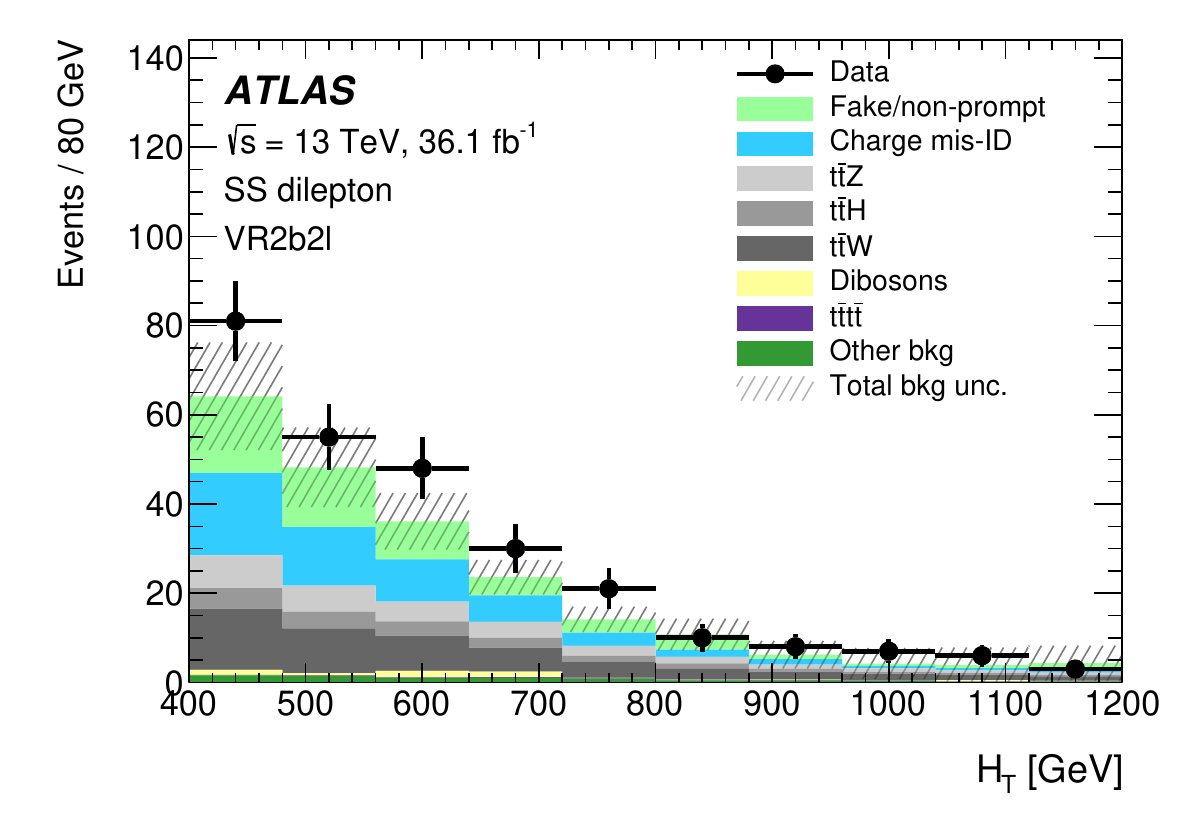}} \hfill
    \subfloat[]{\includegraphics[width=0.46\textwidth]{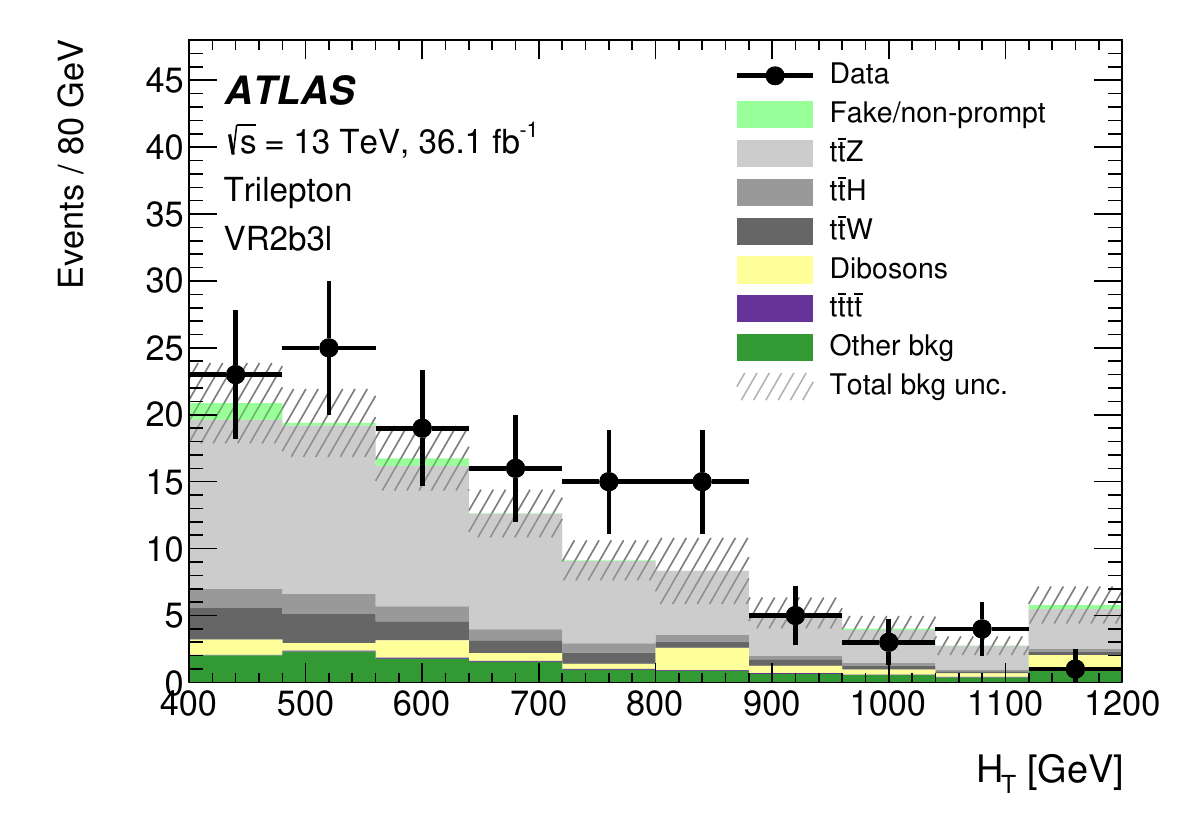}} \\ \vspace{-0.4cm} 
    \subfloat[]{\includegraphics[width=0.46\textwidth]{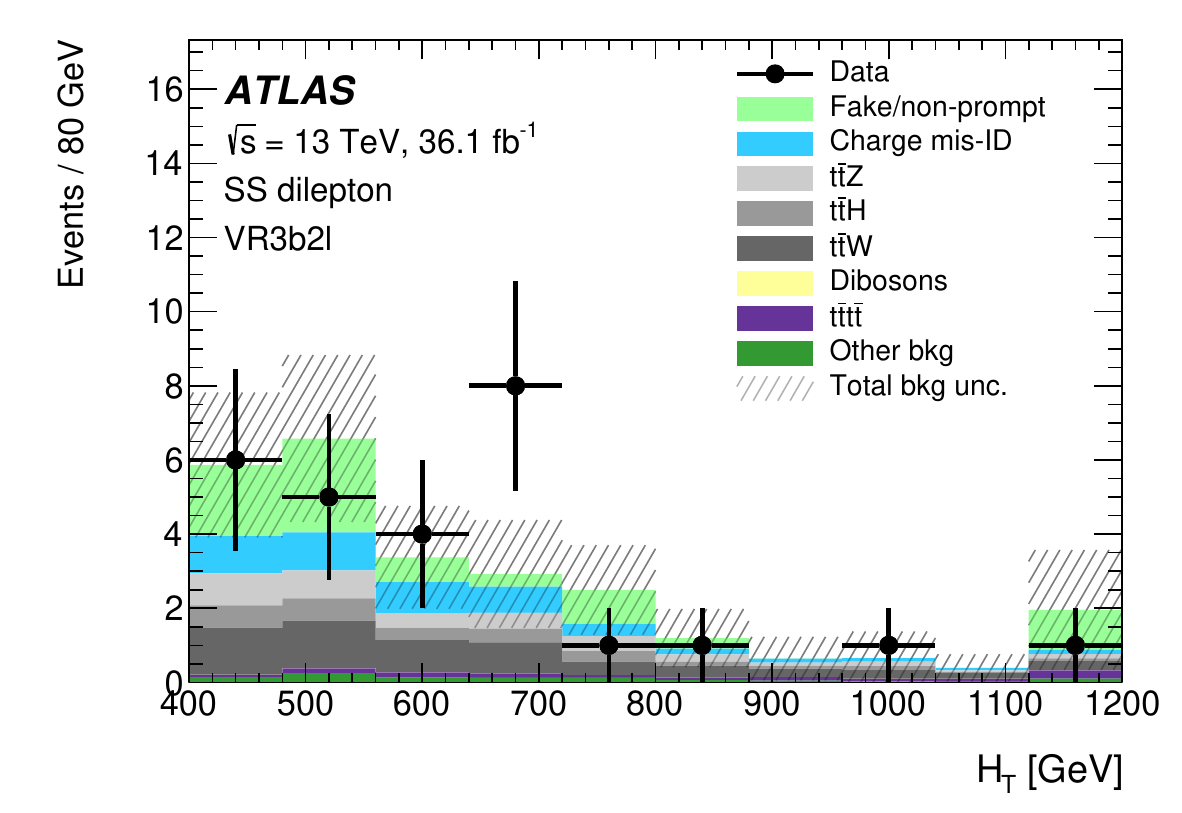}} \hfill
    \subfloat[]{\includegraphics[width=0.46\textwidth]{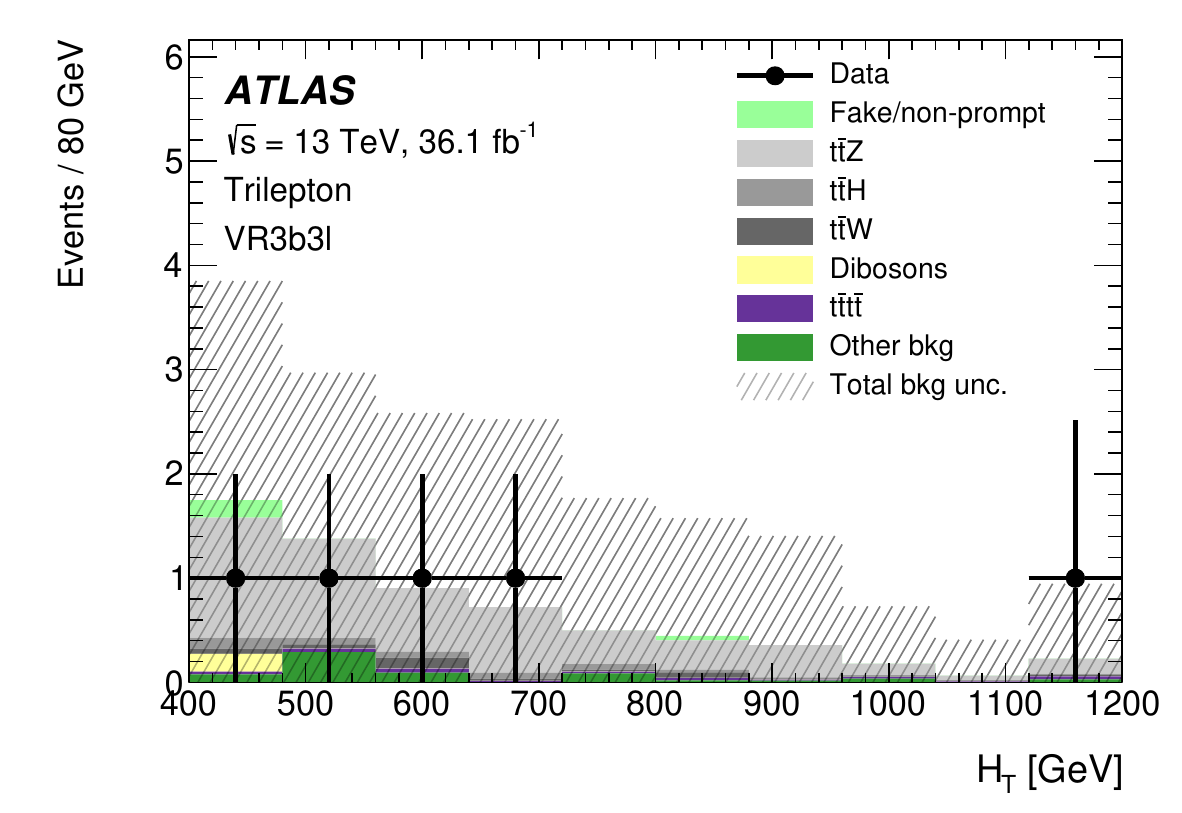}}
    \caption{Distributions of \HT\ in each of the validation regions used for the four-top-quark and VLQ searches. The first (second) column shows distributions of
            dilepton (trilepton) events while each row corresponds to a given $b$-tagged jet multiplicity. The uncertainty, shown as the hashed region,  includes both the statistical and systematic uncertainties from each background source.
         \label{fig:VR_HT_plots}}
  \end{center}
\end{figure}

\begin{figure}[htbp]
  \begin{center}
   \subfloat[]{\includegraphics[width=0.48\textwidth]{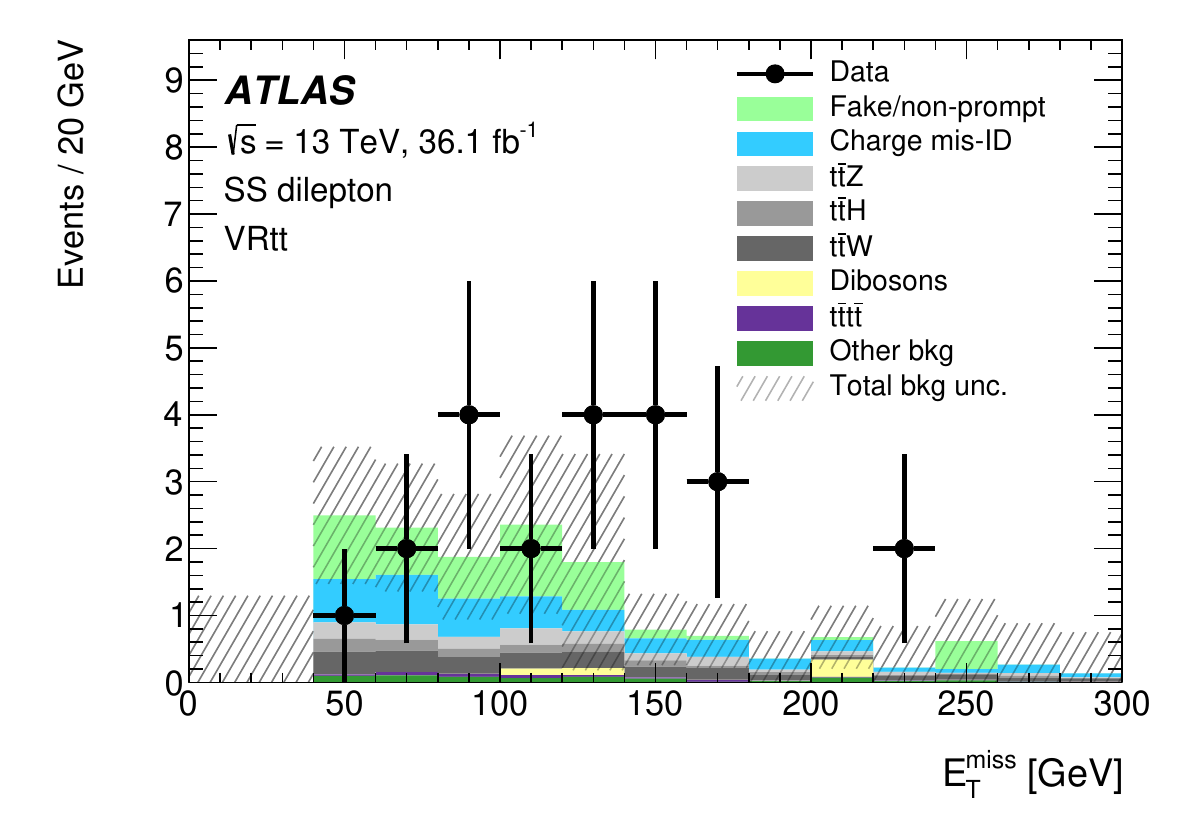}}
     \subfloat[]{\includegraphics[width=0.48\textwidth]{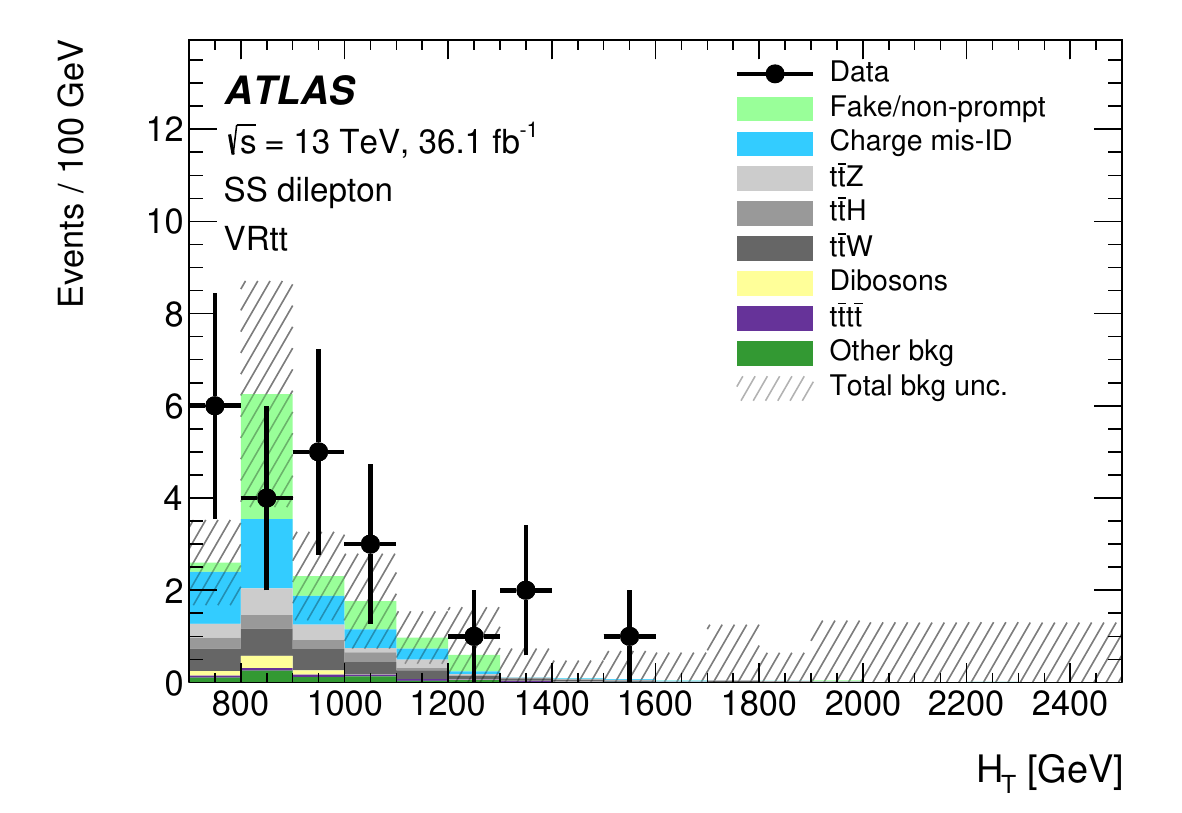}}\\
     \caption{Distributions of (a) \met and (b) \HT\ in the validation region used for the same-sign top-quark pair production search.
       The uncertainty, shown as the hashed region, includes both the statistical and systematic uncertainties from each background source.
         \label{fig:VR_plots_SSTop}}
  \end{center}
\end{figure}

\begin{table}
  \small
\begin{center}
  \caption{Expected background and observed event yields in the validation regions used in the VLQ and four-top-quark searches.  The `Other bkg' category includes contributions from all rare SM processes listed in Section~\ref{sec:sim}. The first uncertainty is statistical and the second is systematic. The $p$-values are the probabilities for the expected background to fluctuate to equal or exceed the observed yield in each region. }
\label{tab:VR_exp_yields}
{\small
\begin{tabular}{ l 
 S[table-format=2.2] @{\,}@{$\pm$}@{\,} S[table-format=2.2]  @{\,}@{$\pm$}@{\,}  S[table-format=2.2]   S[table-format=2.2] @{\,}@{$\pm$}@{\,} S[table-format=2.2]  @{\,}@{$\pm$}@{\,}  S[table-format=2.2]   S[table-format=2.2] @{\,}@{$\pm$}@{\,} S[table-format=2.2]  @{\,}@{$\pm$}@{\,}  S[table-format=2.2]  }
\toprule 
 Source  & \multicolumn{3}{c}{VR1$b$2$\ell$} & \multicolumn{3}{c}{VR2$b$2$\ell$} & \multicolumn{3}{c}{VR3$b$2$\ell$}\\ \midrule 
$t\bar{t}W$ &  49 &  1 &   14   &  48 &  1 &  13   &   5.8 &  0.3 &  2.8  \\ 
$t\bar{t}Z$ &  28.7 &  0.5 &  4.6   &  27.6 &  0.4 &  5.3   &   3.4 &\multicolumn{1}{l}{0.2} & $^{+4.2}_{-3.4}$  \\
Dibosons &  48 &  4 & 35   &   4.9 &  1.3 &  3.5   &    \multicolumn{3}{c}{$< 0.5$} \\ 
$t\bar{t}H$ &  17.7 &  0.4 &  2.4   &  18.3 &  0.4 &  2.6   &   2.6 &  0.2 &  1.9  \\ 
$t\bar{t}t\bar{t}$ & 0.59 & 0.04 & 0.39 & 1.3 & 0.1 & 1.2 & 1.0 &\multicolumn{1}{l}{0.1} & $^{+2.5}_{-1.0}$  \\
Other bkg & 12.3 & 0.5 & 6.4 & 7.3 & 0.3 & 4.0 & 1.1 & 0.2 & 1.1 \\
Fake/non-prompt & 170 & 8 & 87  & 53 & 5 & 28  &  7.8 & 1.6  &  3.8 \\ 
Charge mis-ID &  70 &  1 & 17   &  54 &  1 & 15   &   4.4 &  0.2 &  1.3  \\ 
\midrule 
Total bkg &395 & 9 & 98  & 216 & 5 & 38  & 26 & 2 &  11 \\ \midrule 
Data yield& \multicolumn{3}{c}{407}  & \multicolumn{3}{c}{269}  & \multicolumn{3}{c}{27} \\ \midrule
$p$-value & \multicolumn{3}{c}{0.45}  & \multicolumn{3}{c}{0.10}  & \multicolumn{3}{c}{0.46} \\
\bottomrule 
\end{tabular}\\
\renewcommand*{\arraystretch}{1.2}
\begin{tabular}{ l 
 S[table-format=2.2] @{\,}@{$\pm$}@{\,} S[table-format=2.2]  @{\,}@{$\pm$}@{\,}  S[table-format=2.2]   S[table-format=2.2] @{\,}@{$\pm$}@{\,} S[table-format=2.2]  @{\,}@{$\pm$}@{\,}  S[table-format=2.2]   S[table-format=2.2] @{\,}@{$\pm$}@{\,} S[table-format=2.2]  @{\,}@{$\pm$}@{\,}  S[table-format=2.2]  }
\toprule 
 Source  & \multicolumn{3}{c}{VR1$b$3$\ell$} & \multicolumn{3}{c}{VR2$b$3$\ell$} & \multicolumn{3}{c}{VR3$b$3$\ell$}\\ \midrule 
$t\bar{t}W$ &  10.4 &  0.3 &  3.3   &   9.4 &  0.3 &  2.4   &   0.31 &  \multicolumn{1}{l}{0.09} & $^{+0.57}_{-0.30}$  \\ 
$t\bar{t}Z$ &  70 &  1 &  11   &  66 &  1 & 15   &   4.6 &  \multicolumn{1}{l}{0.2} & $^{+7.4}_{-4.6}$   \\ 
Dibosons &  93 &  7 & 66   &   7.7 &  2.1 &  6.2   &   0.17 &   \multicolumn{1}{l}{0.17} & $^{+0.26}_{-0.00}$  \\ 
 $t\bar{t}H$ &   6.5 &  0.2 &  0.8   &   6.8 &  0.2 &  0.8   &   0.41 &  \multicolumn{1}{l}{0.05} & $^{+0.78}_{-0.41}$   \\ 
$t\bar{t}t\bar{t}$ & 0.21 & 0.02 & 0.14 & 0.64 & 0.03 & 0.37 & 0.21 &  \multicolumn{1}{l}{0.02} & $^{+1.20}_{-0.21}$  \\
Other bkg & 27 & 1 & 14 & 12.0 & 0.5 & 6.1 & 0.7 &  \multicolumn{1}{l}{0.2} & $^{+0.9}_{-0.7}$  \\
Fake/non-prompt & 22 & 4 & 13  & 2.7 & 1.5 &  2.1  &  \multicolumn{1}{l}{\phantom{1}0.21} & $^{+0.31}_{-0.18}$  &  0.12 \\ 
\midrule 
Total bkg & 229 & 8  & 70  & 105 & 3 & 19  & 6.5 &  \multicolumn{1}{l}{0.4} & $^{+10.8}_{-6.5}$ \\ \midrule 
Data yield & \multicolumn{3}{c}{248}  & \multicolumn{3}{c}{126}  & \multicolumn{3}{c}{5} \\ \midrule
$p$-value & \multicolumn{3}{c}{0.40}  & \multicolumn{3}{c}{0.17}  & \multicolumn{3}{c}{0.56} \\
\bottomrule 
\end{tabular}
}
\end{center}
\end{table}

\begin{table}
  \small
\begin{center}
\caption{Expected background and observed event yields in the validation region for the same-sign top-quark pair production search.  The `Other bkg' category includes contributions from all rare SM processes listed in Section~\ref{sec:sim}.  The first uncertainty is statistical and the second is systematic.  The $p$-value is the probability for the expected background to fluctuate to equal or exceed the observed yield. }
\label{tab:VRSStop_yields}
\begin{tabular}{ l 
 S[table-format=1.2] @{\,}@{$\pm$}@{\,} S[table-format=1.2]  @{\,}@{$\pm$}@{\,}  S[table-format=1.2] }
\toprule 
 Source  & \multicolumn{3}{c}{VRtt}  \\  \midrule 
$t\bar{t}W$ & 2.3 & 0.1 & 1.1 \\
$t\bar{t}Z$ & 1.6 & 0.1 & 0.6 \\
Dibosons & 0.5 & 0.4 & 0.3 \\
$t\bar{t}H$ & 1.0 & 0.1 & 0.4 \\
$t\bar{t}t\bar{t}$ & 0.30 & \multicolumn{1}{l}{\hspace{-2.2mm}0.03} & \hspace{-1mm}$^{+0.59}_{-0.30}$ \\
Other bkg & 0.7  & 0.1 & 0.6 \\
Charge mis-ID & 4.0 & 0.2 & 1.4 \\
Fake/non-prompt & \multicolumn{1}{l}{4.7} & $^{+ 1.5}_{- 1.3}$ & \multicolumn{1}{S[table-format=1.2]}{2.5}  \\ 
\midrule 
Total bkg. & \multicolumn{1}{S[table-format=1.2]} {15.1} & $^{+1.6}_{-1.4}$ &  \multicolumn{1}{S[table-format=1.2]}{4.2}  \\ \midrule 
Data yield & \multicolumn{3}{c}{22}  \\ \midrule
 $p$-value &  \multicolumn{3}{c}{0.14}\\
\bottomrule 
\end{tabular}  \\
\end{center}
\end{table}

\clearpage
\section{Systematic uncertainties}
\label{sec:syst}
The expected background yields are subject to several sources of systematic uncertainty.  For the irreducible backgrounds, the uncertainties include those from the background model and from the simulation of the response of the detector.  The background model uncertainties arise from the uncertainty of both the cross-section for a given process and of the acceptance of the signal regions for that process.  For $t\bar{t}W$ and $t\bar{t}Z$ production, these uncertainties are estimated by varying the renormalisation and factorisation scales up and down by a factor of two from their nominal values, and comparing the nominal {\sc MG5\_aMC@NLO}~v2.2.2 with a sample generated with \textsc{Sherpa}~v2.2.1.
For diboson production these uncertainties are estimated by varying the renormalisation, factorisation, and resummation scales up and down by a factor of two from their nominal values, and setting the CKKW merging scale to 15 and 30~\GeV{} (where the nominal value is 20~\GeV{})~\cite{Hoeche:2009rj}. The cross-section uncertainties for $t\bar{t}W$ and $t\bar{t}Z$ are $13\%$ and $12\%$, respectively, $6\%$ for diboson production, and $_{-9\%}^{+6\%}$ for $t\bar{t}H$ production~\cite{Alwall:2014hca}. For other irreducible backgrounds, this uncertainty is set to 50\% of the nominal yield.  The most important detector-related uncertainties are those of the efficiency for identifying $b$-jets (or misidentifying $c$- or light-jets as $b$-jets)~\cite{PERF-2012-04}, the jet energy calibration~\cite{ATLAS:2037613}, and the efficiencies for jets and leptons to satisfy the identification criteria~\cite{ATL-PHYS-PUB-2015-041,PERF-2015-10}.  In addition, there is a global 2.1\% uncertainty of the irreducible background yields due to the uncertainty of the integrated luminosity of the data sample.

Uncertainties of the fake/non-prompt lepton background arise from: $i$) possible differences between the values of $r$ and $f$ in the regions used to measure the efficiencies and in the signal regions, $ii$) statistical uncertainty of the control samples used to measure $r$ and $f$, and $iii$) uncertainties of the normalisation of the MC sample used to subtract the prompt-lepton contribution in the fake control sample used to measure $f$.  The first uncertainty is estimated by modifying the selection criteria for the control samples.  The modified sample for measuring $r$ for electrons requires $\met > 175$~\GeV{}, the modified sample for measuring $f$ for electrons requires $\met < 20$~\GeV{}, the modified sample for measuring $r$ for muons requires $m_{\rm T}(W)> 110$~\GeV{}, and the modified sample for measuring $f$ for muons requires $\met < 20$~\GeV{} and $\met + m_{\rm T}(W)< 60$~\GeV{}.  The second uncertainty is estimated by dividing the control samples randomly into four subsamples, computing the efficiencies in each of them, and observing the variation in the fake/non-prompt lepton yield.  This variation is then divided by two since each of the subsamples has only one fourth of the statistics of the full sample.  This procedure accounts for any correlations in the efficiencies.  The third uncertainty is estimated by varying the normalisation of the MC subtraction in the fake control sample by 10\%. The resulting uncertainty depends on the region the fake/non-prompt lepton background is estimated in, since the fake sample can vary kinematically, but is generally around 40$-$50\% of the expected fake/non-prompt lepton yield for the dominant uncertainty in the signal regions. The first is the dominant uncertainty, particularly from variations in the fake-lepton efficiency when the selection criteria for the control samples are changed.

The uncertainties of the charge mis-ID background arise from uncertainties of the measured rates for electron charge misidentification and uncertainties of the fake/non-prompt lepton background.  The uncertainties of the charge misidentification rates include the following contributions: the statistical uncertainty of the likelihood fit to determine the rates ($\approx 15\%$), the changes in rates observed when the mass windows used to define the $Z \rightarrow ee$ and sideband regions are varied from 0~\GeV{} to 20~\GeV{} ($\approx 6\%$), and the differences observed between the results of the likelihood fit and the true rates when the method is applied to simulation ($\approx 5\%$). These uncertainties sum in quadrature to about 20\% of the expected charge mis-ID yield in the signal regions. The systematic uncertainty of the fake/non-prompt component is estimated as described above, and impacts the charge mis-ID background through a variation in the fake/non-prompt background that is subtracted when calculating the charge misidentification rates ($\approx 10\%$).  This component of the uncertainty is anti-correlated between the fake/non-prompt and charge mis-ID backgrounds.

Since the optimised selection criteria result in small expected background yields in the signal regions, the dominant uncertainty in the analysis is statistical.  Among the systematic uncertainties, the leading contributors are from uncertainties of the fake/non-prompt lepton background estimate, the modelling of the irreducible backgrounds (in terms of both their production cross-sections and acceptance) and uncertainties of the efficiency for identifying $b$-jets. Summaries of the leading sources of systematic uncertainty in each signal region are provided in Tables~\ref{tab:syst_effect_on_total_bkg} and~\ref{tab:syst_effect_on_total_bkg_tt} for the total background yield, and in Tables~\ref{tab:syst_effect_on_TTS_M1000} and~\ref{tab:syst_effect_on_sstops_tt_exclusive_mV2000} for representative signal models (a $T$ vector-like quark with $m_T = 1$~\TeV, and exclusive $tt$ production with $m_V = 2$~\TeV{}, respectively).

\begin{table}[htbp]
  \small
\begin{center}
\caption{Uncertainty of the total background yields in the signal regions for the four-top-quark and VLQ searches due to the leading sources of systematic uncertainty.\label{tab:syst_effect_on_total_bkg}}
\small
\begin{tabular}{l r r r r r r r r }
\toprule
  Uncertainty & SR1$b$2$\ell$ & SR2$b$2$\ell$ & SR3$b$2$\ell$\_L & SR3$b$2$\ell$ & SR1$b$3$\ell$ & SR2$b$3$\ell$ & SR3$b$3$\ell$\_L & SR3$b$3$\ell$\\
\hspace{0.2cm}source                   &      [\%] &      [\%]  &      [\%]       &      [\%]  &      [\%]  &      [\%]  &      [\%]     &      [\%]  \\
\midrule
Jet energy & $3$ & $1$ & $5$ & $6$ & $3$ & $5$ & $3$ & $4$\\
\hspace{0.2cm}resolution  &   &    &     &    &       &   &      &   \\
Jet energy scale & $3$ & $3$ & $9$ & $6$ & $3$ & $5$ & $11$ & $6$\\
$b$-tagging  & $5$ & $3 $ & $6$ & $7$ & $3$ & $4$ & $9$ & $9$\\
\hspace{0.2cm}efficiency  &   &   &   &    &      &  &    &   \\
Lepton ID & $2 $ & $1$ & $1$ & $1$ & $3$ & $3$ & $2$ & $3$\\
\hspace{0.2cm}efficiency  &   &   &   &    &      &  &    &   \\
Pile-up & $5 $ & $2$ & $3$ & $3$ & $3$ & $5$ & $1$ & $6$\\
\hspace{0.2cm}reweighting  &   &   &   &    &      &  &    &   \\
Luminosity & $1$ & $1 $ & $2$ & $2$ & $2$ & $2$ & $2$ & $2$\\
Fake/non-prompt  & $20 $ & $12$ & $13$ & $8$ & $7$ & $2$ & $3$ & $1$\\
Charge mis-ID & $2 $ & $3 $ & $1$ & $2$ & - & - & - & -\\
Cross-section  & $25$ & $13$ & $22$ & $32$ & $32$ & $26$ & $21$ & $24$\\
\hspace{0.2cm}$\times$ acceptance   &  &  & & &  & &  & \\           
\midrule
\end{tabular}
\end{center}
\end{table}

\begin{table}[htbp]
\small
  \begin{center}
 \caption{Uncertainty of the event yields in the signal regions for a representative signal (vector-like $T$ quark, $m_T=1~\TeV$) due to the leading sources of experimental systematic uncertainty. The expected yield for this signal in each region is also given.\label{tab:syst_effect_on_TTS_M1000}}
\begin{tabular}{l r r r r r r r r }
\toprule
  Uncertainty & SR1$b$2$\ell$ & SR2$b$2$\ell$ & SR3$b$2$\ell$\_L & SR3$b$2$\ell$ & SR1$b$3$\ell$ & SR2$b$3$\ell$ & SR3$b$3$\ell$\_L & SR3$b$3$\ell$\\
\hspace{0.2cm}source                               &      [\%] &      [\%]  &      [\%]       &      [\%]  &      [\%]  &      [\%]  &      [\%]     &      [\%]  \\
\midrule
Jet energy & $<1$ & $1$ & $6$ & $4$ & $<1$ & $<1$ & $24$ & $<1$\\
\hspace{0.2cm}resolution  &   &   &   &    &      &  &    &   \\
Jet energy scale & $2$ & $1$ & $23$ & $3$ & $1$ & $1$ & $12$ & $<1$\\
$b$-tagging & $6$ & $3$ & $9$ & $8$ & $5$ & $4$ & $7$ & $8$\\
\hspace{0.2cm}efficiency  &   &   &   &    &      &  &    &   \\
Lepton ID & $2$ & $2$ & $1$ & $2$ & $3$ & $3$ & $2$ & $3$\\
\hspace{0.2cm}efficiency  &   &   &   &    &      &  &    &   \\
Luminosity & $2$ & $2$ & $2$ & $2$ & $2$ & $2$ & $2$ & $2$\\
Pile-up & $3$ & $3$ & $7$ & $3$ & $<1$ & $<1$ & $3$ & $2$\\
\hspace{0.2cm}reweighting  &   &   &   &    &      &  &    &   \\ \midrule
Expected yield &  1.7 & 2.1 & 0.08 & 1.0  & 3.0 & 3.2 & 0.03 & 1.8 \\ \midrule
\end{tabular}
\end{center}
\end{table}

\begin{table}[h]
\small
  \begin{center}
\caption{Uncertainty of the total background yields in the signal regions for the same-sign top-quark pair production search due to the leading sources of systematic uncertainty.\label{tab:syst_effect_on_total_bkg_tt}}
\begin{tabular}{l r r r }
\toprule
Source & SRtt$ee$ & SRtt$e\mu$ & SRtt$\mu\mu$\\
             &    [\%]       &       [\%]       &      [\%]     \\ \midrule
Jet energy resolution         &     $3$     &     $<1$     &     $13$\\
Jet energy scale             &     $2$     &     $2$     &     $9$\\
$b$-tagging efficiency         &     $1$     &     $2$     &     $3$\\
Lepton ID efficiency         &     $<1$     &     $1$     &     $4$\\
Pile-up reweighting             &     $2$     &     $2$     &     $4$\\
Luminosity                     &     $<1$     &     $1$     &     $2$\\
Fake/non-prompt                 &     $36$     &     $17$     &     $5$\\
Charge mis-ID     &     $12$      &     $5$     &     -\\
Cross-section $\times$ acceptance                 &     $10$     &     $15$     &     $25$\\ \midrule
\end{tabular}
\end{center}
\end{table}

\begin{table}[h]
\small
  \begin{center}
\caption{Uncertainty of the event yields in the signal regions for a representative signal of the same-sign top-quark pair production search (exclusive $tt$ production, $m_{V}=2~\TeV$ normalised to $\SI{100}{fb}$) due to the leading sources of experimental systematic uncertainty. In all three channels, the uncertainty due to jet energy resolution is compatible with the statistical uncertainty of the simulated samples.\label{tab:syst_effect_on_sstops_tt_exclusive_mV2000}}
\sisetup{
table-number-alignment = center,
table-figures-integer = 1,
table-figures-decimal = 2,
table-figures-exponent = 2,
}
\begin{tabular}{l r r r }
\toprule
Source & \multicolumn{1}{c}{SRtt$ee$} & \multicolumn{1}{c}{SRtt$e\mu$} & \multicolumn{1}{c}{SRtt$\mu\mu$} \\
             &    [\%]       &       [\%]       &      [\%]     \\ \midrule
Jet energy resolution &  7 & < 1 & < 1\\
Jet energy scale &  1 &  1 & < 1\\
$b$-tagging efficiency &  3 &  2 & < 1\\
Lepton ID efficiency &  5 &  3 &  4\\
Luminosity & 2  &2 &  2\\
Pile-up reweighting &  3 &< 1 &  1\\\midrule
Expected yield &   3.4  &13 & 12  \\ 
\bottomrule
\end{tabular}
\end{center}
\end{table}


\clearpage
\section{Results}
\label{sec:result}

To test for the presence of a BSM signal, the observed numbers of events in a set of signal regions are compared with the expected background yields in those regions. The searches for VLQ and four-top-quark production use the combination of the signal regions defined in Table~\ref{tab:general_SRs_VRs}, while the searches for $tt$ production use the combination of the signal regions defined in Table~\ref{tab:tt_SRs_VRs}. In the case where the SM four-top-quark production is probed, this process is removed from the background contribution. In all other cases, the quoted significances refer to BSM benchmarks.

A Poisson likelihood ratio test is used to assess the probability that the observed yields are compatible with the sum of the expected background and signal, with the nominal signal cross-section scaled by a value $\mu$.  Systematic uncertainties are introduced as nuisance parameters that have Gaussian or log-normal constraints corresponding to their uncertainty values.  For any given choice of $\mu$ the observed likelihood ratio $q_\mu$ is compared with the distribution of values that would be expected under the background-only and signal plus background hypotheses. The probabilities $p_b(\mu)$ of the background fluctuating to be more signal-like than the data, and $p_{s+b}(\mu)$ of the signal plus background fluctuating to be more background-like than the data are both determined by comparing $q_\mu$ with these distributions. The values of $p_b(\mu)$ and $p_{s+b}(\mu)$ are derived using the asymptotic approximation described in Ref.~\cite{Cowan:2010js}.  The quantity $R_{\mathrm{CL}_\mathrm{s}}$~\cite{Read:2002hq} is then defined as

\begin{linenomath*}
\begin{equation*}
R_{\mathrm{CL}_\mathrm{s}}(\mu) \equiv {p_{s+b}(\mu) \over 1-p_b(\mu)}.
\end{equation*}
\end{linenomath*}

If the data are statistically consistent with the background expectation, $R_{\mathrm{CL}_\mathrm{s}}(\mu)$ will tend to decrease as $\mu$ increases.  All values of $\mu$ for which $R_{\mathrm{CL}_\mathrm{s}}(\mu)$ is less
than 0.05 are considered as being excluded at 95\% confidence level (CL).  If, for a particular signal model, $R_{\mathrm{CL}_\mathrm{s}}(\mu = 1)$ is less than 0.05, that model is excluded.

The observed yields in each of the signal regions, along with the expected yields from background sources and some representative BSM physics models are shown in Tables~\ref{tab:SR_yields} and~\ref{tab:SRSStop_yields} and in Figure~\ref{fig:SummaryPlots}.  There are no statistically significant differences between the event yields and the expected background, although in two of the signal regions, SR3$b$2$\ell$\_L and SR3$b$3$\ell$\_L, the event yield exceeds the background by 1.7 and 1.8 standard deviations, respectively. The resulting combined significance depends on the signal being considered, reaching 3.0~standard deviations for SM four-top-quark production (where this contribution is not included among the backgrounds), while a significance of 0.9 standard deviations is expected. More than half of the excess is observed in events with two muons, three $b$-tagged jets and \HT around $700$~\GeV{}. The largest significance for any of the BSM models considered is 2.3~standard deviations, which is obtained for the 2HDM model. Therefore no evidence of BSM signals is found, and limits are set as detailed below. 

Several studies were done to validate the background estimate.  One potential issue was noted when applying the matrix method for muons to the same sample of events used to calculate the fake/non-prompt muon efficiency, where the predicted yield was observed to deviate from data at the level of $1.2$ standard deviations near $\Delta R(\mu, {\text {jet}}) = 1.0$.  Applying a two-dimensional parameterisation of the efficiencies in $p_{{\text T}, \mu} \times  \Delta R(\mu, {\text {jet}})$ substantially improves the level of agreement, and the background in the signal regions was recomputed with this parameterisation.  In addition,  the prompt- and fake-lepton efficiencies used in the estimation of the fake/non-prompt lepton background were recomputed using different requirements for the number of $b$-tagged jets in the control regions (this test is especially important for electrons, where the fraction of candidates arising from photon conversion versus heavy-flavour decay varies strongly with the presence or absence of a $b$-tagged jet), and also using a completely different set of control regions (dilepton events where a tag-and-probe procedure was applied).  The fake/non-prompt lepton background was also estimated using the fake-factor method~\cite{EXOT-2012-25} instead of the matrix method. The level of compatibility between the expected background and observed data yields was similar in all of these variations. 

Further, the events in the signal regions were scrutinised to determine if some of them might have arisen from detector defects or other anomalies.  The distribution of objects in $\eta$, $\phi$, and \pt was found to be consistent with expectations, as was the temporal distribution of the events across the data-taking period.  The reconstructed muon candidates in these events were inspected, and their features (such as the $\chi^2$ of their fitted tracks, and compatibility of the momenta measured in the inner detector and the muon spectrometer) were found to be unremarkable. The three-lepton samples were split between those with and without a lepton pair that formed a $Z$-boson candidate, and the distribution of events in these subsamples was consistent with expectations.  For example, in the subsample of SR3$b$3$\ell$\_L with a $Z$-boson candidate, four events are observed with an expected background of $2.4 \pm 0.6$, while in the subsample without a $Z$-boson candidate, five events are observed with an expected background of $1.7 \pm 0.6$. The composition of $b$-tagged jets (the fractions of such jets that arise from $b$-, $c$-, or light-quarks or gluons) was studied in simulated background events.  It was found that the dominant source of $b$-tagged jets in both the signal and validation regions was in fact $b$-jets, which accounted for 76 -- 95\% of the $b$-tagged jets in each region.  In addition, the kinematic properties of the events were compared with the expectations from the BSM four-top-quark production benchmark models, and found to agree poorly with all of them, particularly in the $b$-tagged jet multiplicity.

\begin{table}
  \small
\begin{center}
\caption{Expected background and observed event yields in the signal regions for the vector-like quark and four-top-quark searches.  The `Other bkg' category contains contributions from all rare SM processes listed in Section~\ref{sec:sim}. The first uncertainty is statistical and the second is systematic. The BSM significance is the number of standard deviations by which a  BSM signal plus background hypothesis is preferred to the background-only hypothesis.  Since this significance depends only on the event yield and expected background in the given signal region, it is independent of the BSM model.  When computing the SM $t\bar{t}t\bar{t}$ significance, the expected SM $t\bar{t}t\bar{t}$ yield is not included in the expected background. Both significances are calculated using the same procedure used to calculate the reported limits. }
\label{tab:SR_yields}
\renewcommand*{\arraystretch}{1.2}
\begin{tabular}{ l 
 S[table-format=1.2] @{\,}@{$\pm$}@{\,} S[table-format=1.2]  @{\,}@{$\pm$}@{\,}  S[table-format=1.2]   S[table-format=1.2] @{\,}@{$\pm$}@{\,} S[table-format=1.2]  @{\,}@{$\pm$}@{\,}  S[table-format=1.2]   S[table-format=1.2] @{\,}@{$\pm$}@{\,} S[table-format=1.2]  @{\,}@{$\pm$}@{\,}  S[table-format=1.2]   S[table-format=1.2] @{\,}@{$\pm$}@{\,} S[table-format=1.2]  @{\,}@{$\pm$}@{\,}  S[table-format=1.2]  }
\toprule 
 Source  & \multicolumn{3}{c}{SR1$b$2$\ell$} & \multicolumn{3}{c}{SR2$b$2$\ell$} & \multicolumn{3}{c}{SR3$b$2$\ell$\_L} & \multicolumn{3}{c}{SR3$b$2$\ell$}\\ \midrule 
$t\bar{t}W$ &   2.04 &  0.14 &  0.49   &   2.68 &  0.15 &  0.55   &   0.95 &  0.11 &  0.31   &   0.40 &  0.06 &  0.10  \\ 
$t\bar{t}Z$ &   0.58 &  0.08 &  0.10   &   0.95 &  0.11 &  0.17   &   0.72 &  0.11 &  0.19   &   0.11 &  \multicolumn{1}{l}{\hspace{-1.7mm}0.05} & $^{+ 0.13}_{-0.10}$  \\ 
Dibosons &   3.2 &  1.5 &  2.4   &     \multicolumn{3}{c}{$<0.5$}  &   0.13 & \multicolumn{1}{l}{0.13} & $^{+0.27}_{-0.00}$  &    \multicolumn{3}{c}{$<0.5$}   \\ 
$t\bar{t}H$ &   0.56 &  0.07 &  0.07   &   0.57 &  0.10 &  0.09   &   0.91 &  0.11 &  0.22   &   0.19 &  0.05 &  0.07  \\ 
$t\bar{t}t\bar{t}$ & 0.10 & 0.01 & 0.05 & 0.44 & 0.03 & 0.23 & 1.46 & 0.05 & 0.74 & 0.75 & 0.04 & 0.38 \\
Other bkg & 0.52 & 0.07 & 0.14 & 0.68 & 0.09 & 0.24 & 0.47 & 0.08 & 0.18 & 0.20 & 0.04 & 0.06 \\
Fake/non-prompt & \multicolumn{1}{c}{4.1\phantom{1.}} & $^{+ 1.6}_{- 1.4}$ &  2.4  & \multicolumn{1}{c}{ 2.5\phantom{11}} & $^{+ 1.0}_{- 0.9}$ &  1.1  & \multicolumn{1}{c}{1.2\phantom{11}} & 
 $^{+0.9}_{- 0.7}$ & 0.6 & \multicolumn{1}{r}{0.20\phantom{1}} & $^{+ 0.46}_{- 0.20}$  & 0.16\\ 
Charge mis-ID &   1.17 &  0.10 &  0.27   &  1.29 &  0.10 &  0.28   &   0.32 &  0.04 &  0.09   &   0.21 &  0.04 &  0.04  \\ 
\midrule 
Total bkg & \multicolumn{1}{S[table-format=1.2]}{12.3} & $^{+ 2.2}_{- 2.1}$ &  3.4  & \multicolumn{1}{c}{ 9.1\phantom{11}} & $^{+ 1.2}_{- 1.1}$ &  1.2  & \multicolumn{1}{c}{ 6.2\phantom{11}} & $^{+ 1.0}_{- 0.8}$ &  1.2  & \multicolumn{1}{c}{ 2.0\phantom{11}} & $^{+ 0.5}_{- 0.2}$ &  0.3 \\ \midrule 
Data yield & \multicolumn{3}{S[table-format=1.2]}{14}  & \multicolumn{3}{S[table-format=1.2]}{10}  & \multicolumn{3}{S[table-format=1.2]}{12}  & \multicolumn{3}{S[table-format=1.2]}{4} \\ \midrule
BSM significance  &  \multicolumn{3}{c}{0.31} &   \multicolumn{3}{c}{0.25} &  \multicolumn{3}{c}{1.7} &   \multicolumn{3}{c}{1.1} \\
SM $t\bar{t}t\bar{t}$ significance  &  \multicolumn{3}{c}{0.33} &   \multicolumn{3}{c}{0.38} &  \multicolumn{3}{c}{2.1} &   \multicolumn{3}{c}{1.6} \\
\bottomrule 
\end{tabular}  \\
\begin{tabular}{ l 
 S[table-format=1.2] @{\,}@{$\pm$}@{\,} S[table-format=1.2]  @{\,}@{$\pm$}@{\,}  S[table-format=1.2]   S[table-format=1.2] @{\,}@{$\pm$}@{\,} S[table-format=1.2]  @{\,}@{$\pm$}@{\,}  S[table-format=1.2]   S[table-format=1.2] @{\,}@{$\pm$}@{\,} S[table-format=1.2]  @{\,}@{$\pm$}@{\,}  S[table-format=1.2]   S[table-format=1.2] @{\,}@{$\pm$}@{\,} S[table-format=1.2]  @{\,}@{$\pm$}@{\,}  S[table-format=1.2]  }
\toprule 
 Source  & \multicolumn{3}{c}{SR1$b$3$\ell$} & \multicolumn{3}{c}{SR2$b$3$\ell$} & \multicolumn{3}{c}{SR3$b$3$\ell$\_L} & \multicolumn{3}{c}{SR3$b$3$\ell$}\\ \midrule 
$t\bar{t}W$ &   0.66 &  0.08 &  0.20   &   0.38 &  0.05 &  0.11   &   0.21 &  0.05 &  0.09   &   0.15 &  0.04 &  0.05  \\ 
$t\bar{t}Z$ &   2.66 &  0.15 &  0.43   &   1.90 &  0.14 &  0.42   &   2.80 &  0.17 &  0.58   &   1.47 &  0.14 &  0.28  \\ 
Dibosons &   2.3 &  0.7 &  1.7   &   0.22 &  0.16 &  0.27   &   \multicolumn{3}{c}{$<0.5$}   &   \multicolumn{3}{c}{$<0.5$}   \\ 
$t\bar{t}H$ &   0.30 &  0.04 &  0.04   &   0.28 &  0.05 &  0.05   &   0.38 &  0.06 &  0.07   &   0.10 &  0.03 &  0.02  \\ 
$t\bar{t}t\bar{t}$ & 0.06 & 0.01 & 0.03 & 0.13 & 0.02 & 0.06 & 0.58 & 0.04 & 0.29 & 0.59 & 0.03 & 0.30 \\
Other bkg. & 1.37 &  0.13 &  0.45 &  0.65 & 0.10 & 0.27 & 0.17 & 0.09 & 0.10 & 0.31 & 0.07 & 0.11 \\
Fake/non-prompt & \multicolumn{1}{c}{ 1.0\phantom{11}} & $^{+ 0.6}_{- 0.5}$ &  0.6  & \multicolumn{1}{c}{ 0.14\phantom{..}} & $^{+ 0.31}_{- 0.12}$ &  0.09  & \multicolumn{1}{c}{ 0.00\phantom{.}} & \multicolumn{1}{c}{ $^{+ 0.38}_{- 0.00}$} &  \multicolumn{1}{c}{$^{+ 0.09}_{- 0.00}$}  & \multicolumn{1}{c}{ 0.03\phantom{.}} & $^{+ 0.15}_{- 0.02}$ &  0.00 \\ 
\midrule 
Total bkg & \multicolumn{1}{c}{ 8.3\phantom{81}} & $^{+ 0.9}_{- 0.8}$ &  1.8  & \multicolumn{1}{c}{ 3.7\phantom{11}} & $^{+ 0.6}_{- 0.3}$ &  0.4  & \multicolumn{1}{c}{ 4.2\phantom{1.}} & $^{+ 0.4}_{- 0.2}$ &  0.7  & 2.7 & 0.2 &  0.5 \\ \midrule 
Data yield & \multicolumn{3}{S[table-format=1.2]}{8}  & \multicolumn{3}{S[table-format=1.2]}{4}  & \multicolumn{3}{S[table-format=1.2]}{9}  & \multicolumn{3}{S[table-format=1.2]}{3} \\ \midrule
BSM significance &  \multicolumn{3}{S[table-format=1.2]}{-0.09} &   \multicolumn{3}{S[table-format=1.2]}{0.14} &  \multicolumn{3}{S[table-format=1.2]}{1.8} &   \multicolumn{3}{S[table-format=1.2]}{0.19} \\
SM $t\bar{t}t\bar{t}$ significance &  \multicolumn{3}{S[table-format=1.2]}{-0.07} &   \multicolumn{3}{S[table-format=1.2]}{0.21} &  \multicolumn{3}{S[table-format=1.2]}{2.1} &   \multicolumn{3}{S[table-format=1.2]}{0.6} \\
\bottomrule 
\end{tabular}
\end{center}
\end{table}

\begin{figure}[htbp]
  \begin{center}
    \subfloat[]{\includegraphics[width=0.50\textwidth]{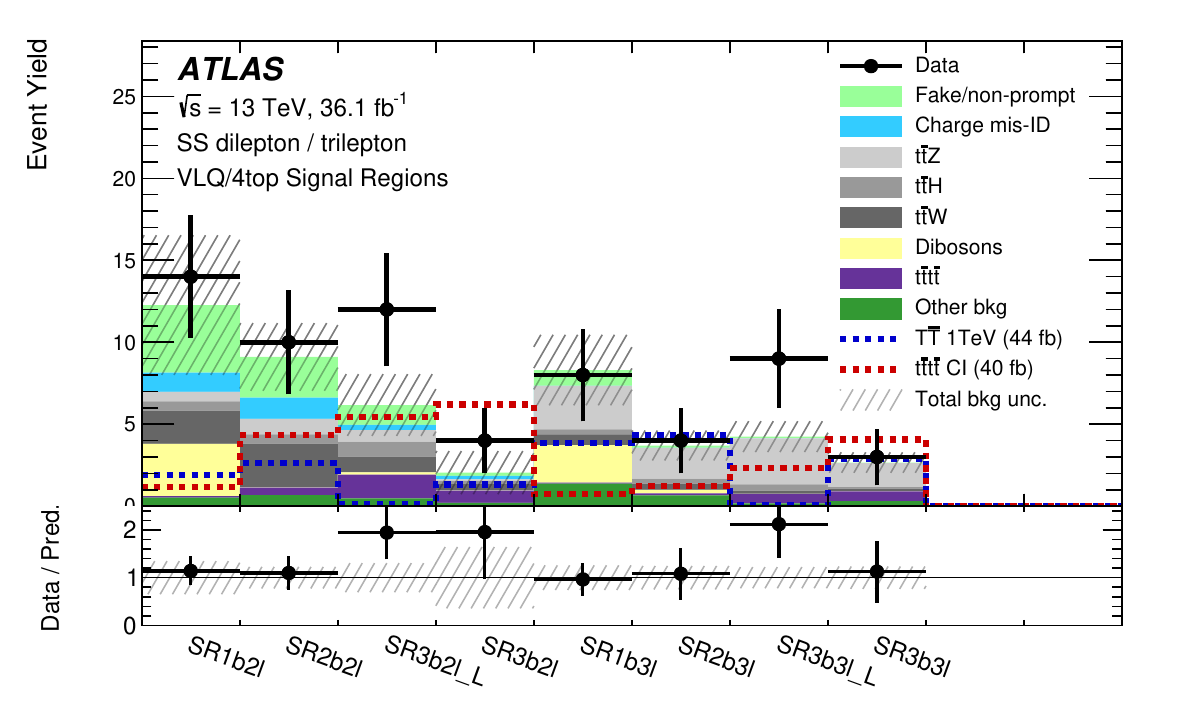}}
    \subfloat[]{\includegraphics[width=0.50\textwidth]{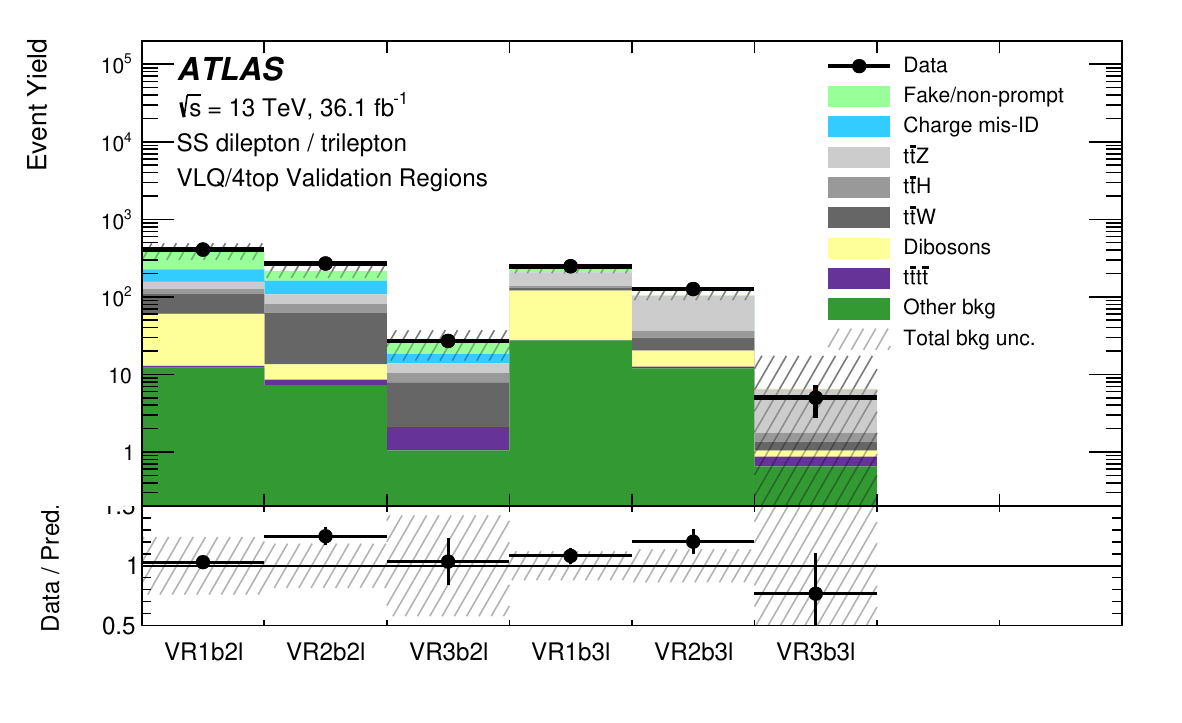}}\\
    \subfloat[]{\includegraphics[width=0.50\textwidth]{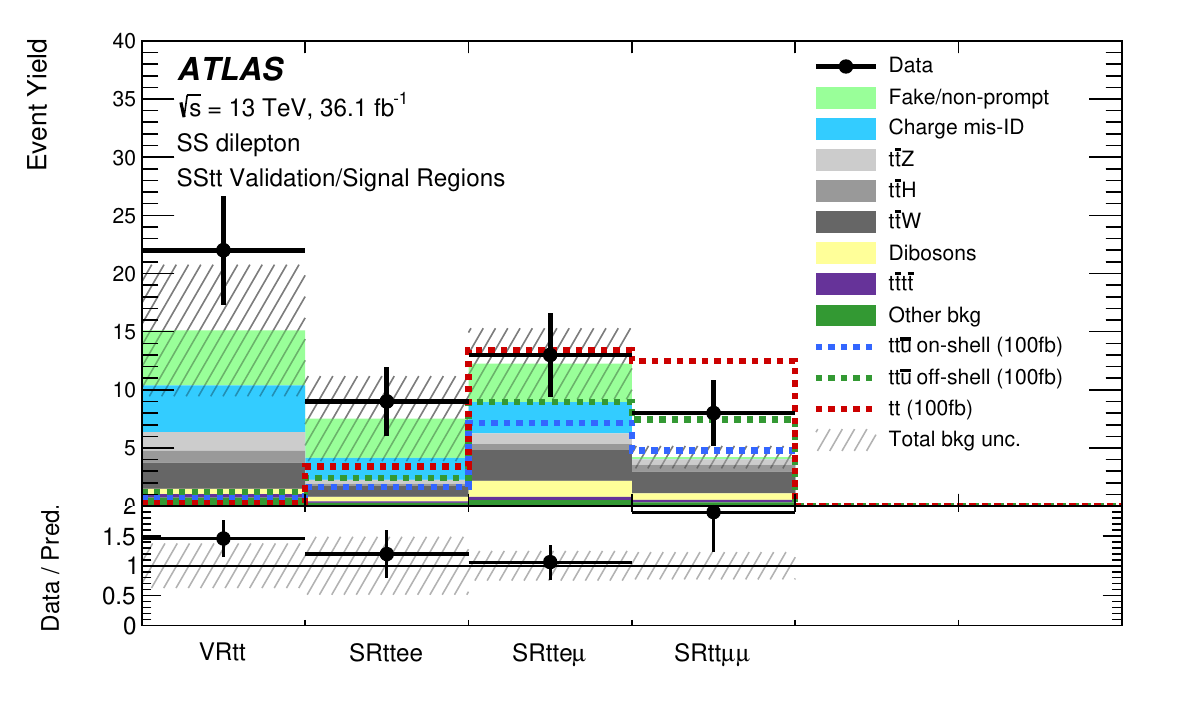}}
    \caption{Predicted background and observed data in (a) the signal regions and (b) validation regions for the vector-like quark and SM four-top-quark searches, and (c) in the signal and validation regions for the same-sign top-quark pair production search,
      along with the predicted yields for typical signals.  The uncertainty, shown as the hashed region, includes both the statistical and systematic uncertainties from each background source.
      \label{fig:SummaryPlots}}
  \end{center}
\end{figure}

\begin{table}
  \small
\begin{center}
\caption{Expected background and observed event yields in the signal regions for the same-sign top-quark pair production search.    The `Other bkg' category includes contributions from all rare SM processes listed in Section~\ref{sec:sim}.  The first uncertainty is statistical and the second is systematic.  The  significance is the number of standard deviations by which the $tt$ signal plus background hypothesis is preferred to the background-only hypothesis. It is calculated using the same procedure used to calculate the reported limits.}
\label{tab:SRSStop_yields}
\begin{tabular}{ l 
 S[table-format=1.2] @{\,}@{$\pm$}@{\,} S[table-format=1.2]  @{\,}@{$\pm$}@{\,}  S[table-format=1.2]   S[table-format=1.2] @{\,}@{$\pm$}@{\,} S[table-format=1.2]  @{\,}@{$\pm$}@{\,}  S[table-format=1.2]   S[table-format=1.2] @{\,}@{$\pm$}@{\,} S[table-format=1.2]  @{\,}@{$\pm$}@{\,}  S[table-format=1.2]    }
\toprule 
 Source  & \multicolumn{3}{c}{SRtt$ee$} & \multicolumn{3}{c}{SRtt$e\mu$} & \multicolumn{3}{c}{SRtt$\mu\mu$} \\ \midrule 
$t\bar{t}W$ &   0.91 &  0.09 &  0.19   &   2.64 &  0.15 &  0.48   &   1.86 &  0.13 &  0.37    \\ 
$t\bar{t}Z$ &   0.35 &  0.07 &  0.09   &   0.91 &  0.09 &  0.12   &   0.47 &  0.08 &  0.09    \\ 
Dibosons &   0.40 &  0.45 &  0.09   &   1.4 & 0.6 & 0.9 &   0.5 &  0.5 &  0.5      \\ 
$t\bar{t}H$ &   0.19 &  0.06 &  0.02   &   0.53 &  0.08 &  0.08   &   0.58 &  0.07 &  0.05    \\ 
$t\bar{t}t\bar{t}$ & 0.12 & 0.02 & 0.06 & 0.30 & 0.02 & 0.15 & 0.22 & 0.03 & 0.11 \\
Other bkg       &   0.29 &  0.06 &  0.13 &  0.51 &  0.08 &  0.16 &  0.33 & 0.08 & 0.12 \\
Fake/non-prompt & \multicolumn{1}{c}{ 3.4\phantom{11}} & $^{+ 2.1}_{- 1.7}$ &  2.5  & \multicolumn{1}{c}{ 3.3\phantom{11}} & $^{+1.2}_{- 1.1}$ &  2.1  & \multicolumn{1}{c}{0.20\phantom{.}} & 
 $^{+0.24}_{- 0.20}$ & 0.18 \\ 
Charge mis-ID &   1.90 &  0.11 &  0.91   &  2.69 &  0.14 & 0.59   &   \multicolumn{3}{c}{N/A}    \\ 
\midrule 
Total bkg. & \multicolumn{1}{c}{7.5\phantom{11}} & $^{+ 2.2}_{-1.8}$ &  2.7  &12.2 &1.3 &  2.5  & \multicolumn{1}{c}{4.2\phantom{11}} & $^{+ 0.6}_{- 0.6}$ &  0.7   \\ \midrule 
Data yield & \multicolumn{3}{S[table-format=1.2]}{9}  & \multicolumn{3}{S[table-format=1.2]}{13}  & \multicolumn{3}{S[table-format=1.2]}{8}  \\ \midrule
Significance &  \multicolumn{3}{S[table-format=1.2]}{0.31} &   \multicolumn{3}{S[table-format=1.2]}{0.16} &  \multicolumn{3}{S[table-format=1.2]}{1.44} \\
\bottomrule 
\end{tabular}  \\
\end{center}
\end{table}

\clearpage
 Limits on $B$- and $T$-quark pair production are set in two scenarios. In the first, it is assumed that the branching ratios are given by the singlet model of Ref.~\cite{AguilarSaavedra:2009es}. These branching ratios vary slightly with the VLQ mass; they are approximately $(0.48, 0.27, 0.25)$ for $B \rightarrow (Wt,Zb,Hb)$ and $(0.49, 0.22, 0.27)$ for $T \rightarrow (Wb, Zt, Ht)$. The resulting 95\% CL upper limits on the production cross-section as a function of the VLQ mass are shown in 
Figure~\ref{Results:VLQ_TT_BB}. Lower limits on the $B$- and $T$-quark masses are extracted from these cross-section limits, resulting in observed (expected) excluded mass $m_B < 1.00$ \TeV{} (1.01 \TeV{}) and  $m_T < 0.98$ \TeV{} (0.99 \TeV{}). The expected and observed limits agree well in spite of the observed excesses in some signal regions because the expected yield of VLQ in those regions is small. In the second scenario, no assumptions are made about the branching ratio, and lower limits on the masses are determined for any possible set of branching ratios, as shown in Figure~\ref{fig:TT_BB_2DLimits}.

\begin{figure}[htbp]
  \begin{center}
    \subfloat[]{\includegraphics[scale=0.4]{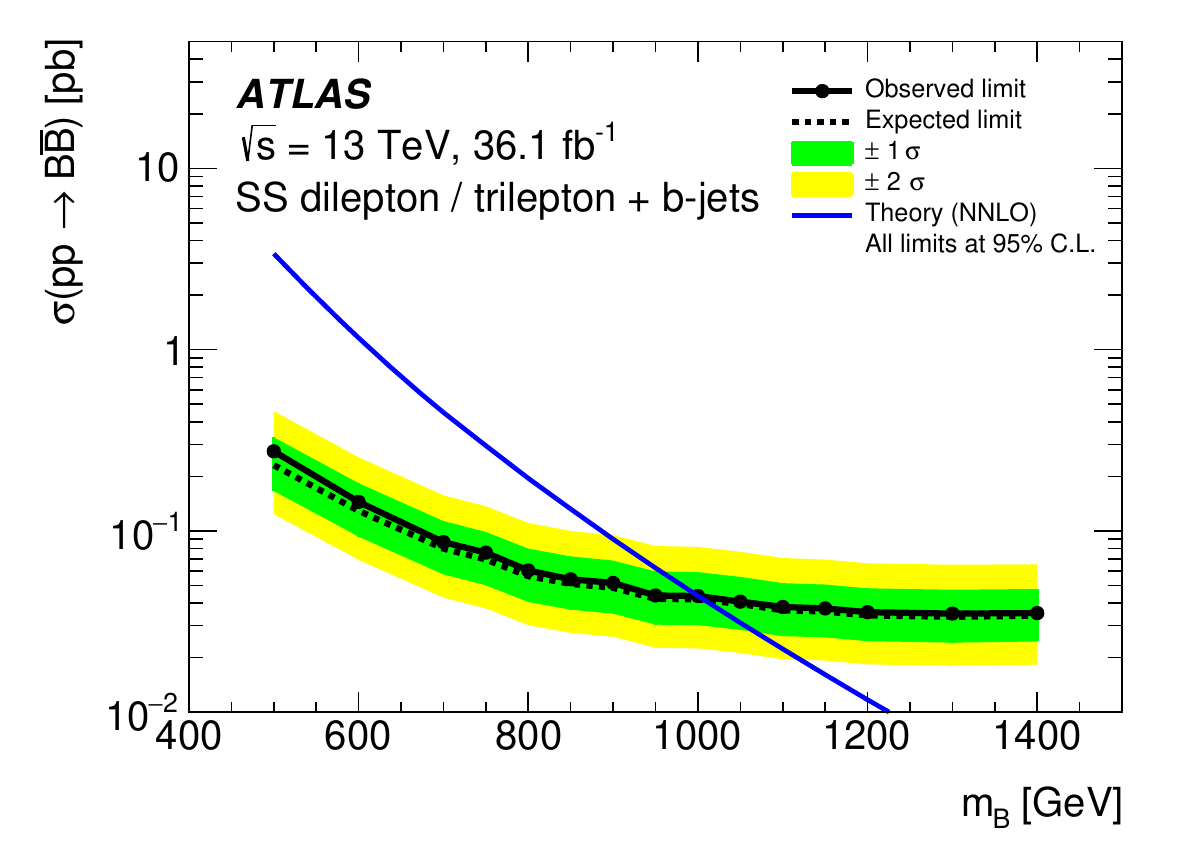}}
    \subfloat[]{\includegraphics[scale=0.4]{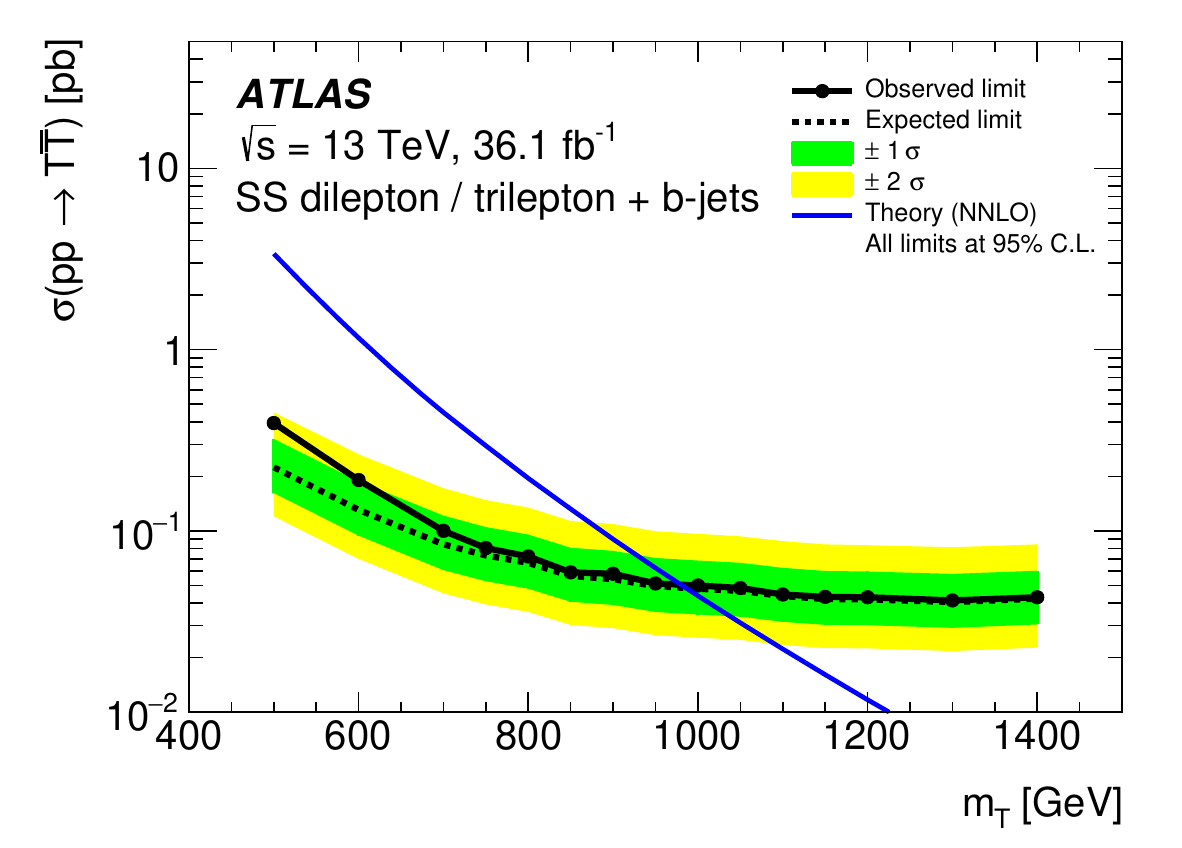}}
    \caption{Expected and observed limits on (a) vector-like $B$- and (b) $T$-quark  pair production as a function of mass, assuming
      the branching ratios expected in the singlet model. The expected 95\% CL limits are shown as a dashed line with its $\pm 1$ and $\pm 2$ standard deviation bands, and the NNLO theory prediction is shown as a continuous line.}\label{Results:VLQ_TT_BB}
  \end{center}
\end{figure}

\begin{figure}[!htb]
  \begin{center}
    \subfloat[]{\includegraphics[width=.49\textwidth]{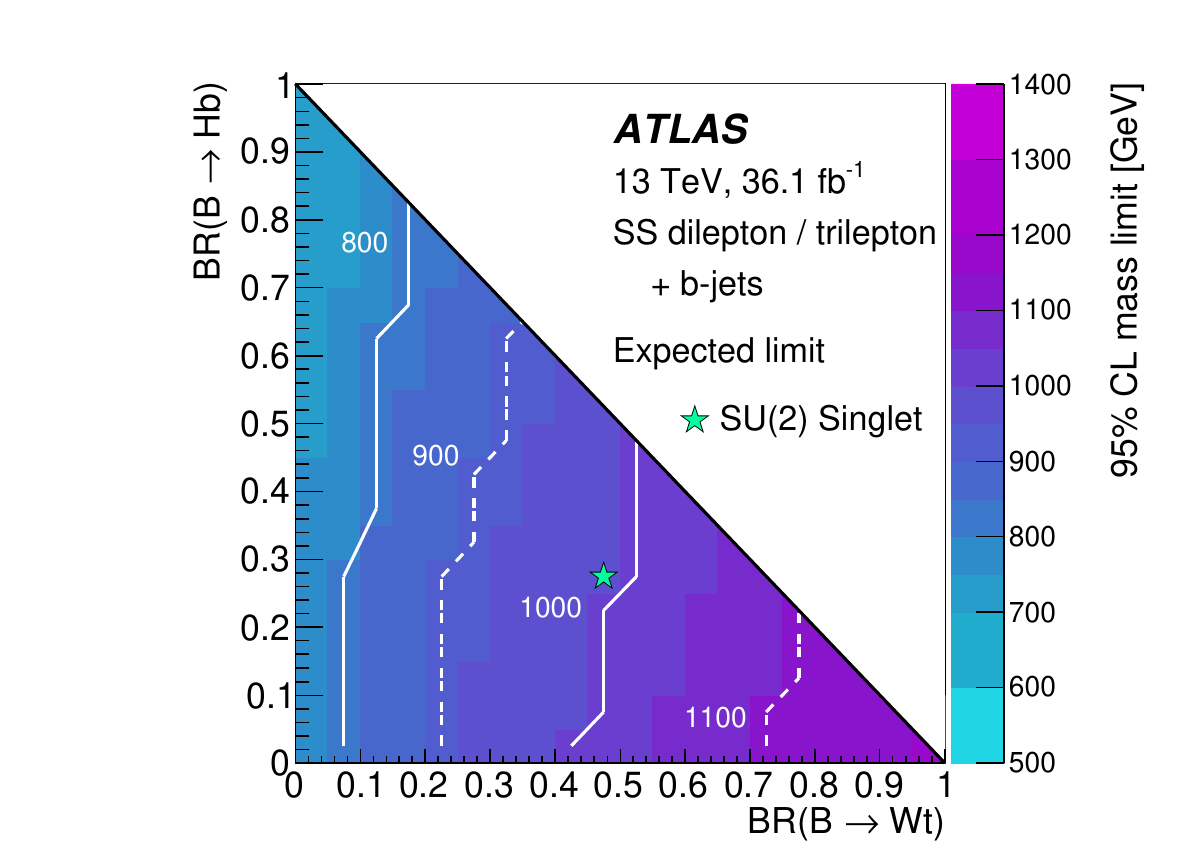}}
    \subfloat[]{\includegraphics[width=.49\textwidth]{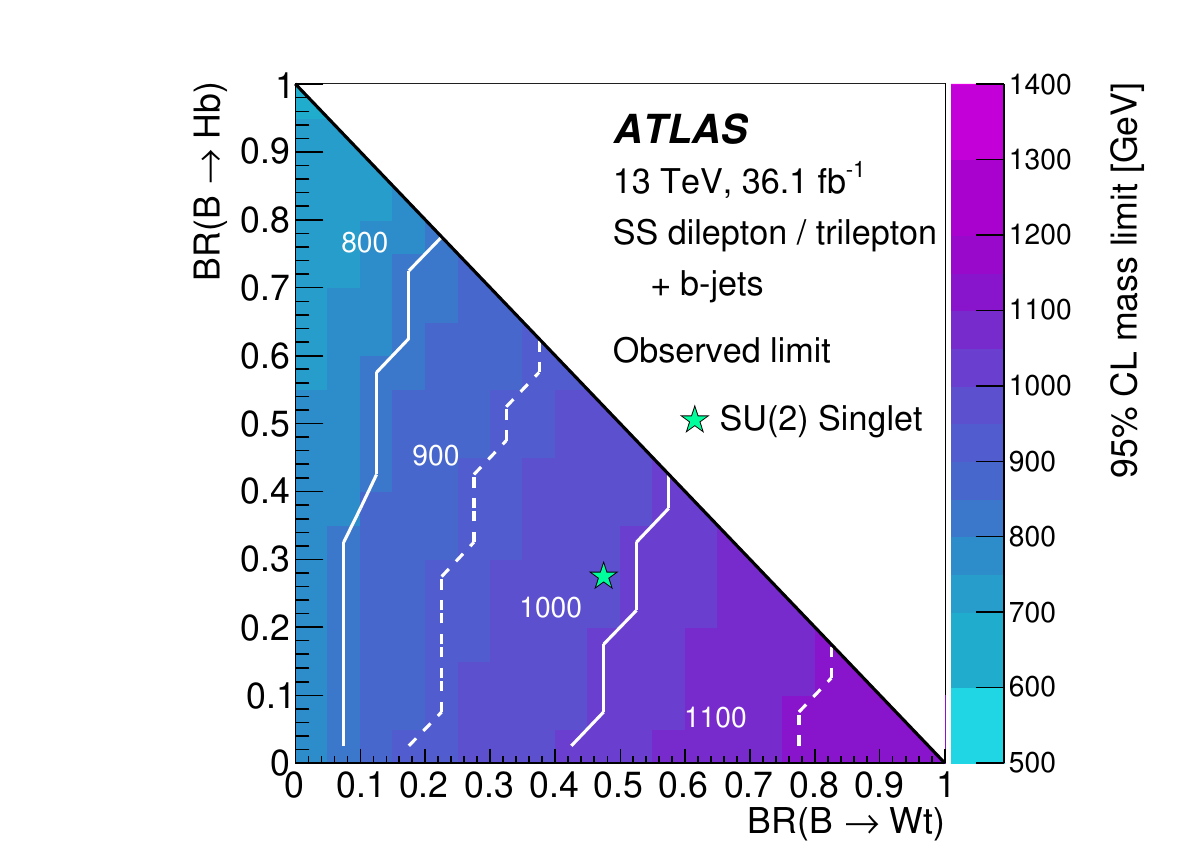}}\\
    \subfloat[]{\includegraphics[width=.49\textwidth]{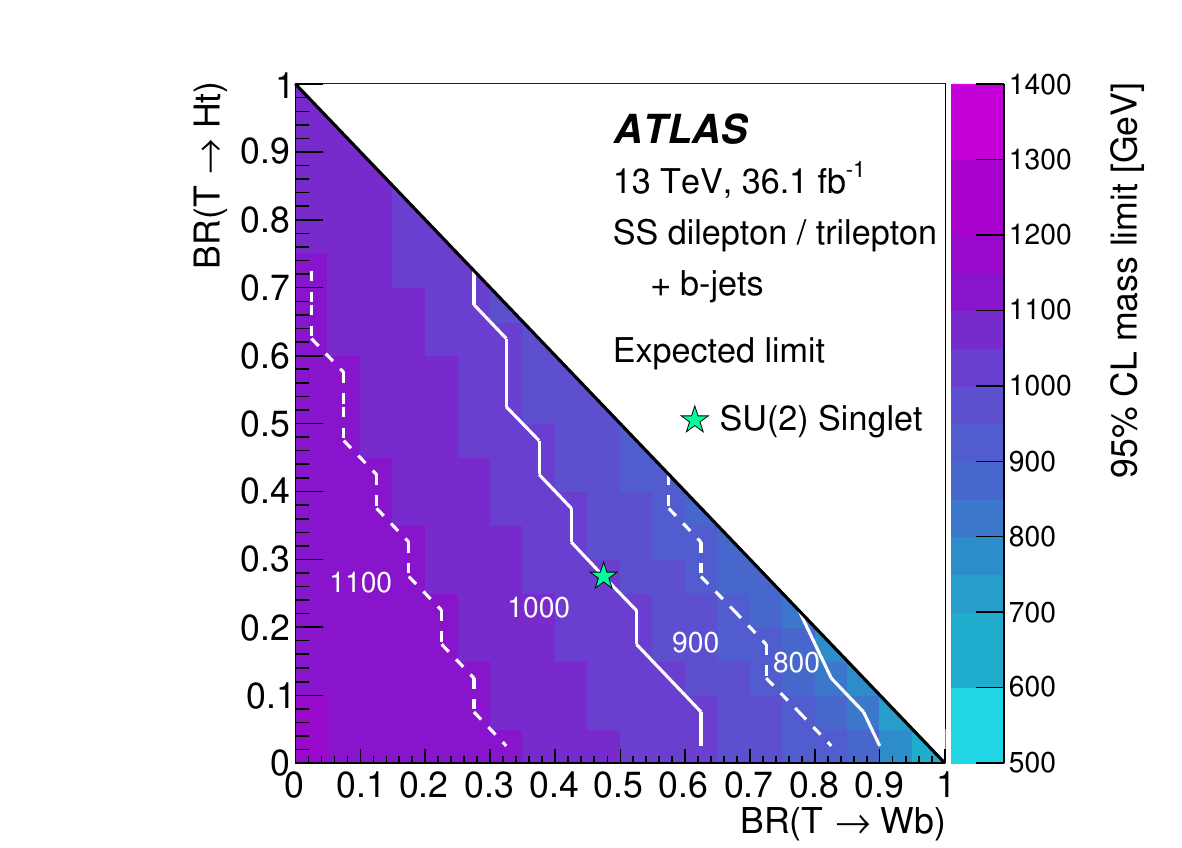}}
   \subfloat[]{\includegraphics[width=.49\textwidth]{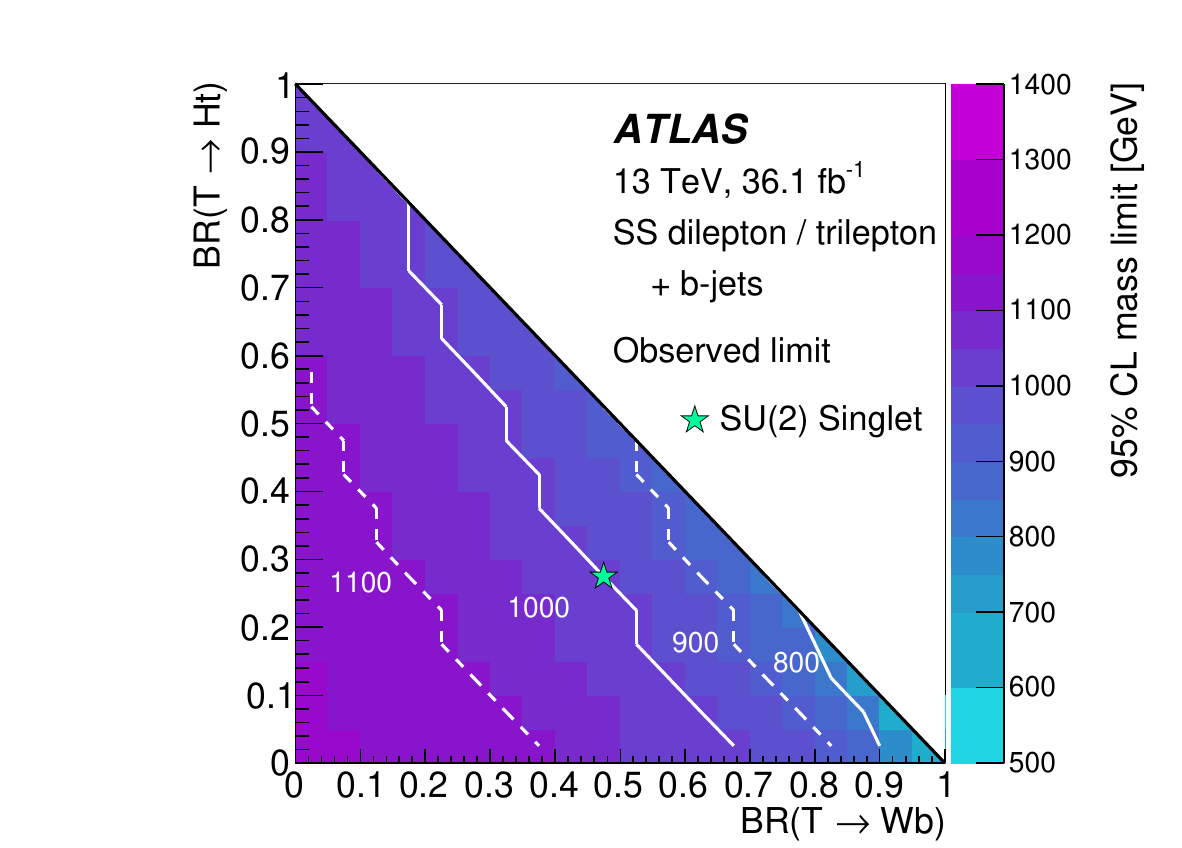}}
    \caption{Mass hypotheses excluded at 95\% CL as a function of the branching ratio: (a) expected and (b) observed limits for a vector-like $B$-quark, (c) expected and (d) observed limits for a vector-like $T$-quark. Contours of constant 95\% CL lower mass limits are shown in white in each plot, and labelled with the mass limit in \GeV{}. The star represents the branching ratios for the SU(2) singlet models of Ref.~\cite{AguilarSaavedra:2009es}.}
    \label{fig:TT_BB_2DLimits}
  \end{center}
\end{figure}

Since a single $T_{5/3}$  quark could decay into a same-charge lepton pair, limits on both single and pair production of $T_{5/3}$ quarks are set.  If only pair production is considered, then the cross-section limit as a function of mass is unambiguous since the only allowed decay channel is $T_{5/3} \rightarrow Wt$, as shown in Figure~\ref{Results:VLQ_XXandX_a}.  The corresponding lower observed (expected) limit on the $T_{5/3}$-quark mass is 1.19 \TeV{} (1.21 \TeV{}). If single production is considered in addition to pair production, the limit depends on the assumed strength of the $T_{5/3}tW$ coupling, as shown in Figure~\ref{Results:VLQ_XXandX_b}. 

\begin{figure}[htbp]
  \begin{center}
    \subfloat[]{\includegraphics[width=.5\linewidth]{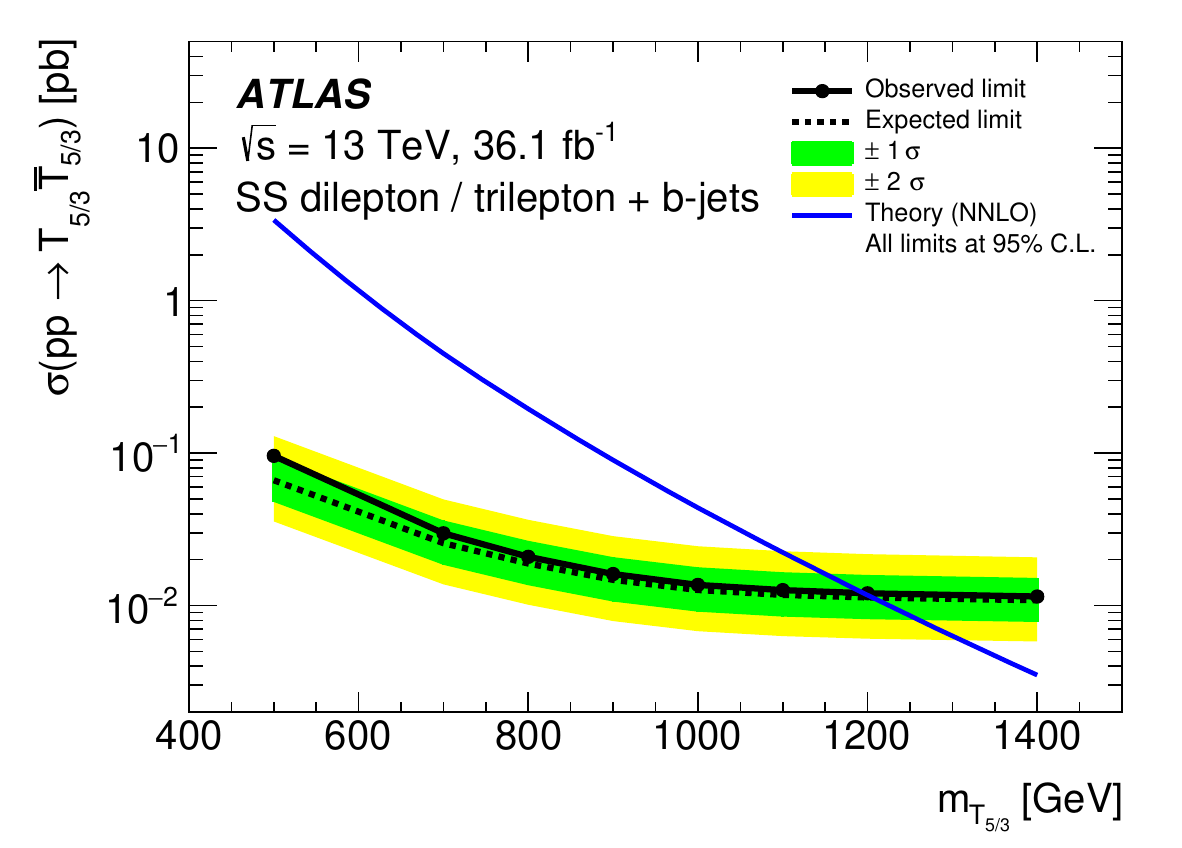}\label{Results:VLQ_XXandX_a}}
    \subfloat[]{\includegraphics[width=.5\linewidth]{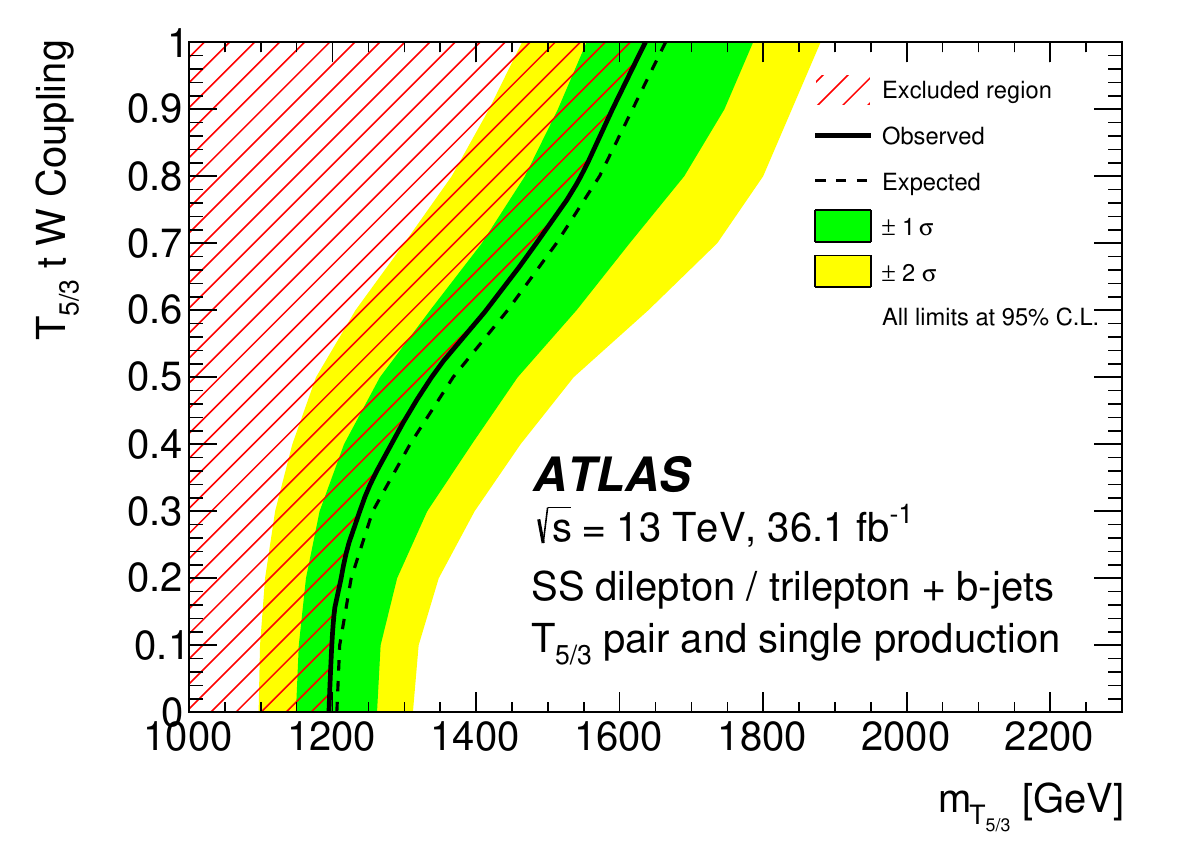}\label{Results:VLQ_XXandX_b}}
    \caption{(a) Expected and observed limits on vector-like $T_{5/3}$ pair production as a function of mass.  The NNLO theory prediction is shown as the continuous line. (b) Expected and observed limits on vector-like $T_{5/3}$ single- plus pair-production as a function of mass and $T_{5/3}tW$ coupling. In both plots, the expected 95\% CL limits are shown with their $\pm 1$ and $\pm 2$ standard deviation bands and it is assumed that the branching ratio $\mathcal{B}(T_{5/3}\to Wt) = 100\%$.}\label{Results:VLQ_XXandX}
  \end{center}
\end{figure}

Limits on four-top-quark production are set in a variety of models. The observed (expected) limit on the cross-section assuming SM kinematics is $69\,$fb ($29\,$fb), and the observed (expected) 
limit assuming kinematics from the EFT model is $39\,$fb ($21\,$fb).  These results are summarised in Table~\ref{tab:limitExp:4t}, along with the limits on the ratio of the  contact interaction strength to the cut-off scale in the EFT model.  The latter limit can also be expressed in the plane of the interaction strength $|\C4t|$ versus cut-off scale $\Lambda$, as shown in Figure~\ref{fig:ued:expected_1D_a}.  Limits on the 2UED/RPP model are shown in Figures~\ref{fig:ued:expected_1D_b} and~\ref{fig:ued:expected_1D_c}. 
The 2UED/RPP limits corresponds to an observed lower limit on the parameter $m_{\mathrm{KK}}$ of 1.45~\TeV{}, where a limit of 1.48~\TeV{} is expected in the background-only model.
Limits on four-top-quark production in the 2HDM interpretation are shown in Figure~\ref{fig:2hdm:expected_1D}. 
Two scenarios are considered: one where only the heavy scalar Higgs boson contributes to the process $pp \to t\bar{t}H$ and $H \to t\bar{t}$, and one where the heavy scalar and pseudo-scalar Higgs bosons
have equal masses and both contribute to the four-top-quark production. Limits are computed and compared with
theoretical predictions at partial NNLO in QCD~\cite{Eriksson:2009ws,Harlander:2012pb,Harlander:2016hcx}. 

\begin{figure}[htbp]
  \begin{center}
    \subfloat[]{\includegraphics[width=0.5\textwidth]{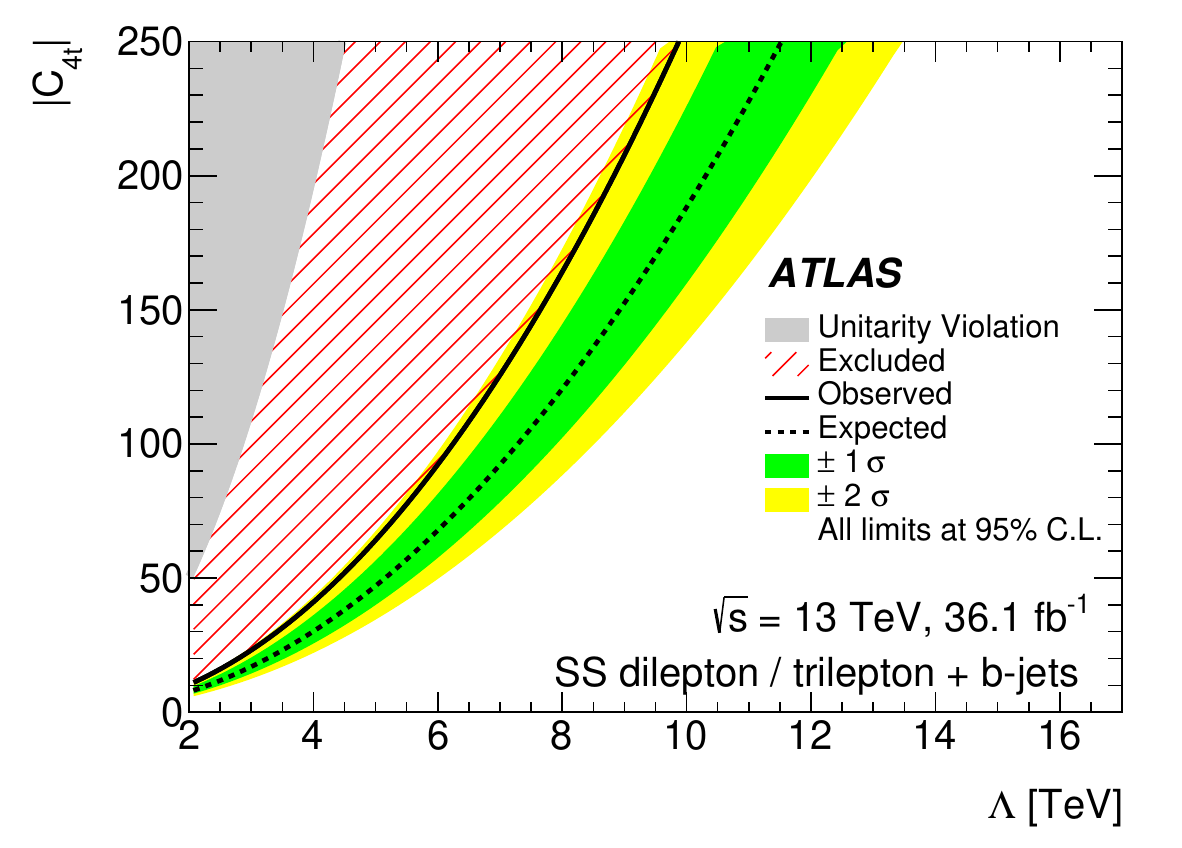}\label{fig:ued:expected_1D_a}}
    \subfloat[]{\includegraphics[width=0.5\textwidth]{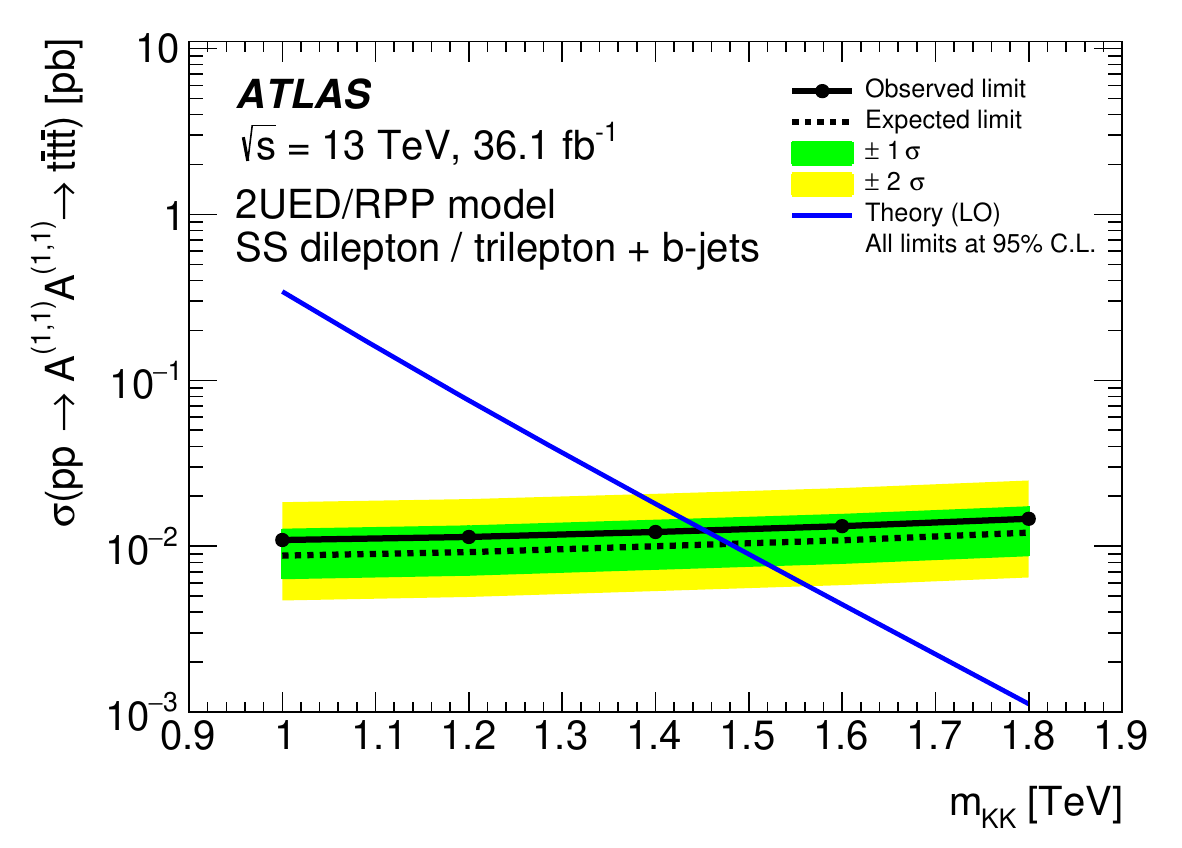}\label{fig:ued:expected_1D_b}} \\
    \subfloat[]{\includegraphics[width=0.5\textwidth]{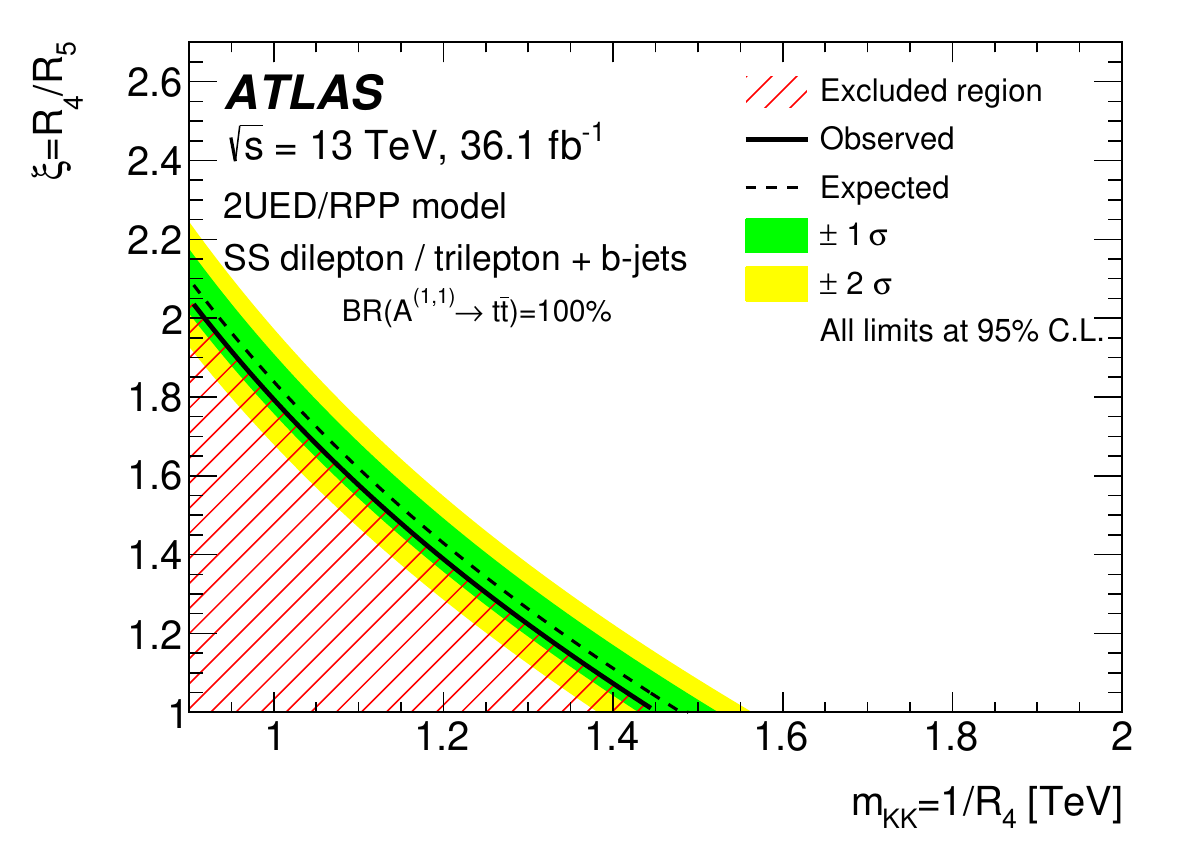}\label{fig:ued:expected_1D_c}}
    \caption{(a) Constraints in the $(|C_{4t}|,\Lambda)$ plane (black line), and
    (b) expected and observed limits on $m_{\mathrm{KK}}$ in the symmetric case. The theory line corresponds to the production of four top quarks 
    by tier $(1,1)$ assuming the branching ratio $\mathcal{B}(A^{(1,1)} \to t\bar{t}) = 100\%$. (c) Expected and observed limit in the $(m_{\mathrm{KK}} = 1/R_4,\xi = R_4/R_5)$
    plane for the 2UED/RPP model. In all plots, the expected 95\% CL limits are shown with their $\pm 1$ and $\pm 2$ standard deviation  bands.}\label{fig:ued:expected_1D}
  \end{center}
\end{figure}

\begin{table}[!htb]
\small
  \begin{center}
  \caption{Expected and observed 95\% CL upper limits on the four-top-quark production cross-section in various models.}\label{tab:limitExp:4t}
    \begin{tabular}{lcrr}
     \toprule\toprule
         Observable &  Expected median with $1\sigma$ range & Observed \\ \midrule
        SM cross-section [fb] &  $29.0^{\;+\;12.2}_{\;-\; 8.1}$  & 69.2 \\ \midrule
        CI cross-section [fb] & $20.8^{\;+\;12.2}_{\;-\; 8.1}$ & 38.6 \\
        \midrule
        CI coupling $|\C4t|/\Lambda^2$ (TeV$^{-2}$) & $1.9^{\;+\;1.2}_{\;-\;0.7}$ & $2.6$ \\     \bottomrule\bottomrule
   \end{tabular}
  \end{center}
\end{table}

\begin{figure}[htbp]
  \begin{center}
    \subfloat[]{\includegraphics[width=0.5\textwidth]{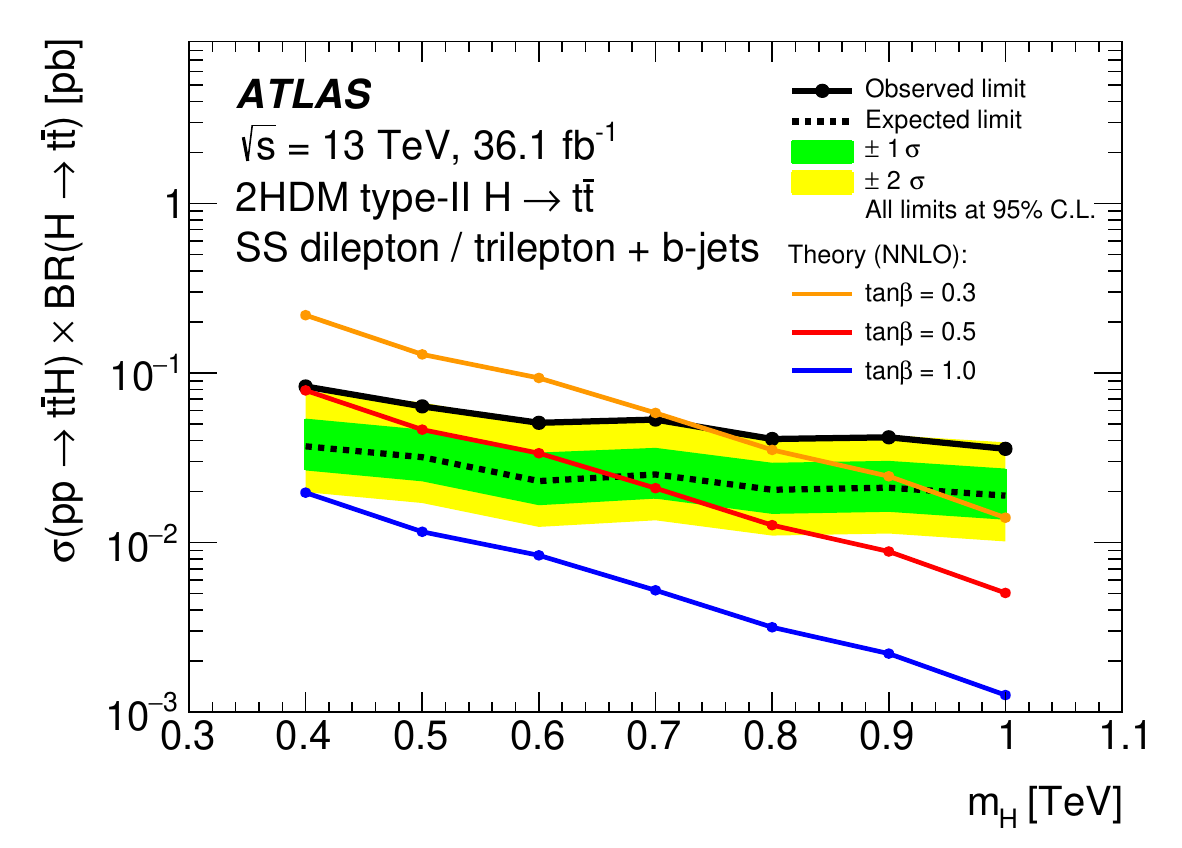}\label{fig:2hdm:expected_1D_a}}\\
    \subfloat[]{\includegraphics[width=0.5\textwidth]{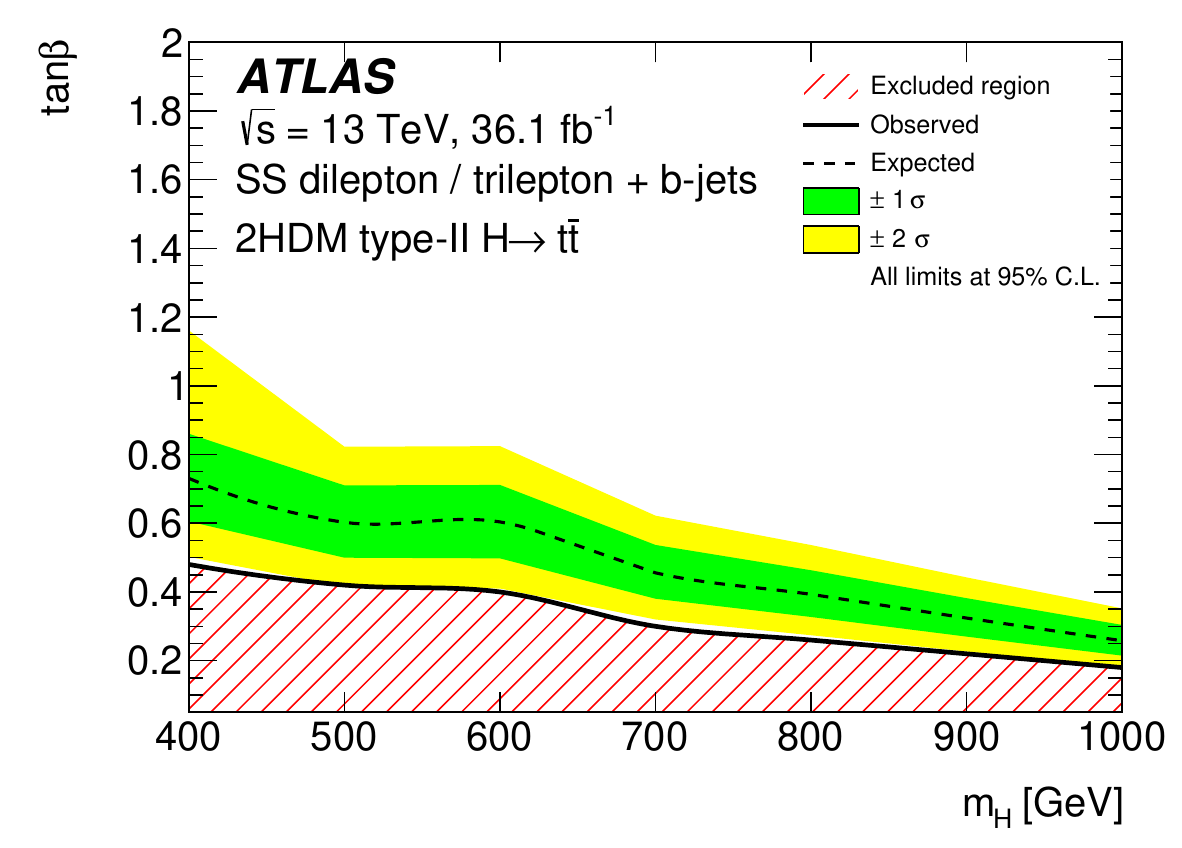}\label{fig:2hdm:expected_1D_b}}
    \subfloat[]{\includegraphics[width=0.5\textwidth]{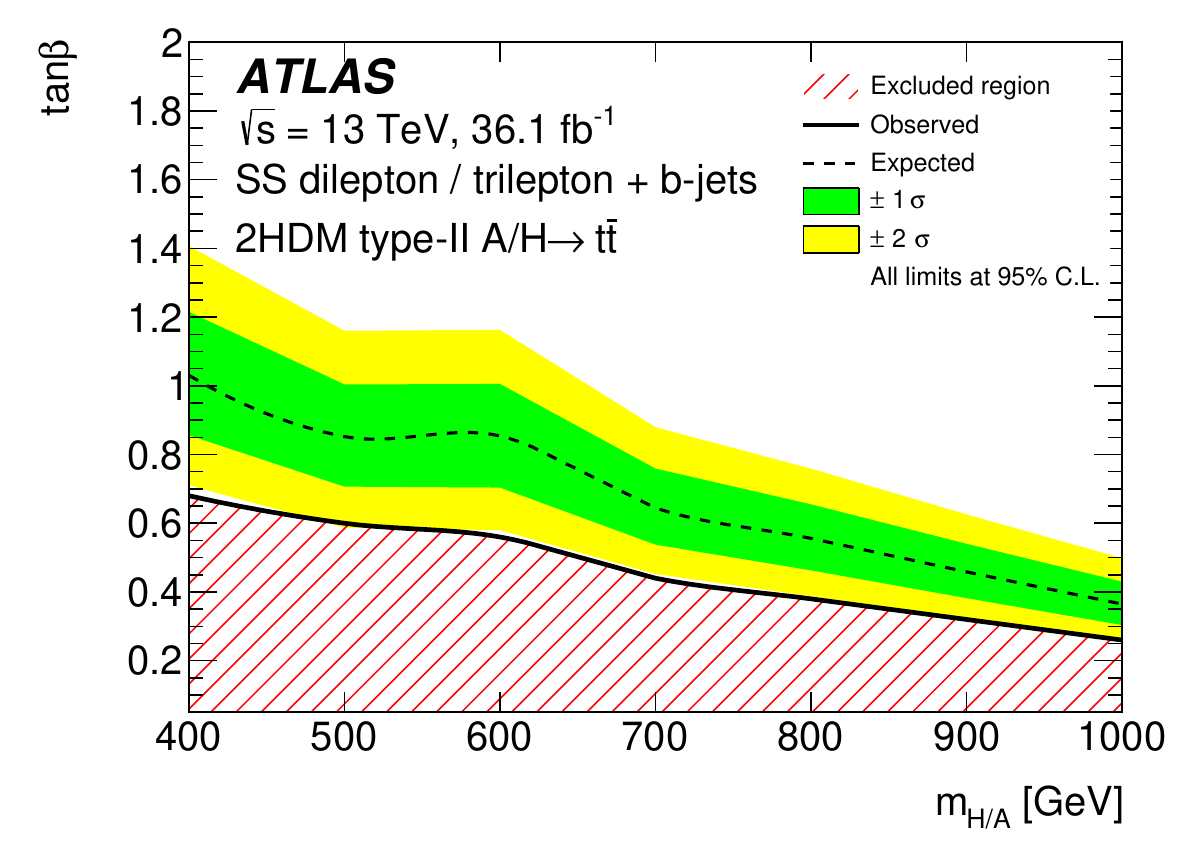}\label{fig:2hdm:expected_1D_c}}
    \caption{Limits on the two-Higgs-doublet model interpretation. (a) Limits on the cross-section of the four-top-quark production through a heavy scalar Higgs boson times the branching ratio for the Higgs boson to decay into \ttbar.
      Theoretical predictions for three values of $\tan\beta$ are shown.
      (b) Limits on four-top-quark production through a heavy scalar Higgs boson in the plane ($m_H$, $\tan\beta$).
      (c) Limits on four-top-quark production considering both a heavy pseudo-scalar and a scalar Higgs boson having the same mass $m_{H/A}$
      in the plane ($m_{H/A}$, $\tan\beta$). In all plots, the expected 95\% CL limits are shown with their $\pm 1$ and $\pm 2$ standard deviation bands.
      In the context of the two-Higgs-doublet model, the Higgs boson width can be large for low $\tan\beta$. In spite of that,
      it was checked that the signal efficiency has a negligible dependence on $\tan\beta$ in the region of interest.
    }\label{fig:2hdm:expected_1D}
  \end{center}
\end{figure}

Limits on same-sign top-quark pair production are set using the signal regions dedicated to this signature (SRtt$ee$, SRtt$e\mu$, and SRtt$\mu\mu$).  These are interpreted in the context of a dark-matter model with three parameters: the mass of the exotic FCNC mediator particle $m_V$, and the couplings $g_{\text{DM}}$ and $g_{\text{SM}}$ of the mediator to dark-matter and SM particles, respectively. In the context of this model, three different same-sign top-quark pair production processes are considered: $i$) production via a $t$-channel mediator, $ii$) production via an on-shell $s$-channel mediator, and $iii$) production via an off-shell $s$-channel mediator. Limits on the production cross-section for each mechanism as a function of $m_V$ are shown in Figure~\ref{fig:sstops:expected_reference_limits}. They are independent of the model parameter values and constrain generic processes such as $uu\to tt$ to a cross-section of less than $89\,$fb, where a limit of $59\,$fb is expected in the background-only model. More generally, the total signal includes the three contributions with a relative importance which depends on the mediator's total width, and thus on the model parameters. Limits in the plane of $g_{\text{SM}}$ versus $m_V$ for three values of $g_{\text{DM}}$ are shown in Figure~\ref{fig:sstops:mu_visible_vs_mV} where the width effects are taken into account. As a reference, the observed upper limit on $g_{\text{SM}}$ is 0.31 when $m_V$ is taken to be 3 \TeV{} and $g_{\text{DM}}$ is taken to be one, where a limit of 0.28 is expected in the background-only hypothesis.  Assuming that $m_V = 1$ \TeV{} and $g_{\text{DM}} = 1$, the observed (expected) upper limit on $g_{\text{SM}}$ is 0.14 (0.13).

\begin{figure}[!htb]
  \begin{center}
          \subfloat[]{\includegraphics[width=0.5\linewidth]{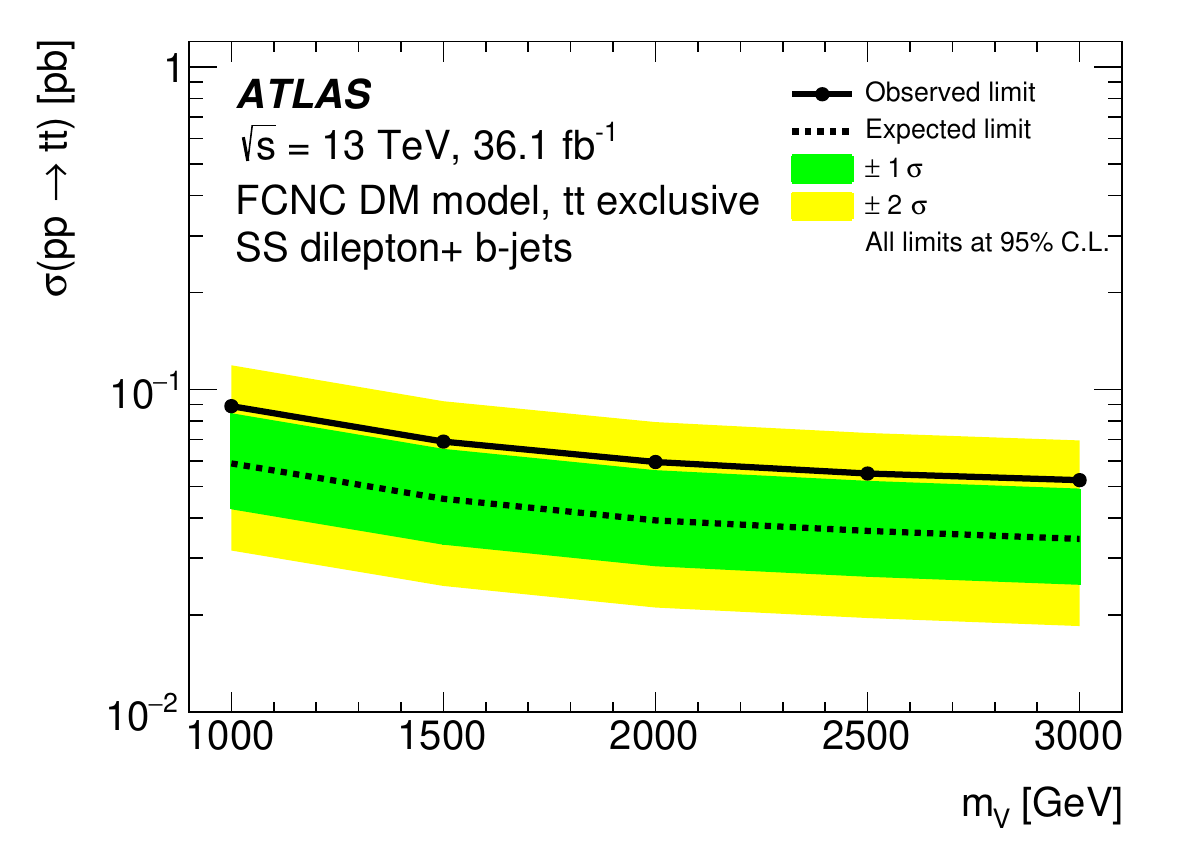}} \hfill
          \subfloat[]{\includegraphics[width=0.5\linewidth]{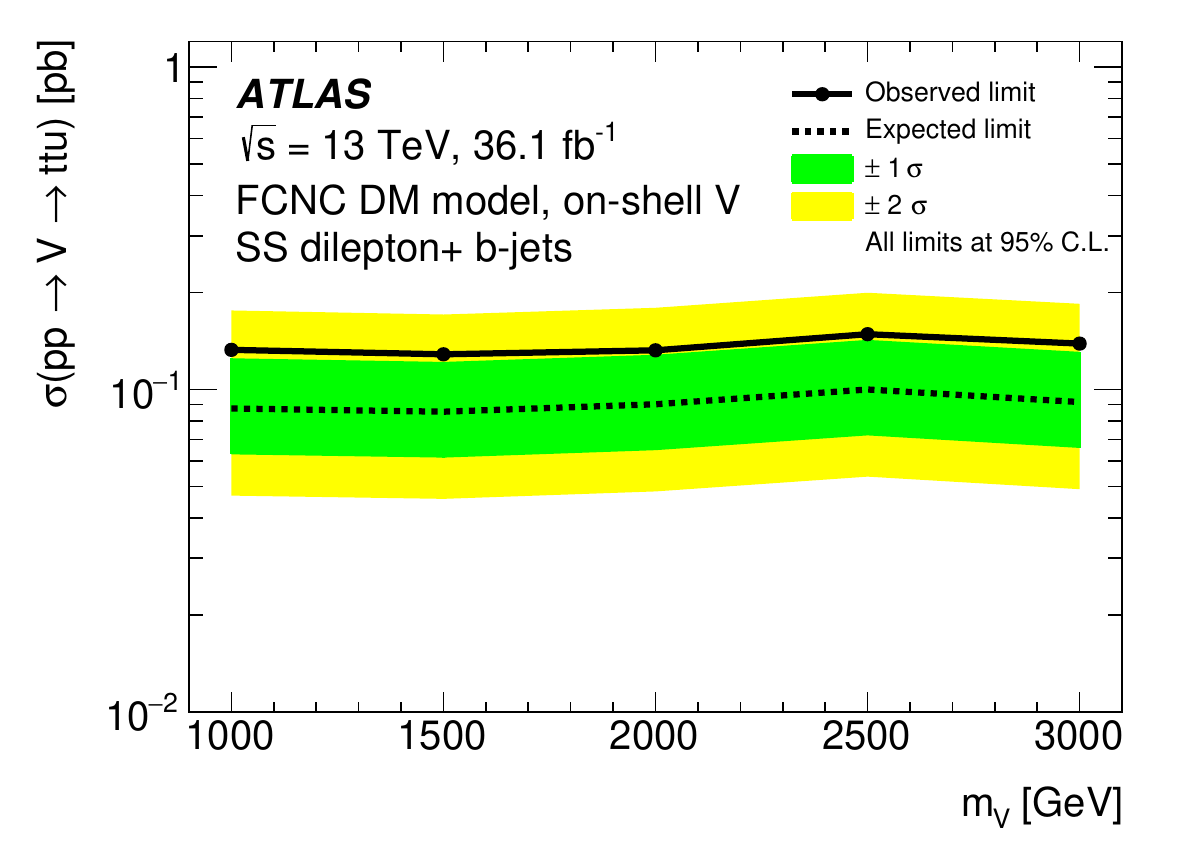}} \\
          \subfloat[]{\includegraphics[width=0.5\linewidth]{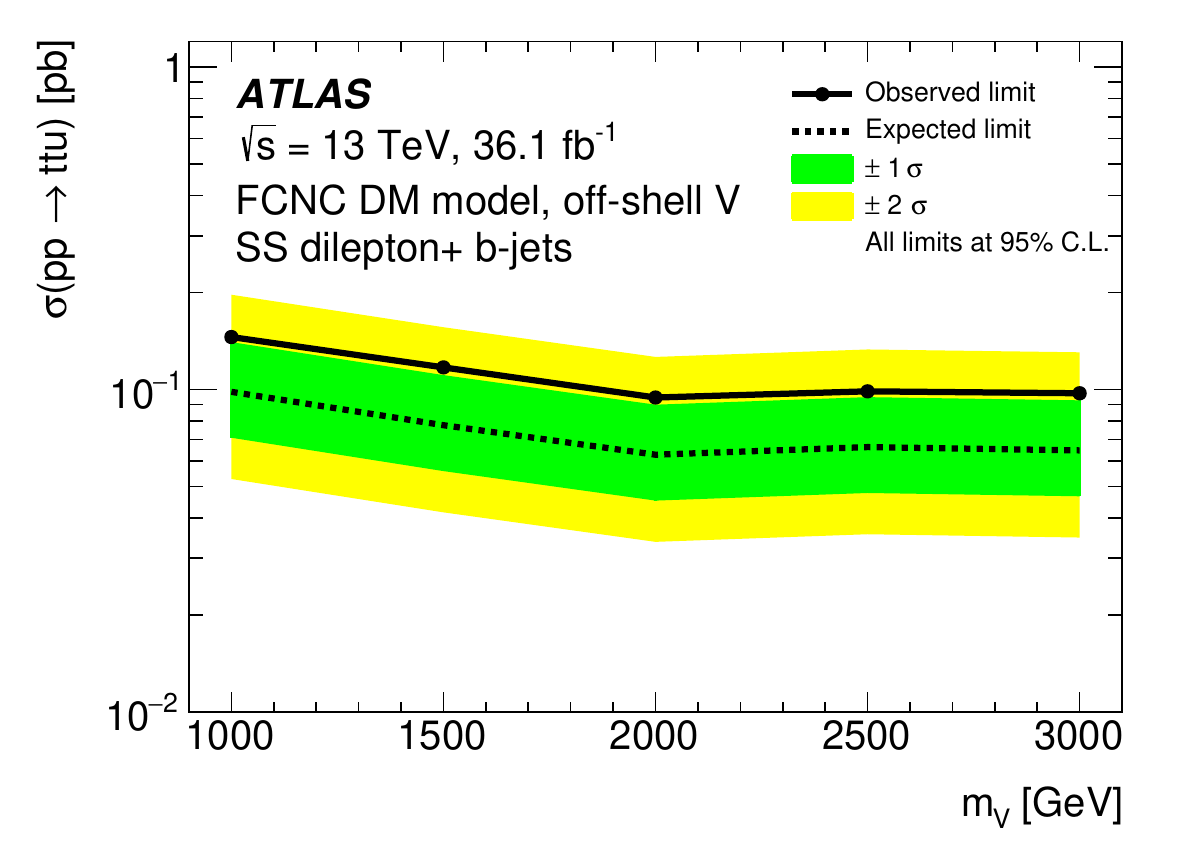}}
          \caption{Expected and observed 95\% CL upper limits on the cross-section for (a) prompt $tt$ production, (b) on-shell mediator, (c) off-shell mediator subprocesses of the same-sign top-quark pair production.
            In all plots, the expected 95\% CL limits are shown with their $\pm 1$ and $\pm 2$ standard deviation bands. Each subprocess is considered as a generic BSM signature and therefore no
          theory prediction is shown. } 
          \label{fig:sstops:expected_reference_limits}
        \end{center}
\end{figure}

\begin{figure}[!htb]
  \begin{center}
          \subfloat[]{\includegraphics[width=0.5\linewidth]{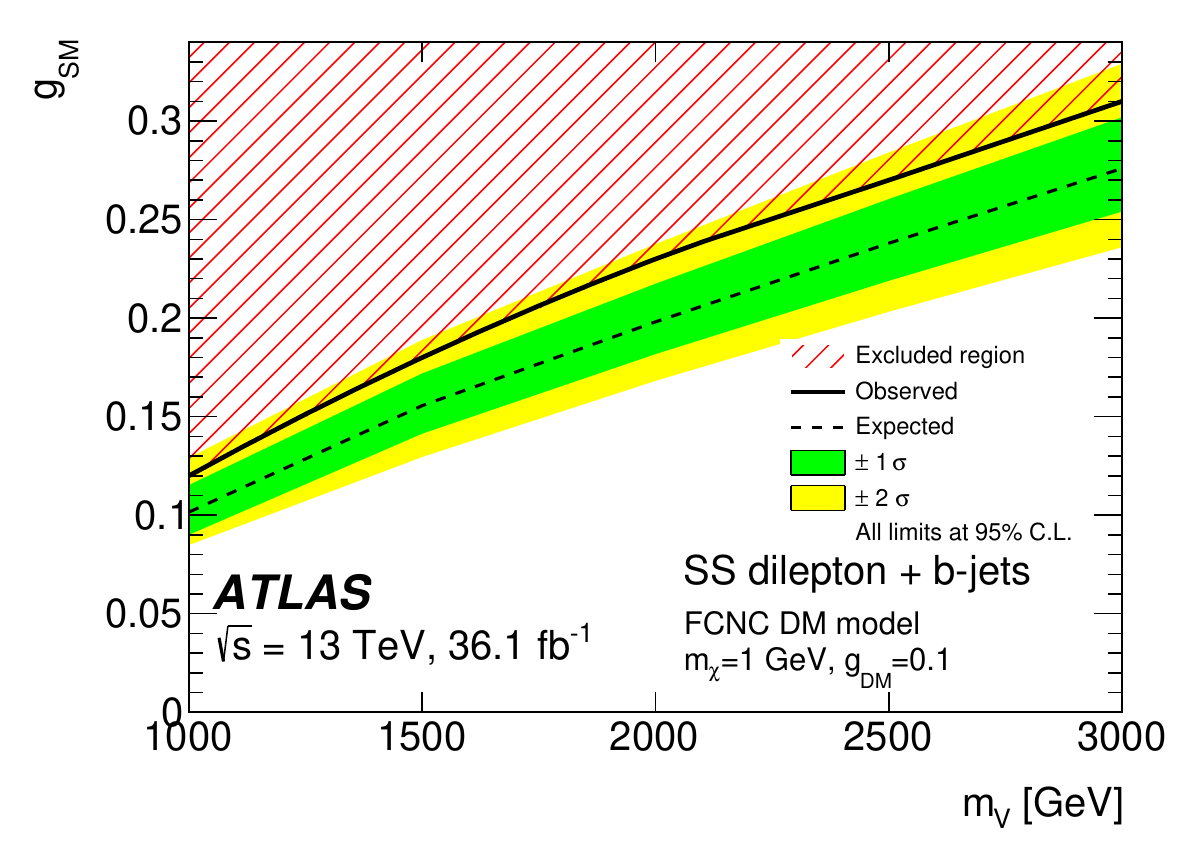}} \hfill
          \subfloat[]{\includegraphics[width=0.5\linewidth]{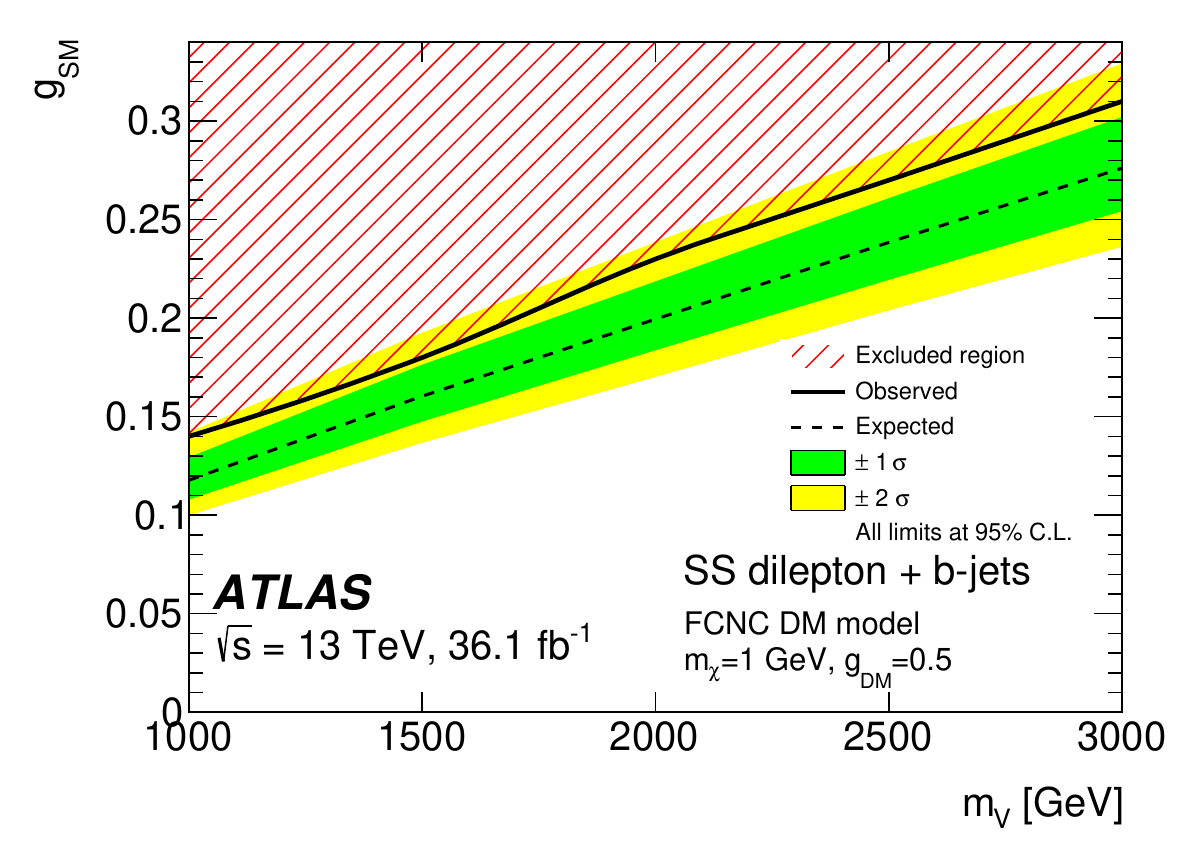}} \\
          \subfloat[]{\includegraphics[width=0.5\linewidth]{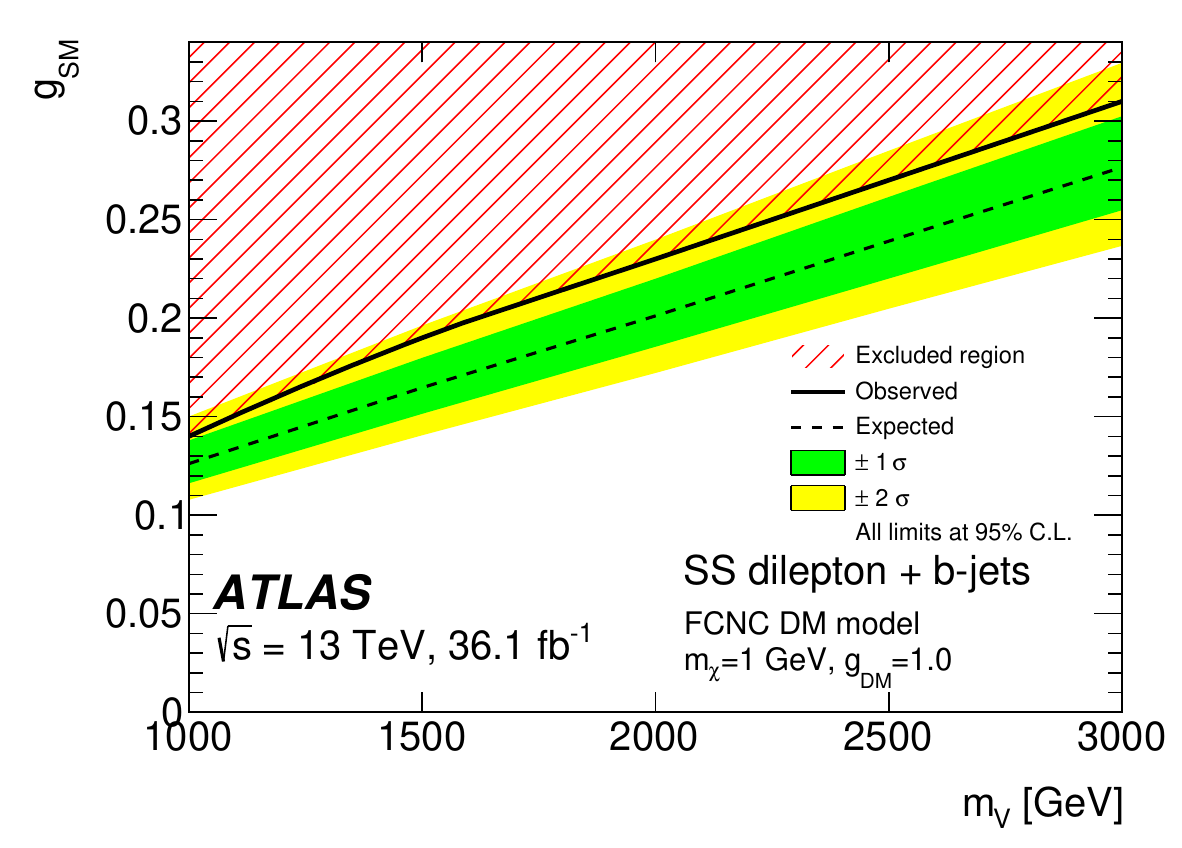}}
          \caption{Expected and observed constraints in the $(\gSM, m_V)$ plane for the combined visible same-sign top-quark pair production,
            (a) $\gDM=0.1$, (b) $\gDM=0.5$, (c) $\gDM=1$. In all plots, the expected 95\% CL limits are shown with their $\pm 1$ and $\pm 2$ standard deviation bands. The mediator width effects are taken into account for every value of the model parameters.}
          \label{fig:sstops:mu_visible_vs_mV}
        \end{center}
\end{figure}

\FloatBarrier

\section{Conclusion}
\label{sec:conclusion}

A search for processes beyond the Standard Model is performed using 36.1~fb$^{-1}$ of proton--proton collisions at
$\sqrt{s} = 13$~\TeV{} recorded by the ATLAS detector at the LHC, based on events with at
least two leptons, including a pair of the same electric charge, at least
one $b$-tagged jet, sizeable missing transverse momentum, and large \HT. Several BSM processes are considered that could enhance the yield of such
events over the small expected background. The search is performed in the
context of BSM models, with signal regions defined for
different models. No significant excess over the expected background is observed.
The regions of parameter space excluded by the data are
quantified by setting limits at 95\% confidence level. The masses of vector-like $T$- and $B$-quarks 
are (expected to be) constrained to $m_T >$ 0.98 \TeV{} (0.99 \TeV{}), $m_B >$ 1.00 \TeV{} (1.01 \TeV{}) assuming
branching ratios of the $W$, $Z$, and $H$ decay modes as predicted by a
singlet model, and the mass of the vector-like $T_{5/3}$ quark is (expected to be) constrained to $m_{T_{5/3}} >$ 1.19 \TeV{} (1.21 \TeV{}) based only on
pair production and assuming a branching ratio $\mathcal{B}(T_{5/3}\to Wt) = 100\%$. With 
single $T_{5/3}$ production included, the observed (expected) lower mass limit is 1.6 \TeV{} (1.7 \TeV{}) for a $T_{5/3}tW$ coupling of $1.0$.
The four-top-quark production cross-section is (expected to be) less than 69~fb 
  (29~fb) assuming SM kinematics and less than 39~fb (21~fb) assuming kinematics from EFT model. The lower limit on the  Kaluza--Klein mass in 
the context of models with two universal extra dimensions, is (expected to be) 1.45 \TeV{} (1.48 \TeV{}).
Finally, limits are set on a dark-matter model based on a flavour changing neutral current producing a pair
of top quarks with the same electric charge. The $uu\to tt$ cross-section is (expected to be) lower than
89~fb (59~fb) for a FCNC mediator mass of $1\,$TeV. Considering a full dark-matter model with a dark-sector coupling $\gDM = 1$, the observed (expected) excluded values for the coupling to SM particles are $\gSM > 0.31$ ($0.28$) for a mediator mass of $m_V=3\,\TeV$, and $\gSM>0.14$ ($0.13$) for $m_V =1\,\TeV $.

\section*{Acknowledgements}


We thank CERN for the very successful operation of the LHC, as well as the
support staff from our institutions without whom ATLAS could not be
operated efficiently.

We acknowledge the support of ANPCyT, Argentina; YerPhI, Armenia; ARC, Australia; BMWFW and FWF, Austria; ANAS, Azerbaijan; SSTC, Belarus; CNPq and FAPESP, Brazil; NSERC, NRC and CFI, Canada; CERN; CONICYT, Chile; CAS, MOST and NSFC, China; COLCIENCIAS, Colombia; MSMT CR, MPO CR and VSC CR, Czech Republic; DNRF and DNSRC, Denmark; IN2P3-CNRS, CEA-DRF/IRFU, France; SRNSFG, Georgia; BMBF, HGF, and MPG, Germany; GSRT, Greece; RGC, Hong Kong SAR, China; ISF, I-CORE and Benoziyo Center, Israel; INFN, Italy; MEXT and JSPS, Japan; CNRST, Morocco; NWO, Netherlands; RCN, Norway; MNiSW and NCN, Poland; FCT, Portugal; MNE/IFA, Romania; MES of Russia and NRC KI, Russian Federation; JINR; MESTD, Serbia; MSSR, Slovakia; ARRS and MIZ\v{S}, Slovenia; DST/NRF, South Africa; MINECO, Spain; SRC and Wallenberg Foundation, Sweden; SERI, SNSF and Cantons of Bern and Geneva, Switzerland; MOST, Taiwan; TAEK, Turkey; STFC, United Kingdom; DOE and NSF, United States of America. In addition, individual groups and members have received support from BCKDF, the Canada Council, CANARIE, CRC, Compute Canada, FQRNT, and the Ontario Innovation Trust, Canada; EPLANET, ERC, ERDF, FP7, Horizon 2020 and Marie Sk{\l}odowska-Curie Actions, European Union; Investissements d'Avenir Labex and Idex, ANR, R{\'e}gion Auvergne and Fondation Partager le Savoir, France; DFG and AvH Foundation, Germany; Herakleitos, Thales and Aristeia programmes co-financed by EU-ESF and the Greek NSRF; BSF, GIF and Minerva, Israel; BRF, Norway; CERCA Programme Generalitat de Catalunya, Generalitat Valenciana, Spain; the Royal Society and Leverhulme Trust, United Kingdom.

The crucial computing support from all WLCG partners is acknowledged gratefully, in particular from CERN, the ATLAS Tier-1 facilities at TRIUMF (Canada), NDGF (Denmark, Norway, Sweden), CC-IN2P3 (France), KIT/GridKA (Germany), INFN-CNAF (Italy), NL-T1 (Netherlands), PIC (Spain), ASGC (Taiwan), RAL (UK) and BNL (USA), the Tier-2 facilities worldwide and large non-WLCG resource providers. Major contributors of computing resources are listed in Ref.~\cite{ATL-GEN-PUB-2016-002}.

\clearpage

\printbibliography


\clearpage 
\input{atlas_authlist}
\clearpage

\end{document}

%% file: EXOT-2016-16-PAPER-metadata.tex
\AtlasTitle{Search for new phenomena in events with same-charge leptons and $b$-jets in $pp$ collisions at $\sqrt{s}= 13~\TeV$ with the ATLAS detector}

\author{The ATLAS Collaboration}

\AtlasRefCode{EXOT-2016-16}

\AtlasNote{EXOT-2016-16}

 \PreprintIdNumber{CERN-EP-2018-171}

 \AtlasJournalRef{JHEP 12 (2018) 039}
 \AtlasDOI{10.1007/JHEP12(2018)039}

\AtlasAbstract{%
   A search for new phenomena in events with two same-charge leptons or three leptons and jets identified as originating from $b$-quarks in a data sample of 36.1~\ifb\ of $pp$ collisions at $\sqrt{s}= 13~\TeV$ recorded by the ATLAS detector at the Large Hadron Collider is reported. No significant excess is found and limits are set on vector-like quark, four-top-quark, and same-sign top-quark pair production. The observed (expected) 95\% CL mass limits for a vector-like $T$- and $B$-quark singlet are $m_T > 0.98~(0.99)$~\TeV{} and $m_B > 1.00~(1.01)$~\TeV{} respectively. Limits on the production of the vector-like $T_{5/3}$-quark are also derived considering both pair and single production; in the former case the lower limit on the mass of the $T_{5/3}$-quark is (expected to be) 1.19~(1.21) \TeV{}. The Standard Model four-top-quark production cross-section upper limit is (expected to be) $69$ ($29$)~fb. Constraints are also set on exotic  four-top-quark production models.  Finally, limits are set on same-sign top-quark pair production. The upper limit on $uu \to tt$ production is (expected to be) 89 (59) fb for a mediator mass of 1~\TeV, and a dark-matter interpretation is also derived, excluding a mediator of 3~\TeV\ with a dark-sector coupling of 1.0 and a coupling to ordinary matter above 0.31. 

}

\AtlasCoverSupportingNote{Search for new physics using events with $b$-jets and a pair of same-charge leptons in $pp$ collisions from 2015-16 at $\sqrt{s}=13$~\TeV{} with the ATLAS detector}{https://cds.cern.ch/record/2224335}
\AtlasCoverCommentsDeadline{21 June 2018}

 \AtlasCoverAnalysisTeam{Simon Berlendis, Thibault Cheval\'erias, Tzu-tien Chung, Frederc Deliot, Cecile Deterre,  Sarah Jones, Romain Kukla, Matt LeBlanc, Romain Madar, Jean-Pierre Meyer, Jianming Qian, Sergey Senkin, Erich Varnes, Zhi Zheng} 

 \AtlasCoverEdBoardMember{Sergey Burdin}
 \AtlasCoverEdBoardMember{Ximo Poveda~(chair)}
 \AtlasCoverEdBoardMember{Giuseppe Salamanna}
 
 \AtlasCoverEgroupEditors{atlas-exot-2016-16-editors@cern.ch}

 \AtlasCoverEgroupEdBoard{atlas-exot-2016-16-editorial-board@cern.ch}

%% file: atlas_authlist.tex
 
\begin{flushleft}
{\Large The ATLAS Collaboration}

\bigskip

M.~Aaboud$^\textrm{\scriptsize 34d}$,    
G.~Aad$^\textrm{\scriptsize 99}$,    
B.~Abbott$^\textrm{\scriptsize 125}$,    
O.~Abdinov$^\textrm{\scriptsize 13,*}$,    
B.~Abeloos$^\textrm{\scriptsize 129}$,    
D.K.~Abhayasinghe$^\textrm{\scriptsize 91}$,    
S.H.~Abidi$^\textrm{\scriptsize 164}$,    
O.S.~AbouZeid$^\textrm{\scriptsize 39}$,    
N.L.~Abraham$^\textrm{\scriptsize 153}$,    
H.~Abramowicz$^\textrm{\scriptsize 158}$
% ...
\bigskip
\\

% ...
$^{13}$Institute of Physics, Azerbaijan Academy of Sciences, Baku; Azerbaijan.\\
% ...
$^{34}$$^{(a)}$Facult\'e des Sciences Ain Chock, R\'eseau Universitaire de Physique des Hautes Energies - Universit\'e Hassan II, Casablanca;$^{(b)}$Centre National de l'Energie des Sciences Techniques Nucleaires (CNESTEN), Rabat;$^{(c)}$Facult\'e des Sciences Semlalia, Universit\'e Cadi Ayyad, LPHEA-Marrakech;$^{(d)}$Facult\'e des Sciences, Universit\'e Mohamed Premier and LPTPM, Oujda;$^{(e)}$Facult\'e des sciences, Universit\'e Mohammed V, Rabat; Morocco.\\
% ...
$^{39}$Niels Bohr Institute, University of Copenhagen, Copenhagen; Denmark.\\
% ...
$^{91}$Department of Physics, Royal Holloway University of London, Egham; United Kingdom.\\
% ...
$^{99}$CPPM, Aix-Marseille Universit\'e, CNRS/IN2P3, Marseille; France.\\
% ...
$^{125}$Homer L. Dodge Department of Physics and Astronomy, University of Oklahoma, Norman OK; United States of America.\\
% ...
$^{129}$LAL, Universit\'e Paris-Sud, CNRS/IN2P3, Universit\'e Paris-Saclay, Orsay; France.\\
% ...
$^{153}$Department of Physics and Astronomy, University of Sussex, Brighton; United Kingdom.\\
% ...
$^{158}$Raymond and Beverly Sackler School of Physics and Astronomy, Tel Aviv University, Tel Aviv; Israel.\\
% ...
$^{164}$Department of Physics, University of Toronto, Toronto ON; Canada.\\
% ...
$^{a}$ Also at Borough of Manhattan Community College, City University of New York, NY; United States of America.\\
$^{b}$ Also at California State University, East Bay; United States of America.\\
$^{c}$ Also at Centre for High Performance Computing, CSIR Campus, Rosebank, Cape Town; South Africa.\\
$^{d}$ Also at CERN, Geneva; Switzerland.\\
$^{e}$ Also at CPPM, Aix-Marseille Universit\'e, CNRS/IN2P3, Marseille; France.\\
% ...
$^{*}$ Deceased

\end{flushleft}
